\documentclass[12pt]{article}

\usepackage[]{inputenc}
\usepackage[T1]{fontenc}
\usepackage{hyperref}
\usepackage{color}
\usepackage{lmodern}
\usepackage[toc,page]{appendix}
\usepackage{fullpage}
\usepackage{authblk}
\usepackage{graphicx}

\usepackage[font=small,labelfont=bf]{caption}

\usepackage{tikz}

\newtheorem{lemma}{Lemma}
\newtheorem{definition}{Definition}
\newtheorem{proposition}{Proposition}

\newcommand{\be}{\begin{equation}}
\newcommand{\ee}{\end{equation}}
\newcommand{\bea}{\begin{eqnarray}}
\newcommand{\eea}{\end{eqnarray}}

\newcommand{\cN}{\mathcal N}

\newcommand{\cS}{\mathcal S}
\newcommand{\cC}{\mathcal C}
\newcommand{\cH}{\mathcal H}

\newcommand{\cW}{\mathcal W}
\newcommand{\cA}{\mathcal A}

\usepackage{amsmath}
\allowdisplaybreaks
\usepackage{amssymb}

\author{Maxime Gadioux and Harvey S. Reall}
\affil{\small Department of Applied Mathematics and Theoretical Physics, University of Cambridge, Wilberforce Road, Cambridge CB3 0WA, United Kingdom \\ mjhg2@cam.ac.uk, hsr1000@cam.ac.uk}

\title{Creases, corners and caustics: properties of non-smooth structures on black hole horizons}

\begin{document}

\maketitle

\begin{abstract}

The event horizon of a dynamical black hole is generically a non-smooth hypersurface. We classify the types of non-smooth structure that can arise on a horizon that is smooth at late time. The classification includes creases, corners and caustic points. We prove that creases and corners form spacelike submanifolds of dimension $2,1$ and that caustic points form a set of dimension at most $1$.
We classify ``perestroikas'' of these structures, in which they undergo a qualitative change at an instant of time. A crease perestroika gives an exact local description of the event horizon near the ``instant of merger'' of a generic black hole merger. Other crease perestroikas describe horizon nucleation or collapse of a hole in a toroidal horizon. Caustic perestroikas, in which a pair of caustic points either nucleate or annihilate, provide a mechanism for creases to decay. We argue that properties of quantum entanglement entropy suggest that creases might contribute to black hole entropy. We explain that a ``Gauss-Bonnet'' term in the entropy is non-topological on a non-smooth horizon, which invalidates previous arguments against such a term.

\end{abstract}

\section{Introduction}

\label{sec:intro}

Consider a smooth spacetime containing a black hole. Only in special circumstances, such as a stationary spacetime, is the event horizon $\cH$ smooth. In general, $\cH$ is a hypersurface that is continuous \cite{Hawking:1973uf} but not everywhere differentiable.  It is non-differentiable at $p$ iff $p$ is an endpoint of at least two horizon generators \cite{Beem:1997uv}. The set of such points is called the {\it crease set}. There exist examples for which the crease set is very complicated \cite{Chrusciel:1996tw}. However, in various simple examples of black hole formation or merger \cite{Hughes:1994ea,Shapiro:1995rr,Lehner:1998gu,Husa:1999nm,Hamerly:2010cr,Cohen:2011cf,Emparan:2016ylg,Bohn:2016soe,Emparan:2017vyp}, it is found that the crease set has a simple structure. In the examples of non-axisymmetric black hole mergers discussed in \cite{Husa:1999nm,Cohen:2011cf,Bohn:2016soe,Emparan:2017vyp}, the crease set consists of a 2-dimensional submanifold of points at which exactly two generators enter the horizon. The boundary of this submanifold is a 1-dimensional set of caustic points. In non-generic examples, the crease set degenerates; to a line in the case of an axisymmetric black hole merger \cite{Lehner:1998gu,Hamerly:2010cr,Emparan:2016ylg}, or to a point in the case of spherically symmetric gravitational collapse.

The first aim of this paper is to prove that certain properties of the crease set in these examples extend to a much wider class of spacetimes, i.e., to identify conditions satisfied by these examples which lead to a fairly simple structure for the crease set. In Section \ref{sec:gen}, we shall review rigorous results concerning properties of the endpoint set $\cH_{\rm end}$ of horizon generators. We shall then add two assumptions that hold for the examples just discussed. Specifically, we shall assume that spacetime is globally hyperbolic, and that $\cH$ is ``smooth at late time''. The latter means that there exists a Cauchy surface $\Sigma$ to the future of $\cH_{\rm end}$ such that $\cH$ is smooth in a neighbourhood of the horizon cross-section $\cH\cap \Sigma$.

\begin{figure}
    \centering
    \includegraphics[width=\textwidth]{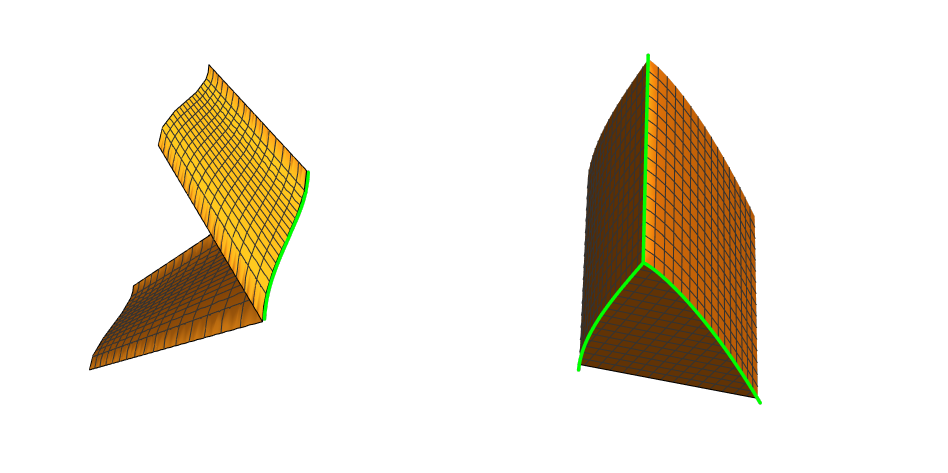}
    \caption{\textit{Left:} Part of a horizon cross-section exhibiting a crease (green). \textit{Right:} Part of a horizon cross-section exhibiting a corner with three creases emanating from it. 
    }
    \label{fig:creaseNcorner}
\end{figure}

We define a {\it normal crease point} to be a non-caustic point of $\cH_{\rm end}$ at which exactly two generators enter $\cH$. We shall show that the set of such points (if non-empty) forms a $2$-dimensional submanifold, the {\it crease submanifold}. At a normal crease point, $\cH$ exhibits a transverse self-intersection so, locally, the crease submanifold resembles the intersection of two null hypersurfaces. On a spatial cross-section of the horizon,  normal crease points form a $1$-dimensional crease at which the horizon looks like a transverse intersection of $2$ surfaces: see Fig. \ref{fig:creaseNcorner}.
This has been seen in various examples. In (non-axisymmetric) black hole mergers, before the merger the two horizons can exhibit ``chisel-like'' structures, with the crease corresponding to the sharp edge of the chisel \cite{Husa:1999nm,Emparan:2017vyp}. After a merger, or in axisymmetric gravitational collapse, the horizon can, in some time-slicings, exhibit a brief period of toroidal (or higher genus) topology. In this case, a crease runs around the inner edge of the hole in the torus \cite{Shapiro:1995rr,Siino:1997ix,Lehner:1998gu,Cohen:2011cf,Bohn:2016soe,Emparan:2017vyp}.

We define a {\it normal corner point} to be a non-caustic point of $\cH_{\rm end}$ at which exactly three generators enter $\cH$. We shall show that the set of such points (if non-empty) forms a $1$-dimensional submanifold, the {\it corner submanifold}. At a normal corner point, $\cH$ exhibits a triple transverse self-intersection and locally resembles the intersection of three null hypersurfaces. A corner on a horizon cross-section is shown in Fig. \ref{fig:creaseNcorner}. Normal corner points are points at which $3$ creases meet, as at a vertex of a tetrahedron or cube.

The set of points of $\cH_{\rm end}$ that are neither normal crease points nor normal corner points consists of (i) caustic points and (ii) non-caustic points at which more than $3$ generators enter $\cH$. We shall prove that this set has (Hausdorff) dimension at most $1$. Thus a generic point of $\cH_{\rm end}$ belongs to the crease submanifold (if non-empty). 

It is natural to focus attention on properties of $\cH_{\rm end}$ that are stable under small perturbations, i.e., properties of $\cH_{\rm end}$ that hold in a {\it generic} spacetime. The results described so far do not assume genericity. However, if one assumes genericity then $\cH_{\rm end}$ exhibits more structure. Siino and Koike used methods of catastrophe theory to classify points of $\cH_{\rm end}$ in a (globally hyperbolic) spacetime, again assuming that $\cH$ is smooth at late time, but now subject to a genericity assumption \cite{Siino:2004xe}. The results of this classification are summarized in Table \ref{SKtable}.
\begin{table}
\begin{center}
\begin{tabular}{ccccc}
& type &  & \# generators & dimension \\
\hline
Non-caustic points & $A_1 $ & regular point& 1  & 3 \\
& $(A_1,A_1)$ & normal crease point &2 & 2 \\
& $(A_1,A_1,A_1)$ & normal corner point &3  & 1 \\
& $(A_1,A_1,A_1,A_1)$ & &4 & 0 \\
\hline
Caustic points & $A_3$ & &1 & 1 \\
& $(A_3,A_1)$ & &2 & 0 
\end{tabular}
\end{center}
\caption{Classification of Siino and Koike \cite{Siino:2004xe} of points on the horizon of a globally hyperbolic 4d black hole spacetime that is smooth at late time, subject to a genericity assumption. The penultimate column indicates how many horizon generators pass through the point. The final column indicates the dimension of the set of points of each type (if non-empty).}
\label{SKtable}
\end{table}
The notation used in this classification is due to Arnol'd \cite{arnold:1976,arnold,arnold_caustics}. The first $4$ rows of the table classify non-caustic points. The first row corresponds to points of $\cH$ that are not endpoints. The next two rows are the normal crease points and normal corner points that we defined above. The fourth row corresponds to a point of quadruple self-intersection of the horizon. Generically such intersections will be transverse and form a set of dimension $0$. (Genericity is important here since in special cases one might have non-transverse quadruple intersections.) 
The final two rows of the table classify caustic points. We shall discuss these in more detail below. All of the endpoints of Table \ref{SKtable} lie in the closure of the crease submanifold (so generically this is non-empty). We emphasize that this work employs a particular mathematical notion of genericity but it is unclear whether this is the same as the physically relevant notion of genericity of the spacetime metric. We shall discuss this point further below. 

In Section \ref{sec:crease_corner} we shall study the time evolution of creases. Given a time function $\tau$ we can foliate spacetime with Cauchy surfaces $\Sigma_\tau$ (level sets of $\tau$). On a horizon cross-section $ \cH \cap \Sigma_\tau$, the qualitative structure of the creases remains unchanged except at special instants of time for which $\Sigma_\tau$ is tangent to the crease submanifold. We refer to such a point of tangency as a {\it pinch point}. A pinch point corresponds to a qualitative change (under time evolution) in the structure of the crease set and hence of $\cH$. Following the terminology of Arnol'd for closely related phenomena arising on wavefronts in flat spacetime \cite{arnold_caustics}, we shall refer to such a change as a crease {\it perestroika}.\footnote{
``Perestroika'' means ``restructuring''.} We emphasize that the definition of a perestroika depends on the choice of a time function; a different choice could shift the location of the pinch point or change its interpretation. 

\begin{figure}
    \centering
    \includegraphics[width=\textwidth]{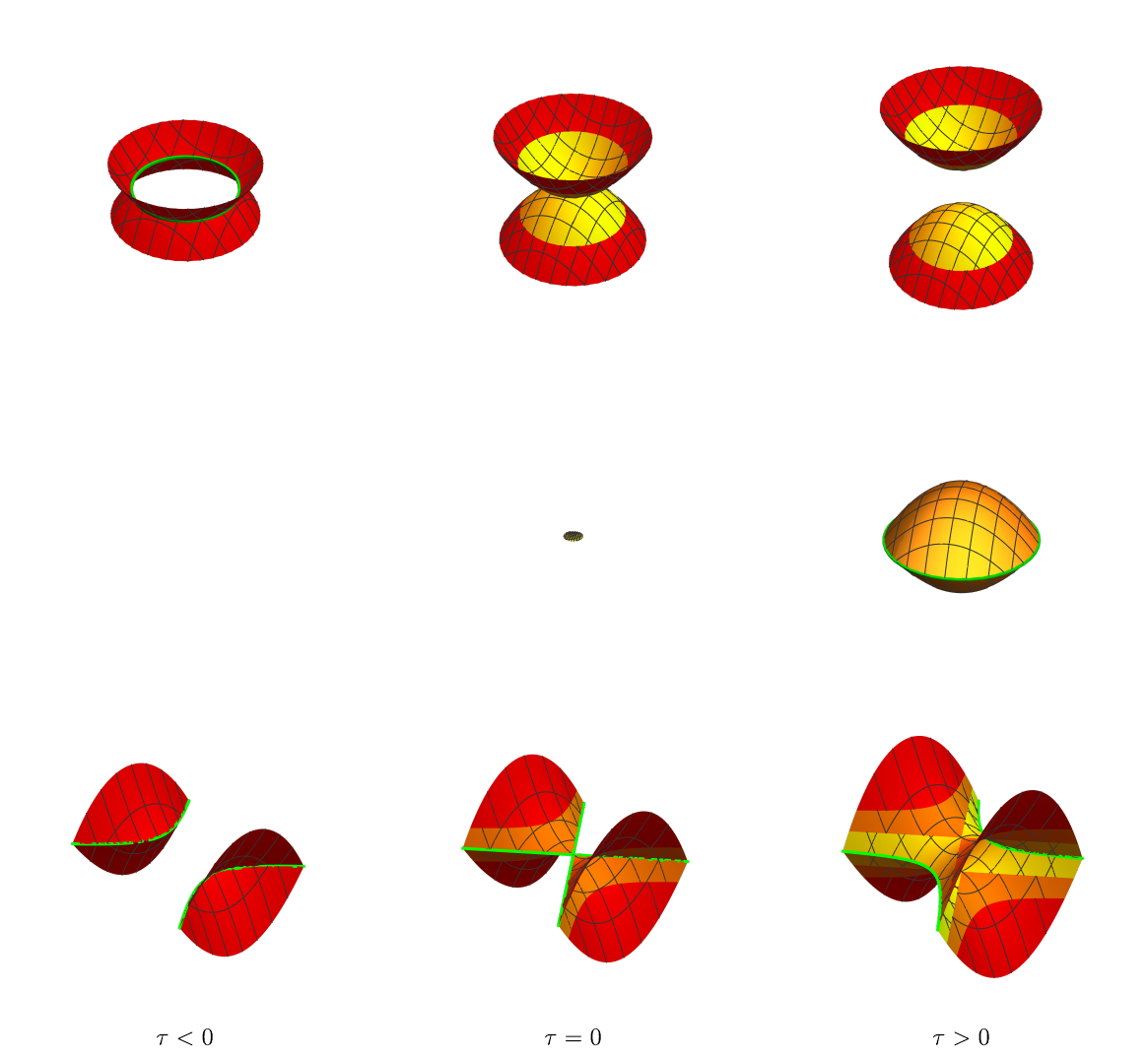}
    \caption{The evolution of the horizon cross-section $\Sigma_\tau \cap \cH$ in the different types of crease perestroika. The time function has been shifted so that the perestroika occurs at $\tau=0$. Creases are highlighted in green. Horizon generators that exist through multiple cross-sections are shown in the same colour. \textit{Top:} Collapse of a hole in the horizon. The black hole region is the exterior of the surface shown on the left and the region between the two surfaces on the right. For $\tau<0$, the horizon has a hole, which closes up as $\tau \to 0$. \textit{Middle:} Nucleation of a component of horizon of spherical topology. The black hole region lies inside the surface shown. At $\tau=0$, the event horizon nucleates and for $\tau>0$ takes the form of a ``flying saucer'', with an elliptical crease around its equator. \textit{Bottom:} Formation of a ``bridge'' between two sections of horizon. For $\tau<0$, there are two (locally) disconnected parts of the horizon, each with a hyperbolic crease. The creases degenerate to a pair of straight lines at $\tau=0$, where the two parts of the horizon merge with sharp tips. For $\tau>0$, the horizon is connected with hyperbolic creases along the edges of the bridge. The black hole region is the interior of the surface shown on the right.}
    \label{fig:pinch_diagrams}
\end{figure}

We shall present an exact local description of the geometry of the horizon around a pinch point associated with a crease perestroika.\footnote{See also  \cite{Brill:2014oya} which describes some of these perestroikas in qualitative terms.} We find that, generically, there are three distinct types of crease perestroika. Examples of these are shown in Fig. \ref{fig:pinch_diagrams}.\footnote{
Because of the teleological nature of an event horizon, it is sometimes helpful to think of these processes in terms of backwards time evolution. However, in order to avoid repetition, we shall only discuss forward time evolution in this paper.
} First, there is a perestroika associated with the ``collapse of a hole in the horizon''. It is well-known that horizons of toroidal (or higher genus) topology can form in gravitational collapse \cite{Hughes:1994ea,Siino:1997ix} or a black hole merger \cite{Bohn:2016soe,Emparan:2017vyp}. These evolve to spherical topology, with the hole in the torus closing up. In such examples, an elliptical crease runs around the inner rim of the hole. The crease perestroika describes the geometry of $\cH$ near the point at which this crease collapses to zero size and the horizon changes topology.  

The second type of crease perestroika describes the nucleation of a topologically spherical component of the horizon, with an elliptical crease running around its rim, so it resembles a ``flying saucer.'' In generic gravitational collapse, this would describe the event horizon at the instant of time at which it first appears (for a generic time function $\tau$). In a black hole merger, flying saucers can nucleate in an intermediate stage, subsequently merging with each other and with the initial black holes. 

The third type of crease perestroika describes the merger of two (locally) disconnected sections of event horizon, for example in a black hole merger. In this case, before the merger each section of horizon exhibits a crease with a hyperbolic shape. At the instant of merger, these creases develop sharp tips and then reconnect so that after the merger there is a ``bridge'' connecting the two sections of horizon, with a crease running along each edge of the bridge. This perestroika provides an exact description of the horizon near the instant of merger of a generic (non-axisymmetric) black hole merger, such as the ones studied in \cite{Husa:1999nm,Bohn:2016soe,Emparan:2017vyp}. 

Similarly to a crease perestroika, a {\it corner perestroika} arises at an instant of time $\tau$ for which $\Sigma_\tau$ is tangent to the corner submanifold. We shall show that there are four types of corner perestroika, each involving either the nucleation, or the annihilation, of a pair of corners. A point of type $(A_1,A_1,A_1,A_1)$ in the classification of \cite{Siino:2004xe} can also be viewed as a perestroika. As above, the nature of this perestroika depends on the choice of time function. The simplest possibility is that such a point describes the nucleation of a component of event horizon of spherical topology, with a tetrahedral arrangement of corners and creases. 

In Section \ref{sec:caustics} we discuss caustic points. The classification of Siino and Koike (Table \ref{SKtable}) contains two types of generic caustic points. As mentioned above, it is unclear whether the notion of genericity/stability used in their work corresponds to the physically relevant notion of stability w.r.t.~perturbations of the metric. We shall give alternative arguments, still based on catastrophe theory, which support their conclusions. We shall highlight the assumptions required to justify these arguments. We shall then go on to study the features of $\cH$ near caustic points of the two generic types according to this classification.  

\begin{figure}
    \centering
    \includegraphics[width=\textwidth]{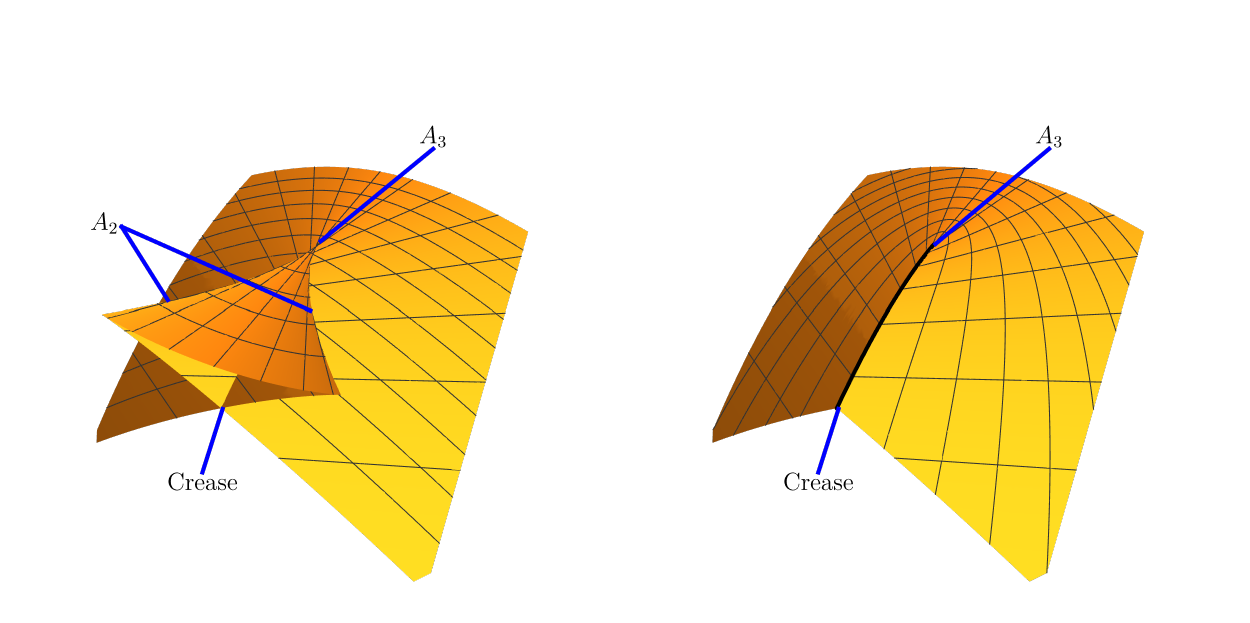}
    \caption{{\it Left}: $A_3$ caustic on a spatial cross-section of the big wavefront obtained by extending generators of $\cH$ beyond their past endpoints. {\it Right}: $A_3$ caustic on a spatial cross-section of $\cH$. 
    }
    \label{fig:A3}
\end{figure}

The first type of generic caustic point, denoted $A_3$, is associated with the famous ``swallowtail'' catastrophe shown on the left in Fig. \ref{fig:A3}. This figure shows an $A_3$ point on a spatial cross-section of the ``big wavefront'' (in the terminology of Arnol'd) obtained by extending the horizon generators beyond their past endpoints as far as possible. In spacetime, $A_3$ points form spacelike lines, and so the intersection with a spacelike hypersurface generically gives isolated $A_3$ points on a cross-section of a wavefront. Emerging from an $A_3$ point on the cross section are two cusp lines, denoted $A_2$ in Arnol'd's notation, and a self-intersection line (i.e., a crease). To obtain a cross-section of $\cH$ from this diagram one must discard the part that corresponds to extending horizon generators beyond their past endpoints (on the crease or $A_3$ point). This gives the diagram on the right of Fig. \ref{fig:A3} where a crease terminates at the $A_3$ point (with the angle at the crease approaching $\pi$ there). Note $A_2$ caustics occur on the big wavefront but not on $\cH$. Siino and Koike do not state a simple reason why $A_2$ caustics are absent in their results. We shall show that an $A_2$ caustic violates achronality and hence cannot occur on $\cH$.

Generically, $A_3$ points form a $1$-dimensional line. We can define an $A_3$ perestroika in the same way as we defined a corner perestroika: it corresponds to an instant of time at which $\Sigma_\tau$ is tangent to the $A_3$ line. In the context of optics, such perestroikas are well-known in the catastrophe theory literature \cite{arnold:1976}. For a horizon cross-section, we shall show that they come in two qualitatively different types. In the first type, a horizon cross-section initially has a section of crease with a pair of $A_3$ endpoints. Under time evolution, the crease shrinks to zero length and the $A_3$ points merge and disappear. In the second type, a horizon cross-section again initially has a section of crease. Under time evolution, an $A_3$ point nucleates on this crease, and immediately splits into two $A_3$ points. These points move apart, ``eating up'' the crease as they go, leaving a smooth section of horizon between the two points. Both perestroikas are processes of ``crease decay'' mediated by $A_3$ points, i.e., they have a smoothing effect on the horizon. 

The non-axisymmetric black hole mergers studied in \cite{Husa:1999nm,Bohn:2016soe,Emparan:2017vyp} exhibit normal crease points and $A_3$ caustics but no other types of endpoint. Given a choice of time foliation, these mergers give rise to a sequence of 
crease and $A_3$ perestroikas of the various types discussed above. These perestroikas might be regarded as the ``elements'', or primitive steps, of a merger. We shall describe this below. 

The second type of generic caustic is denoted $(A_3,A_1)$ and corresponds to a point at which a smooth section of the horizon intersects a line of $A_3$ points transversally. We shall explain how this can describe three different types of perestroika (for different choices of time function) involving a corner and an $A_3$ point.

Section \ref{sec:entropy} is more speculative. We shall discuss whether creases and caustics might play a role in black hole entropy. It has been suggested that at least part of the entropy of a black hole can be attributed to entanglement entropy of quantum fields in the black hole spacetime \cite{Bombelli:1986rw,Srednicki:1993im}. Roughly speaking, a divergence in the entanglement entropy, with coefficient proportional to the horizon area, is absorbed into the Bekenstein-Hawking entropy via a renormalization of Newton's constant \cite{Susskind:1994sm}. It is known that a crease gives rise to a subleading divergence in the entanglement entropy \cite{Klebanov:2012yf,Myers:2012vs}. Combining these ideas suggests that a crease might make a subleading contribution to black hole entropy. Usually such a term would be dominated by the Bekenstein-Hawking term in the entropy. However, using crease perestroikas, we shall show that the second law can be used to constrain this idea. 

We shall also discuss the possibility of a ``Gauss-Bonnet'' term in the entropy. In $4$ spacetime dimensions, a Gauss-Bonnet term in the gravitational action is topological, i.e., it does not affect the equation of motion. However, it does affect black hole entropy, contributing a term proportional to the integral of the Ricci scalar of the induced metric on a horizon cross-section \cite{Jacobson:1993xs,Iyer:1994ys}. For a smooth horizon, this is a topological term, proportional to the Euler number of the cross-section. Since this jumps discontinuously in black hole formation or merger, it has been argued that such a term always leads to a violation of the second law of thermodynamics \cite{Sarkar:2010xp}. However, we shall explain that, for a non-smooth horizon, this term in the entropy is not topological, and instead varies continuously in black hole formation and merger. We find that there is no obvious conflict with the second law if one treats the Gauss-Bonnet term in the sense of effective field theory. 

Finally, we shall discuss the possibility of terms in black hole entropy that are quadratic in the extrinsic curvature of a horizon cross-section. We shall explain why such terms are finite at creases, corners and caustics but, unlike the Gauss-Bonnet term, they diverge at an $A_3$ perestroika and are therefore excluded by finiteness of the entropy in such a process. 

\subsection*{Notation and conventions}

We assume that the spacetime manifold is smooth. We shall sometimes refer to singularities (e.g., ``an $A_3$ singularity''); these are singularities of null hypersurfaces, i.e., caustics, not spacetime singularities. In Section \ref{sec:gen} we shall consider spacetimes of general dimension $d$. We set $d=4$ in Section \ref{sec:crease_corner} onwards. $\cH$ denotes a future horizon, as defined in Section \ref{sec:endpoint_general}. We shall not make use of any equations of motion. $\cW$ denotes the ``big wavefront'' obtained from $\cH$ by extending its generators beyond their past endpoints as far as possible (Section \ref{sec:smooth_late}). If $\Sigma$ is a spacelike Cauchy surface $H \equiv \Sigma \cap \cH$ denotes a cross-section of the horizon and $W\equiv \Sigma \cap \cW$ denotes a ``small wavefront'', i.e., a cross-section of the big wavefront (so $H \subseteq W$). A general time function will be denoted $\tau$ and its level sets as $\Sigma_\tau$, i.e., $\Sigma_{\tau_0}$ is the surface $\tau=\tau_0$. 

The spacetime metric has positive signature. Latin letters $a,b,c,\ldots$ denote abstract tensor indices. Greek letters $\mu,\nu,\rho,\ldots$ are tensor indices referring to a particular basis. 

\section{General results}

\label{sec:gen}

\subsection{Properties of endpoint set}

\label{sec:endpoint_general}

In this section we shall review properties of the endpoint set of an event horizon. We assume that we have a smooth time-oriented spacetime and make the following definitions \cite{Chrusciel:2000cu,Chrusciel:2000gj}:

\begin{definition}
An embedded hypersurface $\cH$ is {\rm future null geodesically ruled} if every $p \in \cH$ belongs to a future-inextendible null geodesic $\Gamma \subset \cH$. Such geodesics are the {\rm generators of $\cH$}. A {\rm future horizon} is an achronal, closed, future null geodesically ruled topological hypersurface. 
\end{definition}

A black hole future event horizon is an example of a future horizon. Another example is a past Cauchy horizon. By applying time reversal one can define a past horizon, which includes a black hole past event horizon or a future Cauchy horizon.

It follows from the definition that generators cannot have future endpoints. (If $p$ were a future endpoint of $\Gamma$, it must belong to $\cH$ since $\cH$ is closed. A generator $\Gamma'$ through $p$ cannot be the extension of $\Gamma$ since $\Gamma$ is inextendible. Therefore we can join $\Gamma$ to $\Gamma'$ and ``round off the corner'' to construct a timelike curve between two points of $\cH$, violating achronality.) 

Let $\cH_{\rm end}\subset \cH$ denote the set of (past) endpoints of generators of $\cH$. For $p \in \cH$ let $N(p)$ be the number of generators through $p$ (which might be $\infty$). Then $\cH$ is differentiable at $p$ iff $N(p)=1$ \cite{Beem:1997uv}. Points with $N(p)>1$ must belong to $\cH_{\rm end}$ \cite{Beem:1997uv} but there may also be points of $\cH_{\rm end}$ with $N(p)=1$.

\begin{definition}
The {\rm crease set} is the set of $p \in \cH_{\rm end}$ with $N(p)>1$, i.e., the set of points at which $\cH$ is non-differentiable.
\end{definition}

We shall now briefly review results of Chru\'sciel {\it et al} \cite{Chrusciel:2000gj} concerning the structure of the crease set. 
Let $\sigma$ be a Riemannian metric and, for $p \in \cH$, let $\cN_p^+$ be the set of future-pointing $\sigma$-unit vectors tangent to a generator of $\cH$ at $p$. The number of such vectors is $N(p)$. Define $\cC_p$ to be the convex cone generated by $\cN_p^+$, i.e., the set $\{\sum_i a_i V_i: a_i \ge 0, V_i \in \cN_p^+ \}$. Now for $k =1 , \ldots, d$ define
\begin{equation}
 \cH[k] = \{ p \in \cH: {\rm dim}(\cC_p) \ge k \}.
\end{equation}
This gives $\cH[1]=\cH$. $\cH[2]$ is the set of points lying on more than $1$ generator, i.e., the crease set. Clearly $\cH[1] \supseteq \cH[2] \supseteq \cH[3] \ldots$ and since ${\rm dim}(\cC_p) \le N(p)$ we also have
\begin{equation}
 \cH[k] \subseteq \{ p \in \cH : N(p) \ge k \}.
\end{equation}
A simple argument \cite{Chrusciel:2000gj} gives 
\begin{equation}
\label{Hkeq}
 \cH[k] = \{ p \in \cH : N(p) \ge k \} \,\, {\rm for} \, \, k=1,2,3.
\end{equation}
Chru\'sciel {\it et al.}~prove that, for $1 \le k \le d$, $\cH[k]$ can be covered, up to a set of zero $(d-k)$-dimensional Hausdorff measure, by a countable collection of $(d-k)$-dimensional $C^2$ submanifolds of $M$. In particular, $\cH[k]$ has dimension at most $d-k$ and $\cH[d]$ is a countable set. 

This result gives some understanding of the size of the set of endpoints with $N(p)>1$. We now discuss endpoints with $N(p)=1$. It can be shown that the following are equivalent \cite{Beem:1997uv}: (1) $\cH$ is differentiable on an open set; (2) $\cH$ is of class $C^r$ on this open set for some $r \ge 1$; (3) this set does not contain any endpoints. It follows that any neighbourhood of an endpoint with $N(p)=1$ must contain an endpoint with $N(p)>1$, i.e., an endpoint with $N(p)=1$ is a limit point of a sequence of endpoints with $N(p)>1$. Furthermore, it can be shown that the set of endpoints with $N(p)=1$ has vanishing $(d-1)$-dimensional Hausdorff measure \cite{Chrusciel:2000gj}. 

We emphasize that the above results follow only from the definition of future horizon given above with no further assumptions. The weakness of these assumptions permits examples exhibiting seemingly pathological behaviour, such as spacetimes for which $\cH_{\rm end}$ is a dense subset of $\cH$ \cite{Chrusciel:1996tw} (of zero measure). The authors of \cite{Chrusciel:1996tw} emphasize that these examples are very artificial and that one would hope that this behaviour cannot occur for event horizons in ``reasonable'' asymptotically flat spacetimes. In other words, by adding extra conditions, such as asymptotic flatness, or global hyperbolicity, one might expect the structure of $\cH_{\rm end}$ to be significantly simpler than the most general possibility discussed above. In the next section we shall introduce further assumptions that result in a much nicer structure for $\cH_{\rm end}$.

\subsection{Horizons smooth at late time}

\label{sec:smooth_late}

We shall study the endpoint set $\cH_{\rm end}$ subject to two further assumptions which are satisfied in the examples discussed in the Introduction:

\medskip

\noindent {\bf Assumptions} {\it (1) Spacetime is globally hyperbolic. (2) There exists a connected future horizon $\cH$ and a smooth spacelike Cauchy surface $\Sigma_\star$ lying to the future of $\cH_{\rm end}$, such that the horizon cross-section $H_\star \equiv \Sigma_\star \cap \cH$ is a smooth oriented compact connected submanifold and $\cH$ is smooth in a neighbourhood of $H_\star$. }

\medskip

Regarding (2), if there are multiple black holes present at arbitrarily late time then the event horizon will be disconnected. In this case we simply define $\cH$ to be a single connected component of the event horizon, corresponding to a single black hole at late time. The smoothness assumption in (2) is made for simplicity; this assumption could be replaced by $C^k$ for sufficiently large $k$ (although see comments after Proposition \ref{caustic_prop} below). Smoothness of $\cH$ certainly fails at $\cH_{\rm end}$. Assumption (2) captures the idea that the horizon is ``smooth at late time'', which is expected to hold in physically relevant situations. For example, consider a black hole formed in gravitational collapse, or through a black hole merger. At late time, it is expected that the black hole will be well-described by a perturbed Kerr black hole. For the simpler case of a (nonlinearly) perturbed Schwarzschild black hole, the smoothness of $\cH$ is related to the smoothness of the perturbation and its behaviour at infinity, as described in \cite{Dafermos:2021cbw}. Similar results are expected for Kerr. These results demonstrate that there is a large class of physically relevant spacetimes for which the horizon is smooth (enough) at late time.

We shall introduce some more terminology for the different types of points in $\cH_{\rm end}$:

\begin{definition}
$p \in \cH_{\rm end}$ is a {\rm caustic point} if $p$ is a focal point of $H_\star$ along a generator of $\cH$. 
\end{definition}
Focal points are defined in \cite{oneill} or \cite{Hawking:1973uf} (where they are called conjugate points). This definition is independent of the choice of $H_\star$, i.e., if one chooses a different $H_\star$ satisfying the assumptions above then the definition of a caustic point doesn't change. Note that if $p$ is a caustic point with $N(p)>1$ (i.e. $p$ belongs to the crease set) then there might be a generator along which $p$ is not a focal point of $H_\star$. We shall prove the following below:

\begin{lemma} 
\label{lem:creaseorcaustic}
Subject to the above assumptions, $\cH_{\rm end}$ is closed and if $p \in \cH_{\rm end}$ then either $p$ is a caustic point or $p$ is a crease point (or both).  
\end{lemma}
In particular, an endpoint with $N(p)=1$ must be a caustic point. It is convenient to refine the classification of non-caustic points as follows:

\begin{definition}
$p$ is a {\rm normal crease point} if it is a non-caustic point with $N(p)=2$. $p$ is a {\rm normal corner point} if it is a non-caustic point with $N(p)=3$. 
\end{definition}
Inspired by results for Riemannian manifolds \cite{itoh_tanaka} and flat spacetime arguments \cite{Husa:1999nm} we shall prove

\begin{figure}
    \centering
    \includegraphics[width=0.5\textwidth]{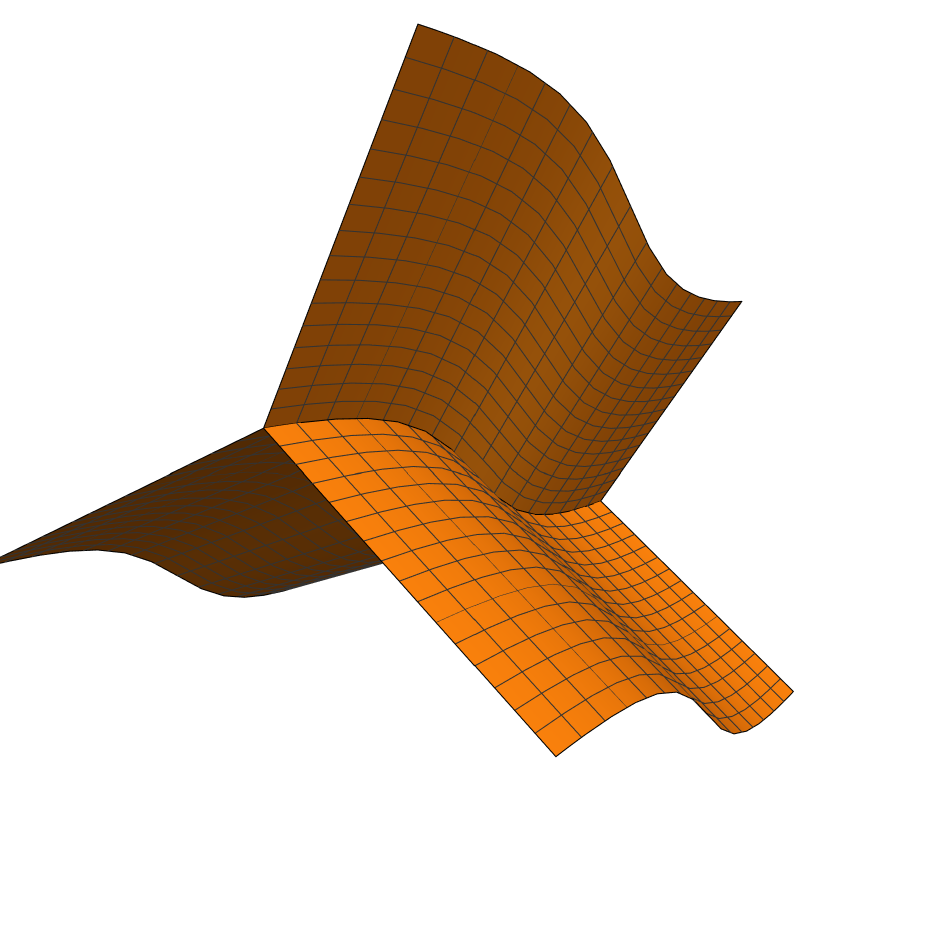}
    \caption{Three components of the crease submanifold meeting at the corner submanifold.}
    \label{fig:corner}
\end{figure}

\begin{proposition} 
\label{normal_prop}
The set of normal crease points (if non-empty) is a smooth spacelike submanifold of dimension $d-2$: the {\rm crease submanifold}. The set of normal corner points (if non-empty) is a smooth spacelike submanifold of dimension $d-3$: the {\rm corner submanifold}. 
\end{proposition}
The intuition behind this result is that near a normal crease (corner) point, $\cH$ looks like a transverse self-intersection of $2$ ($3$) smooth null hypersurfaces. (The result does not generalize to non-caustic points with $N(p)=4$ because transversality might fail, see the comments after the proof of Proposition \ref{normal_prop} below.) These submanifolds might not be connected. If the corner submanifold is non-trivial then it forms part of the closure of the crease submanifold, where three components of the crease submanifold meet, as shown in Fig.~\ref{fig:corner} for $d=4$.

Endpoints not covered by this proposition either have $N(p)>3$ or they are caustic points (or both). From equation \eqref{Hkeq}, the set of endpoints with $N(p)>3$ is a subset of $\cH[3]$ and therefore has Hausdorff dimension at most $d-3$ by the results of \cite{Chrusciel:2000gj} reviewed above (this is true even without the assumptions introduced above). We shall adapt a result from Riemannian geometry \cite{itoh_tanaka} to show that, with the above assumptions, the set of caustic points has the same property:

\begin{proposition}
\label{caustic_prop}
The set of caustic points has Hausdorff dimension at most $(d-3)$.
\end{proposition}

This proposition uses the smoothness of $H_\star$. If $H_\star$ is only $C^k$ then the set of caustic points can have larger Hausdorff dimension. We discuss this briefly after the proof below.

To prove these Propositions, and for later use, we introduce the big and small wavefronts (the terminology is due to Arnol'd \cite{arnold:1976}) which are defined as follows:

\begin{definition}
\label{def:bigwavefront}
The {\rm big wavefront} $\cW$ is the union of the generators of $\cH$, extended as far as possible to the past. A {\rm small wavefront} is an intersection $\Sigma \cap \cW$ where $\Sigma$ is a spacelike Cauchy surface.
\end{definition}

We can pick a smooth past-directed null vector $\ell^a$ on $H_\star$, orthogonal to $H_\star$, such that $-\ell^a$ is everywhere tangent to the generators of $\cH$. $\cW$ is the union of the future and past inextendible null geodesics through $H_\star$ with tangent vector $\ell^a$ on $H_\star$. To the future of $H_\star$, these geodesics coincide with generators of $\cH$. However, to the past of $H_\star$ these generators may have endpoints, in which case $\cW$ corresponds to extending the generators of $\cH$ (as null geodesics) to the past, beyond their past endpoints. Clearly $\cH \subseteq \cW$. More generally, a big wavefront can be defined this way for any smooth orientable codimension-$2$ spacelike submanifold $H_\star$, irrespective of the connection with horizons. 

We define a smooth map $\Phi:\mathbb{R} \times H_\star \rightarrow M$ (where $M$ is the spacetime manifold) as follows. Let $\Phi(\lambda,u)$ be the point affine parameter distance $\lambda$ along the null geodesic starting at the point $u \in H_\star$ with tangent vector $\ell^a$ there. The big wavefront is the image of this map. In a neighbourhood of $H_\star$ this map defines an embedding, i.e., the part of $\cW$ with small $\lambda$ is a smooth submanifold. However, for larger $\lambda$, $\cW$ may exhibit singularities.\footnote{
We emphasize that these wavefront singularities occur in a smooth spacetime, they are unrelated to spacetime singularities.}
For small $\lambda$, the smooth map $\Phi$ is non-singular, i.e., its derivative $d\Phi$ has maximal rank $d-1$. However, there may exist $(\lambda_0,u_0)$ such that $\Phi$ is singular at $(\lambda_0,u_0)$, i.e., $d\Phi$ has rank less than $d-1$. This happens iff $p\equiv\Phi(\lambda_0,u_0)$ is a focal point of $H_\star$ along the null geodesic through $u_0$, i.e., $p$ is a caustic point. The non-singular condition on $d\Phi$ is precisely the condition that $\Phi$ is an immersion. Thus at a caustic point, $\cW$ fails to be an immersed submanifold.  

To prove Lemma \ref{lem:creaseorcaustic}, we shall use the ``null cut locus'' of $H_\star$. This is defined as follows \cite{kemp,kupeli}:

\begin{definition}
Let $\gamma:[0,a) \rightarrow M$ be a null geodesic starting on $H_\star$ and orthogonal to $H_\star$. $\gamma(t_0)$ is a {\rm null cut point of $H_\star$ along $\gamma$} iff for $0 \le t \le t_0$ there does not exist a timelike curve from $H_\star$ to $\gamma(t)$ whereas for $t>t_0$ there does exist such a curve. The {\rm past null cut locus} of $H_\star$ is the set of null cut points along all such past-directed geodesics.
\end{definition}

This is of interest because: 
\begin{lemma}
\label{lem:Hendcutlocus}
$\cH_{\rm end}$ is the set of null cut points of $H_\star$ along the generators of $\cW$. 
\end{lemma}
(There are two families of past-directed null geodesics emanating orthogonally from $H_\star$. Only one of these is $\cW$. So $\cH_{\rm end}$ is not the past null cut locus of $H_\star$ but only a subset of it.)

{\it Proof.} Let $p \in \cH_{\rm end}$ and consider a (past-directed) null geodesic of $\cW$ that passes through $p$. (If $N(p)>1$ then there is more than one such geodesic.) Let $u\in H_\star$ label this geodesic, i.e., $p = \Phi(\lambda_p,u)$ for some $\lambda_p>0$. Consider a point $q$ slightly beyond $p$ along this geodesic. We claim that there is a (past-directed) timelike curve from $H_\star$ to $q$. We justify this as follows. One can introduce normal coordinates at $p$ such that $\cH$ is the surface $x^0 = F(x^i)$ ($i=1,\ldots, d-1$) where $F$ is a Lipschitz continuous function \cite{Hawking:1973uf}. The point $q$ has $x^0 \ne F(x^i)$. Now follow the integral curve of $\partial/\partial x^0$ from $q$ to return to a point $r$ on $\cH$. $qr$ cannot be past-directed because then $pqr$ would be a past directed causal curve from $\cH$ to itself and since this curve is not a null geodesic it can be deformed into a timelike curve, violating achronality. Therefore $qr$ is future-directed. If we now extend $qr$ by attaching it to a future directed generator of $\cH$ through $r$ we obtain a causal curve from $q$ to $H_\star$, which is not a null geodesic so can be deformed into a future-directed timelike curve, establishing our claim. Hence $q$ lies beyond the null cut point on our original geodesic, i.e., $\lambda_q > \lambda_0(u)$ where $\lambda_0(u)$ is the affine parameter of the null cut point on this geodesic. This holds for all $q$ lying beyond $p$ along this geodesic so we must have $\lambda_p \ge \lambda_0(u)$. However, if $\lambda_p > \lambda_0(u)$ then (by the definition of $\lambda_0$) there exists a timelike curve from $H_\star$ to $p$, violating achronality of $\cH$. Therefore we must have $\lambda_p = \lambda_0(u)$, so $p$ is the cut point that lies on this geodesic. This shows that $\cH_{\rm end}$ is a subset of the set of null cut points of $H_\star$ along the generators of $\cW$. Conversely, let $p$ be a point in the latter set, arising from a null geodesic starting at $u \in H_\star$. Then points beyond $p$ along this geodesic are timelike separated from $H_\star$ and so must lie beyond an endpoint $q \in \cH_{\rm end}$. The above argument then shows that $q$ is a null cut point of $H_\star$ along this geodesic, and so we must have $p=q$ as each geodesic has at most one null cut point. This shows that the set of null cut points of $H_\star$ along generators of $\cW$ is a subset of $\cH_{\rm end}$, completing the proof. 

\medskip

{\it Proof of Lemma \ref{lem:creaseorcaustic}}. 
This follows from Lemma \ref{lem:Hendcutlocus} and 
properties of null cut points in globally hyperbolic spacetimes proved in \cite{kemp,kupeli}. Theorem 6.2 of \cite{kemp} or Theorem 4 of \cite{kupeli} assert that if $p$ is a null cut point of $H_\star$ along a past-directed null geodesic orthogonal to $H_\star$ then either (1) $p$ is a focal point of $H_\star$ along this geodesic; or (2) there exist at least two null geodesic segments from $H_\star$ to $p$, both orthogonal to $H_\star$. (Possibly both are true.)
We can apply this to $p \in \cH_{\rm end}$, since Lemma \ref{lem:Hendcutlocus} tells us that $p$ is a null cut point along a generator of $\cW$. In (2) we just need to check that the null geodesics from $H_\star$ to $p$ are generators of $\cW$, rather than belonging to the ``other'' family of past-directed null geodesics emanating orthogonally to $H_\star$. In the latter case, we would have a future-directed null geodesic from $p$ to $q\in H_\star$ that is not a generator of $\cH$. We could extend this to the future by following the generator of $\cH$ through $q$ to reach $r \in \cH$. This gives a causal curve from $p$ to $r$ that is not a null geodesic, so can be deformed into a timelike curve, violating achronality of $\cH$. Hence all the geodesics in (2) must be generators of $\cW$ and hence $N(p)>1$ in this case, i.e., $p$ is a crease point. So either $p$ is a focal point of $H_\star$ along a generator of $\cH$ or $p$ is a crease point. 

Theorem 6 of \cite{kupeli} asserts that the past null cut locus of $H_\star$ is closed. So if $q$ is a limit point of a sequence $p_n \in \cH_{\rm end}$ then $q$ is a null cut point of $H_\star$ along some past-directed null geodesic $\gamma$. The argument of the previous paragraph establishes that $\gamma$ cannot belong to the ``other'' family of null geodesics from $H_\star$, so $\gamma$ must be a generator of $\cW$ and hence $q \in \cH_{\rm end}$ by Lemma \ref{lem:creaseorcaustic}. Therefore $\cH_{\rm end}$ is closed.

\medskip

\begin{figure}
    \centering
    \includegraphics{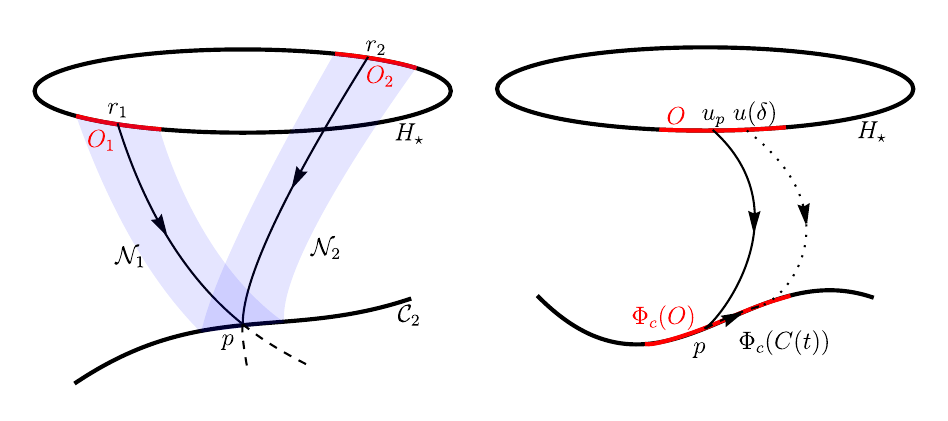}
    \caption{\textit{Left:} Setup for the proof of Proposition~\ref{normal_prop} in the case of a normal crease point. Note that for clarity the null hypersurfaces $\cN_A$ have been drawn up to the crease submanifold $\cC_2$ but they remain smooth in an open set slightly beyond $\cC_2$. \textit{Right:} Setup for the proof of Proposition~\ref{caustic_prop}.}
    \label{fig:proofsv2}
\end{figure}

{\it Proof of Proposition \ref{normal_prop}.} Let $p$ be a normal crease point. There exist exactly two null geodesics $\gamma_1, \gamma_2$ from $H_\star$ to $p$, both belonging to $\cW$, starting at distinct points $r_1,r_2 \in H_\star$. Consider $\gamma_1$. We have $p = \Phi(\lambda_{p1},r_1)$ for some $\lambda_{p1}>0$. There cannot be a focal point of $H_\star$ along $\gamma_1$ for $\lambda \le \lambda_{p1}$ so $\Phi(\lambda,r_1)$ has maximal rank for $\lambda \in [0,\lambda_{p1}]$. By continuity, there exist $\lambda_1>\lambda_{p1}$ and an open neighbourhood $O_1$ of $r_1$ in $H_\star$ such that $\Phi$ has maximal rank on $(0,\lambda_1) \times O_1$. The image of this set under $\Phi$ is an immersed null submanifold $\cN_1 \subset \cW$. By shrinking $O_1$ if necessary we can ensure that $\cN_1$ has no self-intersection, so it is a smooth embedded null hypersurface. The same construction starting from $\gamma_2$ yields a second null hypersurface $\cN_2 \subset \cW$ and, by shrinking $O_1$ and $O_2$ we can arrange that $O_1$ and $O_2$ are disjoint so $\cN_1$ and $\cN_2$ have no generators in common, as shown in the first diagram of Fig.~\ref{fig:proofsv2}. Clearly $p \in \cN_1 \cap \cN_2$. 
We now claim that there exists a neighbourhood $U$ of $p$ such that (a) every $q \in U \cap \cH_{\rm end}$ is a normal crease point; and (b) $U \cap \cH_{\rm end} = U \cap \cN_1 \cap \cN_2$. 

To establish (a), assume the contrary: then there exists a sequence of points $q_n \in \cH_{\rm end}$ with $q_n \rightarrow p$ such that each $q_n$ is not a normal crease point. The properties of $\cN_{1,2}$ imply that there exists $r_n \in H_\star$, $r_n\notin O_1 \cup O_2$ such that there is a null geodesic $\delta_n$ from $r_n$ to $q_n$. (For if $q_n$ is a caustic point then the generator along which $q_n$ is a focal point of $H_\star$ cannot belong to $\cN_{1,2}$ hence its start point must lie outside $O_{1,2}$. If $q_n$ is a non-caustic point then there are at least $3$ null geodesics from $H_\star$ to $q_n$ but at most $2$ of these can belong to $\cN_{1,2}$ so the third must have a start point outside $O_{1,2}$.) Since $q_n \rightarrow p$, the curves $\delta_n$ admit a causal limit curve $\delta$ from $H_\star$ to $p$ \cite{Hawking:1973uf}. This must be a null geodesic orthogonal to $H_\star$ (for otherwise we could deform it into a timelike curve). The start point of $\delta$ lies outside $O_{1,2}$ so $\delta$ differs from $\gamma_{1,2}$ which contradicts $N(p)=2$. 

To establish (b) assume that we cannot find $U$ satisfying (a) such that (b) is also true. Then either (b1) there exists a sequence of normal crease points $q_n \rightarrow p$ such that $q_n \notin \cN_1 \cap \cN_2$ or (b2) there exists a sequence $q_n \in \cN_1 \cap \cN_2$ with $q_n \rightarrow p$ such that $q_n \notin \cH_{\rm end}$. In case (b1) consider the two null geodesics from $H_\star$ to $q_n$, as $n$ varies this gives two sequences of null geodesics that must each admit a limit curve that is a null geodesic from $H_\star$ to $p$. These two limit curves must be $\gamma_{1,2}$. It follows that, for large enough $n$, the null geodesics from $H_\star$ to $q_n$ must belong to $\cN_{1,2}$ so $q_n \in \cN_1 \cap \cN_2$, a contradiction. In case (b2), each $q_n$ is not a cut point, so let $r_{n1}$ and $r_{n2}$ be the cut points along the two null geodesics from $H_\star$ to $q_n$; these must occur strictly before $q_n$ along these geodesics. We have $r_{n1} \in \cN_1$ but $r_{n1} \notin \cN_2$ and similarly for $r_{n2}$. There must exist a null geodesic from $H_\star$ to $r_{n1}$ starting at a point $s_{n1} \notin O_1 \cup O_2$. ($r_{n1}$ is a cut point so by Lemma \ref{lem:creaseorcaustic} either a focal point of $H_\star$ along some null geodesic, which must start outside $O_1\cup O_2$ because $\cN_{1,2}$ are smooth, or there exists a null geodesic not in $\cN_1$ or $\cN_2$ from $H_\star$ to $r_{n1}$.) Taking the limit curve gives a null geodesic from $H_\star$ to $p$ that starts outside $O_{1,2}$, a contradiction. 

Next we show that the intersection $\cN_1 \cap \cN_2$ is transverse, implying that it is a submanifold. Let $q \in \cN_1 \cap \cN_2$ and let $V_A^a$ be tangent to the null geodesic generator of $\cN_A$ passing through $q$. Then $V_1^a$ and $V_2^a$ must be linearly independent for otherwise these two generators would be the same, which contradicts the fact that $O_1$ and $O_2$ are disjoint. Now $(V_A)_a$ is normal to $\cN_A$ at $q$ so we have shown that the normals to $\cN_A$ are linearly independent and hence $\cN_1$ and $\cN_2$ intersect transversally. Since $\cN_1$ and $\cN_2$ are null, the intersection is spacelike. Hence $\cN_1 \cap \cN_2$ is a $(d-2)$-dimensional spacelike submanifold. Any chart of this submanifold can be restricted to $U$ to define a chart on the set of normal crease points. Finally we need to show that these charts are compatible where they overlap. Assume that $p$ belongs to two charts, associated with $U,\cN_1,\cN_2$ and $U',\cN_1',\cN_2'$. From the above construction, $\cN_A$ are locally unique, so we have $U \cap U' \cap\cN_1 \cap \cN_2 = U \cap U'\cap \cN_1' \cap \cN_2'$ and so chart compatibility follows from the compatibility of charts on $\cN_1 \cap \cN_2$. Hence we have shown that the set of normal crease points is a $(d-2)$-dimensional spacelike submanifold.  
 
Now let $p$ be a normal corner point. Arguing as above we can construct three null hypersurfaces $\cN_{1,2,3}$ from $H_\star$ to a neighbourhood of $p$. Let $\cC_3$ be the set of normal corner points. We claim that there exists a neighbourhood $U$ of $p$ such that $U \cap \cC_3 = U \cap \cN_1 \cap \cN_2 \cap \cN_3$. (This is the analogue of statement (b) above, since statement (a) implies that $U \cap \cH_{\rm end} = U \cap \cC_2$ where $\cC_2$ is the set of normal crease points.) As before, we assume that there does not exist such $U$. We have two cases: (1) there exists a sequence of normal corner points $q_n\to p$ such that $q_n \notin \cN_1 \cap \cN_2 \cap \cN_3$, or (2) there exists a sequence $q_n \in \cN_1 \cap \cN_2 \cap \cN_3$ with $q_n\to p$ such that $q_n \notin \cC_3$. To disprove (1), the same argument presented under (b1) above generalises immediately. In case (2), we have three subcases: either (i) $q_n \notin \cH_{\rm end}$, (ii) $q_n$ is a caustic point with $N(q_n)=3$, or (iii) $N(q_n)\geq 4$ ($q_n \notin \cC_2$ since $q_n \in \cN_1 \cap \cN_2 \cap \cN_3$, so there are at least three null geodesics from $H_\star$ to $q_n$). In a general sequence of points, the $q_n$ will fall under different cases for different values of $n$. However, we are only interested in the limit $n \to \infty$, so each case is only relevant if an infinite subset of the $q_n$ falls under it. Hence, we may select a subsequence $\widetilde{q}_n\to p$ consisting of points in the same category. In case (i), the argument for (b2) for normal crease points generalises immediately. Case (ii) implies that there exist caustic points arbitrarily close to $p$ on one of the surfaces $\cN_{1,2,3}$, a contradiction. Case (iii) implies that there are four or more null geodesics from $H_\star$ to each $\widetilde{q}_n$, all but three of which much start from points outside $O_1 \cup O_2 \cup O_3$ for all $n$. Hence, there are at least four distinct limit curves that are null geodesics from $H_\star$ to $p$. So $p \notin \cC_3$, a contradiction.

Finally, we must show that the intersection $\cN_1 \cap \cN_2 \cap \cN_3$ is transverse. Arguing as above implies that $\cN_{1,2,3}$ are {\it pairwise} transverse. If three null vectors are pairwise linearly independent then they are linearly independent. This implies that the three normals to $\cN_{1,2,3}$ are linearly independent. Hence the three null hypersurfaces surfaces intersect transversally at $p$, so the set of normal corner points forms a $(d-3)$-dimensional spacelike submanifold. This completes the proof. 

\medskip

Note that this final step of this proof does not work for an intersection of $4$ null hypersurfaces: a set of $4$ pairwise linearly independent null vectors need not be linearly independent. Hence transversality can fail in this case. So the set of non-caustic points with $N(p)=4$ might not form a $(d-4)$-dimensional submanifold. 

\medskip

{\it Proof of Proposition \ref{caustic_prop}.} This follows closely the proof of the corresponding result in Riemannian geometry \cite{itoh_tanaka}. We have written out the proof in greater detail than \cite{itoh_tanaka} to check that nothing goes wrong in the Lorentzian setting. 

The point $\Phi(\lambda,u) \in \cW$ is a focal point of $H_\star$ (along the generator $\Phi(\cdot,u)$) iff $(d\Phi)(\lambda,u)$ has rank $d-2$ or less. By the Morse-Sard-Federer theorem \cite{morgan}, the image of the set of points $(\lambda,u)$ at which $d\Phi$ has rank $d-3$ or less has Hausdorff dimension $d-3$ or less. So to establish the result we only need to consider the set $\cA$ of caustic points for which $d\Phi$ has rank $d-2$. Let $p=\Phi(\lambda_p,u_p)$ be such a point. Let $\lambda_c$ be the positive function on $H_\star$ defined by the property that $\Phi(\lambda_c(u),u)$ is the first focal point of $H_\star$ along the null geodesic $\Phi(\cdot,u)$ (we define $\lambda_c(u)=\infty$ if there is no focal point along the geodesic; we do not assume that the geodesic is complete). In particular we have $\lambda_c(u_p) = \lambda_p$. 

We shall prove that $\lambda_c$ is smooth in a neighbourhood of $u_p$. To do this we shall study $H_\star$-Jacobi fields (called $P$-Jacobi fields in \cite{oneill}) along the geodesics $\Phi(\cdot,u)$ for $u$ near $u_p$. We recall some standard results about Jacobi fields \cite{Hawking:1973uf}. Introduce a basis $E_\mu^a(\lambda)$ parallelly transported along the geodesic $\Phi(\cdot,u)$ where $E_0^a$ is the (null) tangent to the geodesic $\Phi$, $E_i^a$ ($i=1,\ldots, d-2$) are orthonormal spacelike vectors that are tangent to $H_\star$ at $u$, and $E_{d-1}^a$ is null, orthogonal to $E_i^a$ and satisfies $g_{ab} E_0^a E_{d-1}^a = -1$. Consider the space of $H_\star$-Jacobi fields along $\Phi(\cdot,u)$ that are orthogonal to $E_0^a$. If $S^a$ is such a $H_\star$-Jacobi field then we can write $S^a_i(\lambda) = A_{ij}(\lambda) S^a_j(0)$ where the geodesic deviation equation implies that the matrix $A_{ij}$ satisfies (using a dot for a derivative w.r.t.~$\lambda$)
\be
\label{Aeq}
 \ddot{A}_{ij}(\lambda,u) + R_{0i0k}(\Phi(\lambda,u)) A_{kj}(\lambda,u) = 0
\ee
and this equation admits a conservation law: $\dot{A}_{i[j]}A_{|i|k]}$ is constant along the geodesic. However, the initial conditions satisfied by a $H_\star$-Jacobi field imply that this conserved quantity vanishes on  $H_\star$ (it is proportional to the antisymmetrized extrinsic curvature) and hence vanishes everywhere:
\be
\label{cons}
 \dot{A}_{i[j]}A_{|i|k]}=0.
\ee
The fact that $p$ is a focal point along $\Phi(\cdot,u_p)$ for which $d\Phi$ has rank $d-2$ implies that only a 1-dimensional space of Jacobi fields vanishes at $p$ and so $A_{ij}(\lambda_p,u_p)$ has rank $(d-3)$. We can choose our definition of $E_i^a$ so that the initial direction of a Jacobi field in this 1d space is parallel to $E_1^a(0,u_p)$. Hence $A_{i1}(\lambda_p,u_p)=0$, i.e., the first column of $A$ vanishes at $(\lambda_p,u_p)$. Our rank condition implies that the remaining $d-3$ columns of $A_{ij}(\lambda_p,u_p)$ are linearly independent. Equation \eqref{cons} implies $(\dot{A}_{i1} A_{ij})(\lambda_p,u_p)=0$, i.e., the columns of $A_{ij}(\lambda_p,u_p)$ are orthogonal to $\dot{A}_{i1}(\lambda_p,u_p)$. Now consider
\be
 \det A = \epsilon_{i_1 i_2 \ldots i_{d-2}} A_{i_1 1} A_{i_2 2} \ldots A_{i_{d-2}(d-2)}
\ee
and so
\be
 \partial_\lambda (\det A) (\lambda_p,u_p) = \epsilon_{i_1 i_2 \ldots i_{d-2}} \dot{A}_{i_1 1} A_{i_2 2} \ldots A_{i_{d-2}(d-2)}.
\ee
Assume, to establish a contradiction, that this vanishes. Then $\dot{A}_{i_1 1}(\lambda_p,u_p)$ is a linear combination of the (linearly independent) non-zero columns of $A_{ij}(\lambda_p,u_p)$ but we have just seen that $\dot{A}_{i 1}(\lambda_p,u_p)$ is orthogonal to these columns. Hence $\dot{A}_{i 1}(\lambda_p,u_p)$ must vanish. But setting $j=1$ in \eqref{Aeq} gives a linear ODE for $A_{i1}(\lambda,u_p)$ and we have shown that this quantity and its derivative both vanish at $\lambda=\lambda_p$, hence $A_{i 1}(\lambda,u_p)$ vanishes for all $\lambda$, in particular at $\lambda=0$, which is not possible. We conclude that $\partial_\lambda \det A$ is non-zero at $(\lambda_p,u_p)$. Since $\partial_\lambda \det A$ depends smoothly on $(\lambda,u)$ near $(\lambda_p,u_p)$ we can apply the implicit function theorem to deduce that there exists a neighbourhood $O$ of $u_p$ in $H_\star$ such that $\det A(\lambda,u)=0$ admits a smooth solution $\lambda=\lambda_c(u)$. By continuity we can choose $O$ so that $\partial_\lambda \det A(\lambda_c(u),u) \ne 0$ in $O$, which implies that the singular matrix $A_{ij}(\lambda_c(u),u)$ has $d-3$ linearly independent columns and hence has rank $d-3$ throughout $O$, i.e., the space of Jacobi fields vanishing at the focal point $\Phi(\lambda_c(u),u)$ is 1-dimensional for $u \in O$, so $(d\Phi)(\lambda_c(u),u)$ has rank $d-2$ for $u \in O$. 

We now define a smooth map $\Phi_c:O \rightarrow M$ by $\Phi_c(u) = \Phi(\lambda_c(u),u)$. We claim that $d\Phi_c$ is singular at $u_p$ and hence has rank at most $d-3$. Since $p$ is an arbitrary point of $\cA$, the Morse-Sard-Federer theorem implies that the set $\cA$ has Hausdorff dimension at most $d-3$. To justify the claim, introduce coordinates $x^\mu$ on $M$ so that $\Phi(\lambda,u)$ has coordinates $x^\mu(\lambda,u)$ and $\Phi_c(u)$ has coordinates $x^\mu_c(u) = x^\mu(\lambda_c(u),u)$. Then since $d\Phi$ is singular at $(\lambda_c(u),u)$ there exists $(z^\lambda(u),z^A(u)) \ne (0,0)$ in its kernel, where $A=1,\dots,d-1$.\footnote{We shall use capital Latin letters as indices in several different sections of this paper. The range of these indices is not the same in different sections.} In coordinates this means that
\be
\label{xz}
 0 = x^\mu_{,\lambda}(\lambda_c(u),u) z^\lambda(u) + x^\mu_{,A}(\lambda_c(u),u) z^A(u).
\ee
$x^\mu_{,\lambda}$ is tangent to the geodesic $\Phi(\cdot,u)$ and hence non-zero. It follows that $z^A(u) \ne 0$. The kernel of $(d\Phi)(\lambda_c(u),u)$ is $1$-dimensional which implies that $(z^\lambda(u),z^A(u))$ may be assumed to depend continuously on $u$.
Now consider $(d\Phi_c)(u)$ evaluated on $z^A(u)$. In coordinates this is
\be
\label{xmuc}
 x^\mu_{c,A}(u) z^A(u) = x^\mu_{,\lambda}(\lambda_c(u),u) \lambda_{c,A}(u) z^A(u)+ x^\mu_{,A}(\lambda_c(u),u) z^A (u) = \alpha(u) x^\mu_{,\lambda}(\lambda_c(u),u) 
\ee
where the second equality uses \eqref{xz} and we have defined
\be
\alpha(u) = \lambda_{c,A}(u) z^A(u) - z^\lambda(u).
\ee
We shall show that $\alpha(u_p)=0$, so \eqref{xmuc} implies that $z^A(u_p)$ is in the kernel of $(d\Phi_c)(u_p)$, establishing the claim. So assume $\alpha(u_p) \ne 0$. By reversing the sign of $z^A(u)$ if necessary we can arrange $\alpha(u_p)<0$. View $z^A(u)$ as a vector field on $H_\star$ and let $C(t)$ be the integral curve of this vector field through $u_p$, with $C(0)=u_p$. The LHS of equation \eqref{xmuc} is the tangent vector to the curve $\Phi_c(C(t))$, this equation shows that for each $t$ this curve is tangent to the null geodesic $\Phi(\cdot,C(t))$, i.e., it is an envelope curve of these null geodesics. However, for small enough $t$, it has the opposite sense (as $\alpha(t)<0$) to these geodesics, i.e., it is future- instead of past-directed. So for small $\delta>0$ consider the past-directed causal curve defined by following the generator $\Phi(\cdot,u(\delta))$ from $H_\star$ to $\Phi_c(u(\delta))$, then following $\Phi_c(C(t))$ backwards (decreasing $t$) to $\Phi_c(C(0))=\Phi_c(u_p) = p$, as illustrated in the second diagram of Fig.~\ref{fig:proofsv2}. This is a past-directed causal curve from $H_\star$ to $p$. Therefore it must be an unbroken null geodesic orthogonal to $H_\star$ (for otherwise we could deform it into a timelike curve, contradicting $p \in \cH$). But then $\Phi_c(C(\delta))$ is a focal point on this null geodesic that occurs before $p$, so again we can deform into a timelike curve, again a contradiction. We conclude therefore that $\alpha(u_p)=0$, completing the proof.

\medskip

The above proof makes use of our assumption that $H_\star$ is smooth, which implies that the map $\Phi$ is smooth. If $\Phi$ is only $C^k$ then the first application of the Morse-Sard-Federer theorem implies that the set of caustic points for which $d\Phi$ has rank $d-3$ or less has Hausdorff dimension at most $d-3+2/k$. Similarly if $\Phi_c$ is $C^l$ then the second application of Morse-Sard-Federer implies that the set of caustic points for which $d\Phi$ has rank $d-2$ has Hausdorff dimension at most $d-3+1/l$.

We have formulated the above Propositions to apply to the endpoint set $\cH_{\rm end}$ of a future horizon. However, in view of Lemma \ref{lem:Hendcutlocus} one might expect similar results to apply to the past (or future) null cut locus of any smooth, compact, spacelike, acausal, oriented codimension-$2$ submanifold $H_\star$ in a smooth globally hyperbolic spacetime. This is indeed the case: for a point $p$ in this cut locus we can define $N(p)$ to be the number of null geodesics from $H_\star$ to $p$ and define the notions of caustic, normal crease and normal corner points as above. The proofs of the Propositions are slightly modified because there are two families of past-directed null geodesics orthogonal to $H_\star$, which we can label arbitrarily as the $+$ family and the $-$ family.  
Instead of a single map $\Phi$ there are two maps $\Phi^\pm$. In the proof of Proposition \ref{normal_prop} we have to allow for the fact that two geodesics from $H_\star$ to $p$ might start at the same point of $H_\star$ but belong to different families. We can do this by adding an extra label, e.g. referring to $(O_1,+)$ instead of $O_1$.
This makes only minor changes to the proof. Similarly the proof of Proposition \ref{caustic_prop} works with only minor changes. 

\section{Creases and corners}

\label{sec:crease_corner}

\subsection{Transverse self-intersections}

At a normal crease (corner) point, the big wavefront $\cW$ (Definition \ref{def:bigwavefront}) is locally an intersection of $2$ ($3$) null hypersurfaces. As explained in the proof of Proposition \ref{normal_prop}, such an intersection is always {\it transverse}, i.e., the normals to the hypersurfaces are linearly independent. Transversality can fail for a self-intersection involving $4$ sections of $\cW$; however, {\it generically}, one would expect such an intersection to be transverse and this corresponds to a point of type $(A_1,A_1,A_1,A_1)$ in the classification of \cite{Siino:2004xe} summarized in Table \ref{SKtable}. A self-intersection involving more than $4$ sections of $\cW$ is non-generic.

Locally we can describe the geometry of $\cH$ near a point $p$ of transverse self-intersection by discarding the points of $\cW$ that ``lie beyond the self-intersection''. To do this, let $\cN_A$ ($A=1,2,\ldots$) be null hypersurfaces corresponding to the different intersecting sections of $\cW$. Then, locally, $\cW$ is the union of these surfaces. To construct $\cH$ we retain only the portion of $\cN_A$ that contains the future-directed null geodesic generator of $\cN_A$ that starts at $p$. These geodesics are the generators of $\cH$ with a past endpoint at $p$.

\begin{figure}[t]
    \centering
    \includegraphics[width=0.6\textwidth,trim={2cm 3cm 3cm 2cm},clip]{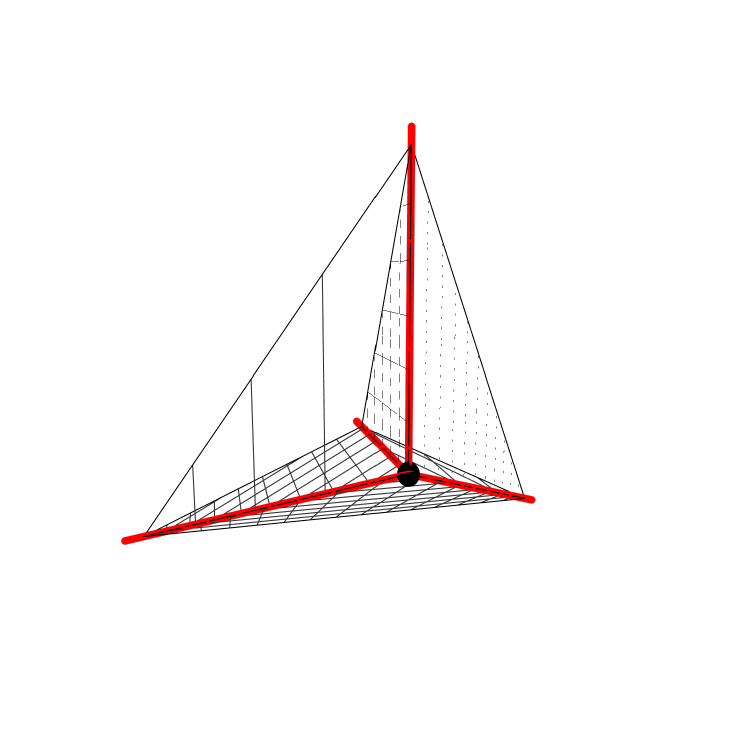}
    \caption{Crease set near a point of quadruple self-intersection $(A_1,A_1,A_1,A_1)$ (black point). Shown in red is the corner submanifold. The remaining six surfaces are sections of the crease submanifold.}
    \label{fig:A1_4}
\end{figure}

The main aim of this section is to use this construction to provide an exact local description of the geometry near a crease or corner perestroika. Before doing this, we shall briefly discuss points of type $(A_1,A_1,A_1,A_1)$, corresponding to a point of quadruple transverse self-intersection of $\cW$. Generically, such points will be isolated. Emanating from each such point will be $4$ sections of the corner submanifold and $6$ sections of the crease submanifold. The behaviour of the crease set near such a point $p$ is shown in Fig.~\ref{fig:A1_4}. Consider a time function\footnote{
Recall that $\Sigma_{\tau_0}$ is the Cauchy surface $\tau=\tau_0$ (see end of Section \ref{sec:intro}).}
$\tau$ such that $p \in \Sigma_0$ and $\Sigma_\tau$ does not intersect $\cH_{\rm end}$ near $p$ for $\tau<0$. For $\tau>0$, $\Sigma_\tau$ will intersect all of the components of the crease and corner submanifolds emanating from $p$. The result is that $\Sigma_\tau \cap \cH$ has a topologically spherical component with a tetrahedral arrangement of creases and corners. Hence, for this choice of time function, $p$ describes the nucleation of such a section of the horizon. However, if $\Sigma_\tau$ does intersect $\cH_{\rm end}$ for $\tau<0$ then the interpretation will be different e.g., one possibility appears to describe a process in which a corner is present on $\Sigma_\tau \cap \cH$ for $\tau<0$ and for $\tau>0$ the tip of this corner has been ``sliced off'' (removing a tetrahedron) to produce a configuration with $3$ corners. The time reverse of this process also appears possible. We shall not attempt to describe all of the other possible interpretations of a point of type $(A_1,A_1,A_1,A_1)$.

\subsection{Crease perestroikas}

\label{sec:crease_perestroikas}

Let $\tau$ be a time function. For a generic value of $\tau$, if $\Sigma_\tau$ intersects the crease submanifold then it will do so transversally. However, as $\tau$ varies there will be special values $\tau=\tau_\star$ such that $\Sigma_{\tau_\star}$ intersects the crease submanifold {\it tangentially}. For a generic time function, such a tangential intersection will occur only at isolated points of $\Sigma_{\tau_\star}$. We shall call such a point a {\it pinch point}. At such points, as we shall explain, there is a qualitative change in the structure of the creases. Motivated by the nomenclature of Arnol'd, we shall refer to such a change as a {\it crease perestroika}. We emphasize that the definition of a pinch point depends on the choice of time function. Different time functions give different pinch points. In this subsection we shall present a local description of the event horizon near a pinch point and investigate the physical interpretation of the resulting crease perestroikas.

Our approach is partly motivated by the final section of \cite{Emparan:2017vyp} which presents a model for the local behaviour of the horizon in an axisymmetric black hole merger \cite{Emparan:2017vyp}. We shall discuss the axisymmetric case, and comment on this model, in Section \ref{sec:axisym}.

Let $p$ be a pinch point and, without loss of generality, assume that this occurs at $\tau=0$. We shall determine how the local geometry of $\cH \cap \Sigma_\tau$ changes as $\tau$ increases from small negative values to small positive values. Near $p$ we can describe the big wavefront $\cW$ as the union of two null hypersurfaces $\cN_A$, $A=1,2$, that intersect transversally. $\cH$ corresponds to the subset of $\cW$ obtained by discarding those parts of generators of $\cN_A$ that lie in the past of the intersection. In particular the portions of the generators of $\cN_A$ through $p$ that lie to the future of $p$ are generators of $\cH$ (i.e., two generators enter $\cH$ at $p$). 

In a neighbourhood of $p$ we can introduce Riemannian normal coordinates $X^\mu=\{t,x^i\}$, so $p$ is the point $t=x^i=0$, such that $t=0$ is the tangent plane to $\Sigma_0$ at $p$. These coordinates are unique up to rotations of $x^i$. In these coordinates, $\Sigma_\tau$ has equation $t=T(\tau,x^i)$ for some smooth function $T$ with $\partial_i T(0)=0$. Taylor expanding $T$ gives the equation of $\Sigma_\tau$ as
\begin{equation}
\label{ttau}
    t = a \tau + c\tau^2 + d_i \tau x^i + \frac{1}{2}K_{ij} x^i x^j + \ldots
\end{equation}
where $a,c,d_i$ and $K_{ij}$ are constants, $a >0$ and $K_{ij}$ is the extrinsic curvature tensor of $\Sigma_0$ at $p$. The ellipsis denotes terms of cubic or higher order in $(\tau,x^i)$.
 
Let the equation of $\cN_A$ be $f_A(t,x^i)=0$ where $f_A$ is smooth with $f_A(0,0)=0$. We choose $f_A$ so that the null vector $(df_A)^a$ is future-directed. 
Locally the crease submanifold has equation $f_1=f_2=0$. At $p$, $\Sigma_0$ is tangent to the crease submanifold so the normal to $\Sigma_0$ must be a linear combination of $df_1$ and $df_2$. This implies there exist $\alpha_A$ such that $\alpha_1 df_1+\alpha_2 df_2 = dt$ at $p$, which is equivalent to $\alpha_1 \partial_i f_1+ \alpha_2 \partial_i f_2=0$ at $p$. This is the statement that the normals to $\Sigma_0\cap \cN_1$ and $\Sigma_0\cap \cN_2$ are either parallel or antiparallel. (The equation of $\Sigma_0 \cap \cN_A$ is $f_A(T(0,x^i),x^i)=0$, with normal proportional to $\partial_i T \partial_t f_A + \partial_i f_A$, which reduces to $\partial_i f_A$ at $p$.) If the normals were parallel then $df_1$ and $df_2$ would be parallel, contradicting the fact that $\cN_A$ intersect transversally. Hence at $p$, the normals to the two sections of the small wavefront $\Sigma_0 \cap \cW$ are antiparallel: $p$ can be visualized as an event at which a pair of wavefronts moving in opposite directions touch.

We shall now consider the geometry of an arbitrary smooth null hypersurface through $p$, with the aim of applying the results to the surfaces $\cN_A$. Such a surface has equation $f=0$ for some smooth function $f$. Smoothness implies that $f$ can be expanded in our Riemmanian normal coordinates as
\be
\label{nulleq}
 f = a_\mu X^\mu + b_{\mu\nu} X^\mu X^\nu +  c_{\mu\nu\rho} X^\mu X^\nu X^\rho + {\cal O}(X^4)
 \ee
for certain constant coefficients $a_\mu$, $b_{\mu\nu}$ etc. The condition that the surface is null is that $g^{\mu\nu} \partial_\mu f \partial_\nu f \propto f$. Using $g_{\mu\nu} = \eta_{\mu\nu} + O(X^2)$ this implies
\begin{equation}
\label{nullcond}
    \eta^{\mu\nu} a_\mu a_\nu=0 \qquad \qquad a^\mu b_{\mu\nu} \propto a_\nu 
\end{equation}
where $a^\mu = \eta^{\mu\nu} a_\nu$. 
Smoothness implies that the null surface has a unique generator passing through $p$. This has equation $X^\mu = a^\mu \lambda$ where $\lambda$ is an affine parameter. 

The function $f$ is not unique: locally $gf$ describes the same null hypersurface where $g$ is any smooth function non-vanishing at $p$. Expanding $g= A + B_\mu X^\mu + \ldots$ gives
\begin{equation}
    gf= a'_\mu X^\mu +  b'_{\mu\nu} X^\mu X^\nu + \ldots
\end{equation}
where
\begin{equation}
\label{nullgauge}
    a'_\mu = A a_\mu \qquad \qquad b'_{\mu\nu} = A b_{\mu\nu} +  B_{(\mu} a_{\nu)}.
\end{equation}
We shall use this freedom to simplify the form of the equation for the surface. We can arrange that $a^\mu$ is future-directed, as assumed above. This restricts us to transformations with $A>0$. A rotation of the spatial coordinates $x^i=(x,y,z)$, and an appropriate choice of $A$, allows us to set $a_\mu = (-1,0,0,1)/\sqrt{2}$, so $a^\mu = (1,0,0,1)/\sqrt{2}$. Now define null coordinates (w.r.t.~$\eta_{\mu\nu}$) $u = (t-z)/\sqrt{2}$, $v=(t+z)/\sqrt{2}$ so $a_\mu = -(du)_\mu$ and $a^\mu = (\partial/\partial v)^\mu$. The second equation of \eqref{nullcond} implies $b_{vv}=b_{vi}=0$. One can then choose $B_\mu$ in \eqref{nullgauge} to set $b_{u\mu}=0$. The result is that we have simplified $f$ to
\begin{equation}
    f=-u +  b_{AB} x^A x^B + O (X^3)
\end{equation}
where $A,B \in \{1,2\}$ (corresponding to the $xy$ directions). A final rotation of the coordinates can be used to set $b_{AB}={\rm diag}(b_{1},b_{2})$ so our null hypersurface has equation
\begin{equation}
\label{null_canonical}
    t = z + b_1 x^2 + b_2 y^2 + O(X^3).
\end{equation}
Consider the intersection of this surface with a surface of constant $t$. Generically $b_1$ and $b_2$ will be non-zero so this surface is a paraboloid (modulo corrections of order $X^3$). The axis of the paraboloid lies along the $z$-axis. If $b_1$ and $b_2$ have the same sign then it is an elliptic paraboloid, if they have opposite sign then it is a hyperbolic paraboloid. As $t$ varies, we obtain a paraboloid moving at the speed of light in the positive $z$-direction. 

We can now return to the problem of describing the behaviour near a pinch point. We can apply the above analysis to the surface $\cN_1$, bringing its equation to the form \eqref{null_canonical}. Now $\cN_2$ has an equation of the form \eqref{nulleq}, with constants $\hat{a}_\mu$, $\hat{b}_{\mu\nu}$, etc. The condition that the intersections of $\cN_2$ and $\cN_1$ with $\Sigma_0$ have anti-parallel normals at $p$ implies that $\hat{a}_\mu \propto (-1,0,0,- 1)$. By rescaling as in \eqref{nullgauge} we can then take $\hat{a}_\mu = -(dv)_\mu$, so $\hat{a}^\mu = (\partial/\partial u)^\mu$. Repeating the analysis above we find that we can bring the equation for $\cN_2$ to the form $\hat{f}=0$ where
\begin{equation}\label{null_canonical_2}
\hat{f} =-v + \hat{b}_{AB} x^A x^B + O(X^3)
\end{equation}
where $\hat{b}_{AB}$ is generically non-degenerate. This surface is another paraboloid (modulo $X^3$ terms), elliptic if $\hat{b}_{AB}$ is positive/negative definite and hyperbolic otherwise. 

To recap, we have introduced Riemannian normal coordinates around the pinch point, with the surface $t=0$ tangent to the Cauchy surface $\Sigma_0$. In these coordinates, the two sections of the horizon which intersect are a pair of paraboloids (up to $X^3$ corrections) whose axes are both along the $z$-axis. The first paraboloid (with parameters $b_{AB}$) moves at the speed of light in the positive $z$-direction. The second paraboloid (with parameters $\hat{b}_{AB}$) moves at the speed of light in the negative $z$ direction. At $t=0$ they are tangent to each other at the origin (the pinch). 

Recall that $p$ is a normal crease point so there are precisely two generators of $\cH$ that pass through $p$ (and end there). These are the generators $z =\pm t$, $x^A=0$ of $\cN_1$ and $\cN_2$ (respectively) with $t \ge 0$. Locally, only the parts of the null hypersurfaces $\cN_A$ lying to the future of their intersection belong to $\cH$. These parts can be identified by the fact that they contain the two generators just described.

Usually we shall be interested in situations for which $\cH$ satisfies the area theorem. This implies that the expansion of $\cN_A$ must be non-negative near these generators; in particular it must be non-negative at $p$. This implies that $b_{AA} \ge 0$, i.e., $b_1 + b_2 \ge 0$, and $\hat{b}_{AA} \ge 0$. Generically these inequalities will be strict, i.e., $b_{AA} >0$, $\hat{b}_{AA}>0$. 

We can now consider the intersection of $\cN_1$ and $\cN_2$, corresponding to (part of) the crease submanifold. Taking the sum and difference of the equations of the two surfaces gives equations for the intersection:
\begin{equation}
\label{int_t}
    2t =  \left( b_{AB} + \hat{b}_{AB} \right) x^A x^B + O(X^3)
\end{equation}
and
\begin{equation} 
\label{int_z}
2z =  \left( \hat{b}_{AB} - b_{AB} \right) x^A x^B + O(X^3).
\end{equation}
These equations give an exact local description of the crease submanifold near the pinch point.\footnote{In flat spacetime, equation \eqref{int_t} (but not \eqref{int_z}), neglecting $O(X^3)$ terms, was written down in \cite{Emparan:2017vyp} as a local model for the crease set in a non axisymmetric merger.} The part of $\cN_A$ that belongs to $\cH$ is the part lying to the future of the intersection, which has $t \ge (1/2)(b_{AB} + \hat{b}_{AB})x^A x^B$. 

Equation \eqref{int_t} indicates that the intersection is (generically) an ellipse or hyperbola in the $x^A$ plane, and $\sqrt{|t|}$ sets the scale for this curve. We are interested in the behaviour for small $|t|$, say $|t| < \epsilon$. Then
the interesting region near the pinch point has $x^A =O(\epsilon^{1/2})$ and from \eqref{int_z}, also $z =O(\epsilon)$. Equation \ref{ttau} implies $|\tau| = O(\epsilon)$ in this region. The ``height'' of this region in the $z$-direction is much smaller than its ``width'' in the $x^A$ directions. This can be ascribed to the fact that the evolution of $\cH$ in the $z$-direction arises from the surfaces $\cN_A$, which describe propagation at the speed of light, but the evolution in the $x^A$ direction arises from the crease, which propagates superluminally (because the crease submanifold is spacelike). 

We can now study the geometry of $\cH$ on a Cauchy surface $\Sigma_\tau$ by writing out the equations for $\cN_{1,2}$ in terms of $\tau$, using $x^i$ as coordinates on $\Sigma_\tau$. Using \eqref{ttau} and focusing on the region just described gives the equations for $\cN_{1,2}$ as
\be
\label{Ntau}
 a\tau = z+ \left( b_{AB} -\frac{1}{2}K_{AB} \right)x^A x^B + \ldots \qquad a\tau = -z+ \left( \hat{b}_{AB} -\frac{1}{2}K_{AB} \right)x^A x^B + \ldots
\ee
where the ellipses denote terms that are $O(|x^A|^3)$ or $O(|\tau x^A|)$ or $O(\tau^2)$ (we eliminate $z$ from these correction terms by iterating the equations). Thus $\Sigma_\tau \cap \cN_1$ is locally a paraboloid, which is elliptic or hyperbolic according to the signature of $b_{AB}-K_{AB}/2$, and similarly for $\Sigma_\tau \cap \cN_2$. Taking the sum and difference of these equations (or using \eqref{ttau}, \eqref{int_t}, \eqref{int_z}) gives the equations of the crease on $\Sigma_\tau$ (i.e., the intersection of $\Sigma_\tau$ with the crease submanifold):
\begin{equation}
\label{reduced_intersection}
    2a\tau  = \left( b_{AB} + \hat{b}_{AB} -K_{AB}\right) x^A x^B +\ldots \qquad 2z = \left( \hat{b}_{AB} - b_{AB} \right) x^A x^B +\ldots
\end{equation}
For a generic time function, $b_{AB}+\hat{b}_{AB}-K_{AB}$ will be non-degenerate. So, to leading order, the crease is either an ellipse or a hyperbola (with $2$ branches) in the $x^A$ plane.  We shall discuss the elliptical and hyperbolic cases separately.

{\it Elliptical intersection.} This corresponds to $b_{AB} + \hat{b}_{AB}- K_{AB}$ being either positive or negative definite. We consider first the negative definite case. Since the area theorem implies $b_{AA} + \hat{b}_{AA} \ge 0$, this case requires that $K_{AA} > 0$, in particular it excludes the choice $\tau=t$ (which gives $K_{ij}=0$). For constant $\tau<0$, the surfaces $\cN_A$ have an elliptical intersection, i.e., there is an elliptical crease. The ellipse shrinks to zero size at the pinch point at $\tau=0$, and the surfaces do not intersect for $\tau>0$. The union of $\cN_1$ and $\cN_2$ describes (part of) the big wavefront $\cW$. To construct $\cH$ (locally) we need to discard the parts of the big wavefront corresponding to horizon generators extended beyond their past endpoints. To do this, we just discard the parts of $\cN_1$ and $\cN_2$ which have not yet entered the intersection. An example is shown on the top row of Fig.  \ref{fig:pinch_diagrams}. The first diagram shows $\Sigma_\tau \cap \cH$ for $\tau<0$, where the black hole region lies outside the surface shown. This is a horizon with an elliptical ``hole'' in it, i.e., a horizon of toroidal (or higher genus) topology. The second diagram on the top row of Fig. \ref{fig:pinch_diagrams} shows the behaviour at $\tau=0$ where the hole in the horizon collapses to zero size and the horizon cross-section has two sections that meet tangentially at the pinch point. 
The behaviour for $\tau>0$ is shown in the third diagram on the top row of Fig. \ref{fig:pinch_diagrams}, where we now have two paraboloidal sections of horizon moving apart, with the black hole region between them.\footnote{In this figure the $z$-axis is vertical, we have set $b_{AB}-K_{AB}/2=\hat{b}_{AB}-K_{AB}/2$ and taken this quantity to be diagonal with values $(-1/4,-1/4)$ on the first row, $(1,1)$ on the second row and $(1,-1)$ on the third row.}

For $\tau<0$, the elliptical intersection of $\cN_1$ and $\cN_2$ is a crease running around the circumference of the hole. We shall now calculate some geometrical properties of this crease. We have $x^A = O(\sqrt{-\tau})$ and $z = O(-\tau)$ on the crease. So for small $\tau<0$, the length of the ellipse scales as $\sqrt{-\tau}$, i.e., the ``circumference of the hole'' tends to zero as $\sqrt{-\tau}$. To work out the angle $\Omega$ at which the two smooth sections of horizon meet at the crease, we proceed as follows. First determine the induced metric on a surface of constant $\tau$, finding it is $h_{ij} = \delta_{ij} + O(\tau)$ at the crease. Second, use \eqref{Ntau} to determine the unit normal $n$ to the surface $\Sigma_\tau \cap \cN_1$ within $\Sigma_\tau$. Repeat to determine the normal $n'$ to $\Sigma_\tau \cap \cN_2$. Finally calculate the angle $\Omega$ using $\cos(\pi-\Omega) = h^{ij} n_i n'_j$. The result is $\Omega = O(\sqrt{-\tau})$. (This can also be understood more heuristically using $\tan \Omega \sim z/|x^A|$.)

Next we consider the case where $b_{AB} + \hat{b}_{AB} - K_{AB}$ is positive definite. Now the surfaces $\cN_A$ do not intersect for $\tau<0$ and there is an elliptical intersection for $\tau>0$. In this case, $\cH$ is obtained by discarding the part of $\cW$ that lies {\it outside} the intersection. For $\tau<0$ this removes everything, so $\cH$ is (locally) empty for $\tau<0$. 
For $\tau>0$ we have a ``flying saucer''-shaped horizon,  with an elliptical crease running around its equator. An example is shown in the second row of Fig. \ref{fig:pinch_diagrams}. The height of the saucer scales as $\tau$, its circumference as $\sqrt{\tau}$, its area as $\tau$, and the angle at the crease as $\sqrt{\tau}$ as above. This case describes the nucleation of an event horizon of spherical topology. It is easy to visualize how this arises: the surfaces $\Sigma_\tau$ ``bulge upwards'' towards the crease submanifold. They initially start to the past of this submanifold. At $\tau=0$, the bulge of $\Sigma_0$ touches the crease set at the pinch point, and for $\tau>0$ the intersection is a flying saucer. One can choose a time foliation with multiple bulges so one can arrange for arbitrarily many of these tiny black holes to nucleate (and subsequently grow and merge).\footnote{
Similarly, the previous case of a hole in the horizon arises when the surfaces $\Sigma_\tau$ ``bulge downwards''. With many downward bulges one can arrange that the horizon cross-section has arbitrarily many holes, i.e., arbitrarily high genus \cite{Siino:1997ix}.} This possibility of adjusting the time function to obtain arbitrarily many black holes has been noted previously \cite{Siino:1997ix} and explicit examples have been found numerically \cite{Bohn:2016soe}.

{\it Hyperbolic intersection.} In this case $b_{AB} + \hat{b}_{AB}-K_{AB}$ is non-degenerate with indefinite signature. Near $p$ the intersection of $\Sigma_\tau$ with the crease submanifold is a hyperbola with $2$ branches. At $\tau=0$ the hyperbola degenerates to a pair of straight lines through the origin. This describes a pair of creases which intersect and then reconnect. We need to determine which sections of $\cN_1$ and $\cN_2$ belong to $\cH$. These sections must include the generators  $x^A=0$, $z = \pm t$, $t \ge 0$, i.e.,  $z =\pm a\tau + \ldots$. This implies that for $\tau>0$, $\Sigma_\tau \cap \cH$ contains the parts of $\cN_1$ and $\cN_2$ lying between the two branches of the hyperbola. This is a connected surface with two creases (the two branches of the hyperbola), width scaling as $\sqrt{\tau}$ and height scaling as $\tau$. For $\tau<0$ we must take the parts of $\cN_1$ and $\cN_2$ that lie outside the two branches of the hyperbola. This gives, at least locally, two disconnected parts of the horizon (e.g., two black holes), each with a crease with hyperbolic shape. These sections of horizon merge to form a ``bridge'' connecting the two sections of horizon. The bridge has hyperbolic creases along its two edges. See the bottom row of Fig. \ref{fig:pinch_diagrams}. This is in good agreement with the behaviour seen in the numerical simulations of black hole mergers in \cite{Bohn:2016soe} (compare the top rows of Figs 15 and 16 of \cite{Bohn:2016soe}.) At the instant of merger, the crease on each section of horizon degenerates to a pair of straight lines, so each section of horizon has a sharp tip at the instant of merger, with the tips of the two horizon sections touching. 
For small $|\tau|$, the angle along each crease, at the point where the creases are closest, scales as $\sqrt{|\tau|}$. This vanishes at the sharp tips, i.e., the horizon flattens out at these tips. 

It should be emphasized that this is a {\it local} description of a merger, valid only near the pinch point. In particular, whether or not the horizon is disconnected cannot be determined locally. In Section \ref{sec:elements} we shall describe how two crease perestroikas can describe the formation of a horizon of toroidal topology in a black hole merger. An elliptic perestroika describes the subsequent collapse of the hole to form a horizon of spherical topology.  

We emphasize that these results depend on the choice of time function. If we fix a normal crease point $p$ and restrict to time functions such that $\Sigma_0$ is tangent to the crease submanifold at $p$ then we still have the freedom to adjust the extrinsic curvature of $\Sigma_0$ at $p$. 
All three of the possible behaviours in Fig. \ref{fig:pinch_diagrams} can arise from the same pinch point $p$ by adjusting $K_{ij}$. Conversely, if we have enough control over $K_{ij}$ (e.g. in a numerical simulation) to arrange that it is negative definite at $p$ then the first row of \ref{fig:pinch_diagrams} cannot arise without violating the area law. 

There are many similarities between our results above and the results of Arnol'd and collaborators for wavefront perestroikas involving caustics. The pictures in Fig. \ref{fig:pinch_diagrams} and the scaling of geometrical quantities with $\tau$ are the same as for a perestroika associated with a ``Legendrian singularity of type $A_2$''. See for example Fig. 48 of \cite{arnold_caustics}. However, we are not studying the same thing: the $A_2$ singularity is a caustic, rather than a transverse self-intersection. The similarity arises because (as we shall discuss in Section \ref{sec:a2}) the $A_2$ singularity is of codimension $2$, just like the crease submanifold so the form of its intersection with surfaces $\Sigma_\tau$ is qualitatively similar.

\subsection{Axisymmetry}

\label{sec:axisym}

We can relate the above discussion to the case of a horizon in a 4d axisymmetric spacetime by considering a reduction to $3$ dimensions. This can be done when the Killing vector field associated with axisymmetry is hypersurface-orthogonal (e.g., a head-on merger of non-rotating black holes). In this case let $S$ be a hypersurface orthogonal to the Killing field. In adapted coordinates, $S$ is the union of surfaces $\phi=0$ and $\phi=\pi$ together with the axis of symmetry. The axisymmetry reduces to a reflection symmetry in 3d which interchanges the sections with $\phi=0$ and $\phi=\pi$. The metric $h$ induced on $S$ is Lorentzian and so $(S,h)$ is a 3d spacetime. 
The intersection $S\cap \cH$ (or $S\cap \cW$) is null w.r.t.~$h$.  The above analysis applies straightforwardly to study a pinch point of $S\cap \cH$ in this 3d spacetime; we wish to understand the 4d interpretation of such a point. We shall assume that the pinch point is invariant under the 3d reflection symmetry so that it lifts to a point $p$ on the axis of symmetry in 4d.\footnote{ 
If the pinch point were not invariant under the 3d reflection symmetry then we would need $2$ such points, related by this symmetry. These would lift to a circle of points in 4d. We shall not consider this case.}

We assume that the time function $\tau$ respects axisymmetry. In 4d, we can introduce Riemannian normal coordinates $X^\mu$ at $p$ as described above. We then transform the spatial coordinates to cylindrical polar coordinates to make the axisymmetry manifest. The metric becomes $g = -dt^2 + dr^2 + r^2 d\phi^2 + dy^2 + O(X^2)$. Restricting to a surface orthogonal to $\partial/\partial \phi$ gives the 3d metric $h = -dt^2 + dr^2 + dy^2 + O(X^2)$. In these coordinates we must allow $r$ to become negative: $r>0$ corresponds to $\phi=0$ and $r<0$ corresponds to $\phi=\pi$ in 4d. The 3d reflection symmetry is $r \rightarrow -r$. The 3d pinch point is $t=r=y=0$. 

In 3d the pinch point is associated with a transverse self-intersection. We denote the two intersecting sections of $S \cap \cW$ as $\cN_A$ as above. 
The general analysis above shows that we can perform a rotation of the 3d spatial coordinates $(r,y)$ to new coordinates $(x,z)$ such that the two surfaces may be taken to have equations $t =  z + bx^2+O(X^3)$ and $t=-z+\hat{b}x^2 + O(X^3)$ and the horizon generators entering at $p$ are $x=0$ and $z =\pm t$ respectively, for $t \ge 0$. We now need to determine the rotation relating the $(x,z)$ coordinates to the $(r,y)$ coordinates. To do this, we use the reflection symmetry (inherited from axisymmetry), which must preserve $S \cap \cH$ and hence preserve $\cN_1 \cup \cN_2$. There are two cases. (1) $\cN_1$ and $\cN_2$ are each invariant under the reflection; (2) the reflection interchanges $\cN_1$ and $\cN_2$.

In case (1), the reflection must act as $x \rightarrow -x$ so the $z$-axis is the axis of reflection symmetry and we can identify $(x,z)=(r,y)$.
In 4d, $\cN_A$ become a pair of wavefronts moving in opposite directions along the axis of symmetry. This is simply the axisymmetric version of the elliptic intersections discussed above, i.e., it corresponds to the first two rows of Fig. \ref{fig:pinch_diagrams}. (In the 4d coordinates, it corresponds to taking $b_{AB}$, $\hat{b}_{AB}$ and   $K_{AB}$ proportional to $\delta_{AB}$ which excludes the hyperbolic case.)

\begin{figure}[t]
    \centering
    \includegraphics[width=0.95\textwidth]{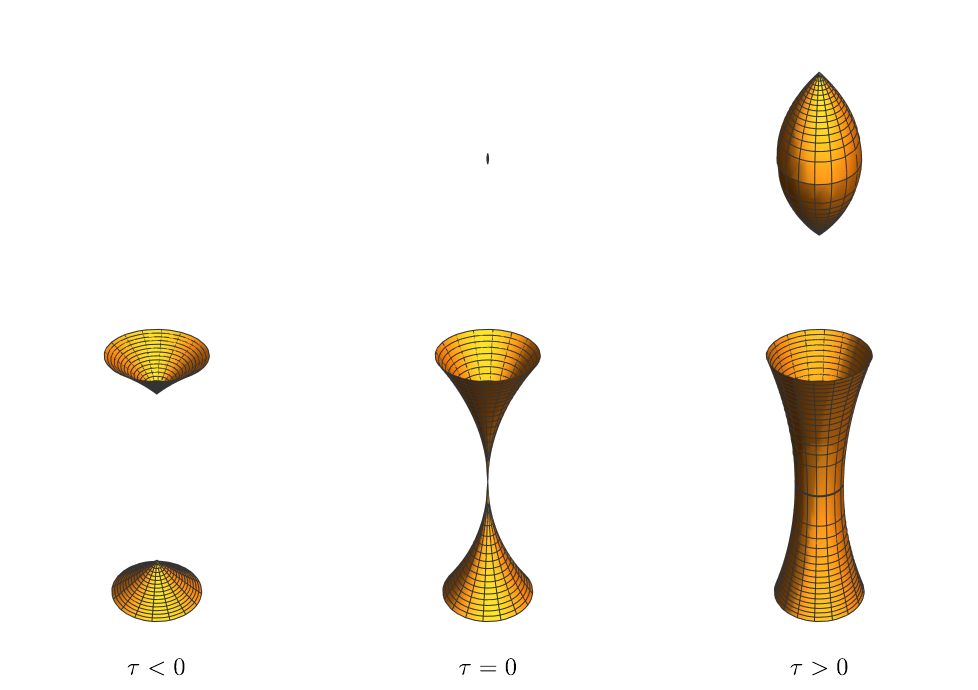}
    \caption{The evolution of the horizon in some 4d axisymmetric spacetimes arising from the 3d versions of crease perestroikas. \textit{Top:} Nucleation of a ``spindle'': a portion of horizon of spherical topology, with conical singularities at the poles. The nucleation occurs at $\tau=0$. \textit{Bottom:} Merger of two portions of the horizon. For $\tau<0$, there are two disconnected parts of the horizon, each with a conical singularity. The singularities touch at $\tau=0$, and the two sections of the horizon merge to form a smooth connected surface for $\tau>0$. 
}
    \label{fig:axisymmetric_pinch_diagrams}
\end{figure}

In case (2) we must have $\hat{b}=b$ and the reflection acts as $z \rightarrow -z$. We can identify $(x,z)=(y,r)$. The 4d lift of the surfaces $\cN_A$ is a single surface with equation $t=r+by^2 + O(X^3)$. The generators entering $\cH$ at $p$ have equation $y=0$, $r=t \ge 0$, $\phi={\rm const}$ and there are infinitely many of them, related by the rotational symmetry. In 4d $p$ is a caustic point rather than a transverse self-intersection. This type of caustic is non-generic outside of axisymmetry so it does not appear in Table \ref{SKtable}. We now consider the intersection with surfaces $\Sigma_\tau$. In axisymmetry, \eqref{ttau} becomes 
\begin{equation}
    t = a \tau + c \tau^2 + d\tau y + \frac{1}{2} (K_{yy} y^2 + K_{rr} r^2) + O(X^3)
\end{equation}
Repeating the analysis leading to \eqref{reduced_intersection} we find that for small $\tau$, the small wavefront has $r \sim \tau$ and $y \sim \sqrt{|\tau|}$ and equation
\begin{equation}
    a \tau  = r +(b- K_{yy}/2) y^2 +O(\tau^2)
\end{equation}
If  $b-K_{yy}/2>0$ this describes the nucleation of a ``spindle'': an axisymmetric portion of event horizon of spherical topology, with a conical singularity at the poles. This is shown in the first row of Fig. \ref{fig:axisymmetric_pinch_diagrams}. The angle at the conical singularity scales as $\sqrt{\tau}$ and the horizon area scales as $\tau$. In the case $b-K_{yy}/2<0$ it describes the merger of two disconnected axisymmetric sections of horizon, each with a conical singularity, to form a smooth section of horizon. The angle at the conical singularities scales as $\sqrt{|\tau|}$. This is shown in the second row of Fig. \ref{fig:axisymmetric_pinch_diagrams}.

Previous studies have found that in an axisymmetric black hole merger, $\cH_{\rm end}$ is a 1-dimensional spacelike line \cite{Lehner:1998gu,Hamerly:2010cr,Emparan:2016ylg}. By adjusting $\tau$ one can intersect this set in different ways: as in the ``flying saucer'' examples, a spindle is produced when $\Sigma_\tau$ ``bulges upwards'' towards the crease set. By including multiple bulges, one can arrange for arbitrarily many spindles to appear, along the axis of symmetry, at an intermediate stage of an axisymmetric merger. Each subsequent merger of these spindles with each other, or with the initial black holes, is described locally by the above results.

We shall now discuss the relation to \cite{Emparan:2017vyp}, which presented a flat space model for the behaviour of the event horizon near the merger point in an axisymmetric black hole merger. The event horizon was modeled by the surface $\tilde{f} =0$ in flat spacetime where (in our cylindrical polar coordinates) $\tilde{f} = -t+r + by^2$ and $b<0$. Clearly this is very closely related to what we have just discussed: it corresponds to case (2) with time function $\tau=t$ (so $K_{yy}=0$) and neglecting the higher order terms in the equation for the surface and in the metric (i.e., the metric is described as exactly flat). A drawback of neglecting such terms is that it gives $\eta^{\mu\nu} \partial_\mu \tilde{f} \partial_\nu \tilde{f} = 4b^2 y^2$, which implies that the surface is null only at $y=0$; everywhere else it is timelike. This seems unsatisfactory as a model of an event horizon! However, there is no compelling reason to use this truncated equation in Minkowski spacetime. We have shown that one can perform an exact curved spacetime treatment using Riemannian normal coordinates as above (and allowing for a non-vanishing $K_{ij}$) to obtain exactly the same results as in \cite{Emparan:2017vyp}, i.e., that the angle at the conical singularities scales as $\sqrt{|\tau|}$.

\subsection{Corner perestroikas}

Recall that a normal corner point is a non-caustic point with $N(p)=3$ and the set of such points forms the corner submanifold. Locally, this submanifold is a transverse intersection of $3$ null hypersurfaces $\cN_A$, $A=1,2,3$. Let these have equations $f_A=0$ where $f_A$ are smooth functions and $df_A$ are null and linearly independent.

We define corner perestroikas similarly to crease perestroikas. Given a generic time function $\tau$, $\Sigma_\tau$ generically intersects the corner submanifold transversally, in isolated (corner) points. However, for special values of $\tau$, $\Sigma_\tau$ may intersect the corner submanifold {\it tangentially} at $p$. As above, we shall call such $p$ a pinch point. We shall shift our time function such that $\tau(p)=0$, so $\Sigma_0$ is tangent to the corner submanifold at $p$. Generically, the corner submanifold either ``bends upwards'' or ``bends downwards'' from $\Sigma_0$ at $p$. In the former case, $\Sigma_\tau$ does not intersect the corner submanifold (locally, near $p$) for $\tau<0$ and intersects it at two points for $\tau>0$, and vice-versa in the latter case. 
Thus a corner perestroika describes a process in which a pair of corners either nucleates or disappears.


Let $V^a$ be tangent to the corner submanifold at $p$ (and hence also tangent to $\Sigma_0$). Then $V^a$ is also tangent to each surface $\cN_A$. 
This implies that different sections of the small wavefront $\Sigma_0 \cap \cN_A$ have a common tangent vector $V^a$ at $p$. Each pair of surfaces defines a crease. On $\Sigma_0$ these are $\Sigma_0 \cap \cN_A \cap \cN_B$ for $A \ne B$. The three crease lines are tangential to $V^a$ at $p$, where they meet. We need to determine which of these lines belong to $\Sigma_0 \cap \cH$. 

\begin{figure}[t]
    \centering
    \includegraphics[width=0.9\textwidth]{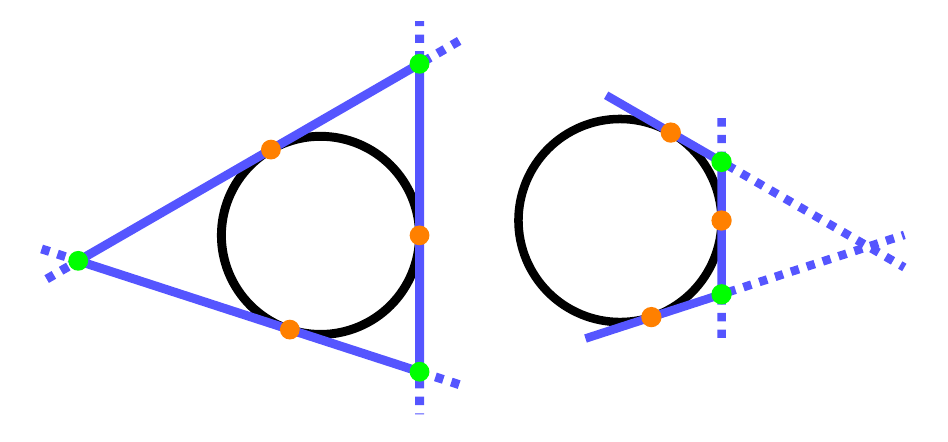}
    \caption{Wavefront shortly after a corner perestroika in the 3d spacetime $S$. This figure shows the wavefront in $\Sigma_\tau \cap S$ for infinitesimal positive $\tau$. The black circle is a cross-section of the future light cone at $p$. The orange dots indicate the $3$ generators entering $\cH$ at $p$. The blue lines indicate $\cN_A$ and the green dots are the intersections of the crease lines with $S$. 
    The dashed segments of $\cN_A$ are in $\cW$ but not $\cH$. \textit{Left:} The three lines that belong to $\cH$ close up to form a triangle. \textit{Right:} The three lines do not close up locally. In both cases, the local geometry of $\Sigma_\tau \cap \cH$ can be visualized by translating the solid blue lines in the direction $V^a$ normal to the plane of the figure.}
    \label{fig:cornerTimelikeCrossSection}
\end{figure}

At $p$, one generator of each of $\cN_A$ must enter $\cH$. Hence for small positive $\tau$, $\Sigma_\tau \cap \cH$ must have three smooth sections corresponding to the three $\cN_A$. Their intersections are crease lines. To visualize the geometry, we can, for infinitesimal $\tau$, take a cross-section of $\Sigma_\tau \cap \cH$ transverse to the vector $V^a$. More precisely, consider a timelike surface $S$ through $p$ with normal $V^a$ at $p$. Since $V^a$ is tangent to $\cN_A$ we have $0=V^a (df_A)_a$ and so $(df_A)^a$ is orthogonal to $V^a$. Hence the generator of $\cN_A$ through $p$ is tangent to $S$. If we regard $S$ as a 3d spacetime (using the induced metric) then the generators entering $\cH$ at $p$ lie on the future null cone of $p$ in this spacetime. So $S \cap \cN_A$ is tangent to this null cone. A surface of infinitesimal positive $\tau$ (i.e. $\Sigma_\tau \cap \cS$) corresponds to taking a cross-section through this future null cone, which gives a circle, on which the three generators are three points and, locally, $S\cap \cN_A$ are straight lines tangent to the circle at these points (see Fig. \ref{fig:cornerTimelikeCrossSection}). The intersections of these tangent lines are the intersections of creases with $S$. As usual, parts of these lines correspond to portions of $\cW$ that do not belong to $\cH$.
To construct $\Sigma_\tau \cap \cH$ we must retain the three portions of the lines containing the three generators entering at $p$. This leads to two cases. (1) The three lines form a triangle (Fig. \ref{fig:cornerTimelikeCrossSection} left). When we reinstate the direction perpendicular to $S$, this implies that $\Sigma_\tau \cap \cH$ locally resembles a triangular prism, whose cross-section shrinks to zero size as $\tau \rightarrow 0+$. (2) There is one short line with two other lines extending from its endpoints and, locally these other lines do not intersect each other (Fig. \ref{fig:cornerTimelikeCrossSection} right). In this case, $\Sigma_\tau \cap \cH$ locally resembles an ``open prism'' with one narrow face that joins onto two other faces. The width of this narrow face shrinks to zero as $\tau \rightarrow 0+$.

For either of these cases there are two subcases to consider: either (a) $2$ corners are present for $\tau>0$ and none for $\tau<0$ (the corner submanifold ``bends up'' from $\Sigma_0$) or (b) $2$ corners are present for $\tau<0$ and none for $\tau>0$ (the corner submanifold ``bends down'' from $\Sigma_0$).

\begin{figure}
    \centering
    \includegraphics[width=\textwidth,trim={0 0 0 1cm},clip]{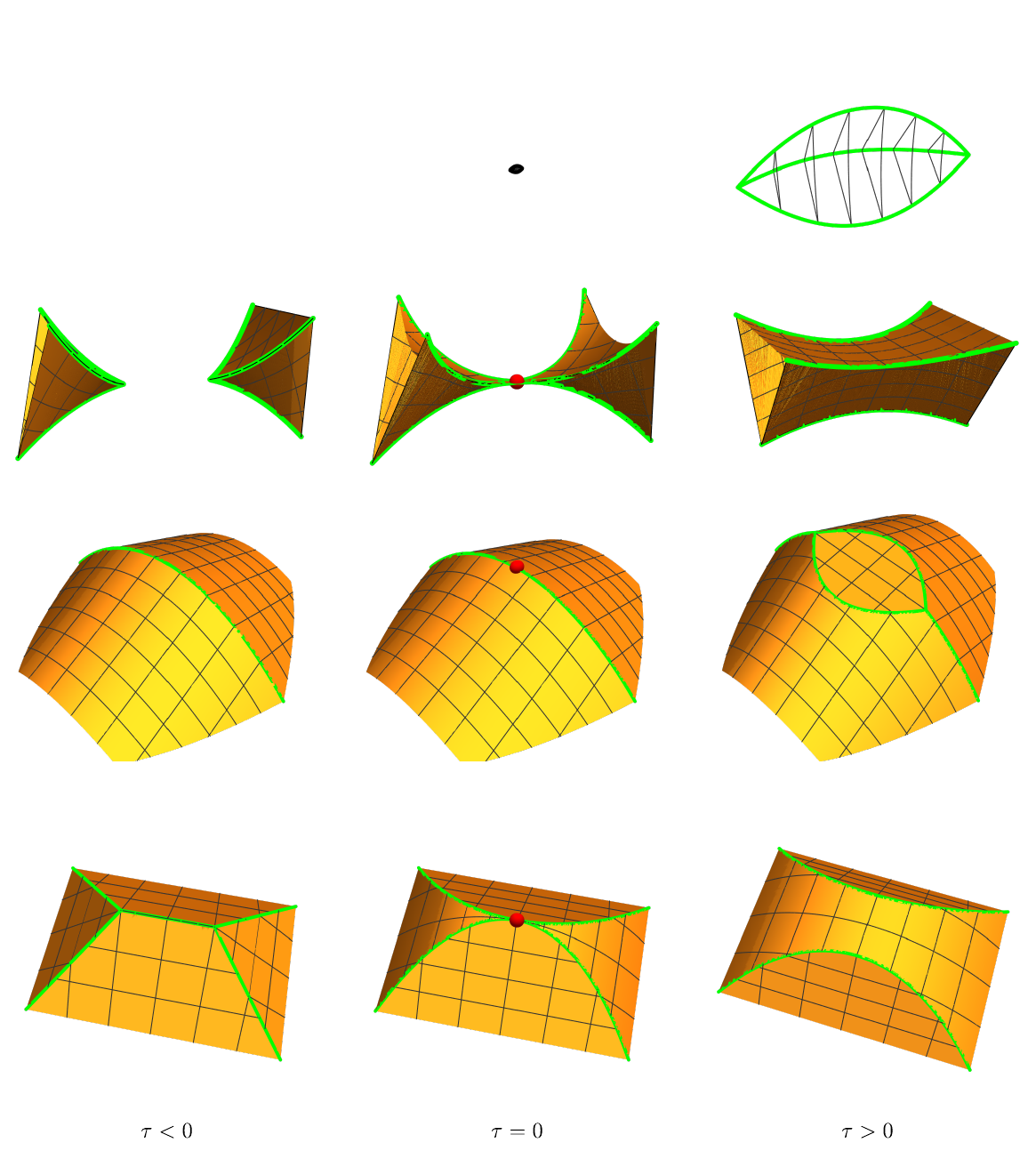}
    \caption{Corner perestroikas, with creases highlighted in green. Corners that are not obvious are shown by red points. \textit{Top:} Nucleation of a surface of spherical topology, with two corners connected by three crease lines for $\tau>0$. (Only the edges are shown for clarity.) \textit{Upper Middle:} Two locally disconnected surfaces, each containing a corner, touch and merge as $\tau\to0$. For $\tau>0$ the corners have annihilated and there remain three disconnected crease lines, forming an expanding triangular prism. \textit{Lower Middle:} Nucleation of two corners on a crease. The corners are connected by two creases. \textit{Bottom:} Annihilation of two corners, leaving two locally disconnected creases for $\tau>0$.}
    \label{fig:CornerPerestroika}
\end{figure}

In case (1a) the corner perestroika describes the nucleation of a topologically spherical section of event horizon, with corners at the ``poles'' and three crease lines connecting these corners. The horizon has an expanding triangular cross-section. See the top row of Fig. \ref{fig:CornerPerestroika}. In case (1b), for $\tau<0$, $\Sigma_\tau \cap \cH$ exhibits two (locally) disconnected sections, each with a corner from which three crease lines emanate. The corners approach each other and merge at $p$ to form a connected section of horizon with three crease lines, and an expanding triangular cross-section. See the second row of Fig. \ref{fig:CornerPerestroika}. At $p$, the corners degenerate to sharp ``spikes'', with vanishing solid angle. For small $\tau<0$, the solid angle at each corner scales as $\Omega \sim O(|\tau|)$ and the angle at the corners between each pair of crease lines is $O(\sqrt{-\tau})$ as $\tau \rightarrow 0-$.

These corner perestroikas are analogous to the crease perestroikas shown in the second and third rows of Fig. \ref{fig:pinch_diagrams}. In particular they provide an alternative mechanism for horizon nucleation or merger. However, they only occur if the corner submanifold is non-empty. 

In case (2a), the perestroika describes the nucleation of a pair of corners on a crease. For $\tau<0$, $\Sigma_\tau \cap \cH$ has (locally) two smooth sections meeting at a single crease line. At $\tau=0$ a corner point appears on this line, and immediately splits into two corners which are connected by two new crease lines bounding a new smooth section of the horizon. See the third row of Fig. \ref{fig:CornerPerestroika}. In case (2b) the perestroika describes the annihilation of a pair of corners. For $\tau<0$, a pair of corners is present,
each with $3$ crease lines emanating from it, with one of these crease lines connecting the two corners. At $\tau=0$ this crease line shrinks to zero size and the corners disappear, leaving a horizon with two (locally) disconnected crease lines for $\tau>0$, as depicted in the bottom row of Fig. \ref{fig:CornerPerestroika}.

\section{Caustic points}

\label{sec:caustics}

\subsection{Generic wavefront singularities}

In this section we shall discuss caustics for the class of horizons defined in Section \ref{sec:smooth_late} with the added assumption of genericity (stability). A caustic is a singularity of the big wavefront $\cW$ (Section \ref{sec:smooth_late}) at which it fails to be an immersed submanifold. Catastrophe theory aims to classify {\it stable} wavefront singularities, i.e., singularities whose qualitative properties are unchanged by a small perturbation in the wavefront. In our case, the big wavefront is uniquely determined once the metric is fixed. However, we can perturb the spacetime metric. More precisely, if we have equations of motion that admit a well-posed initial value problem then we can perturb the initial data on a Cauchy surface and ask how this affects properties of the big wavefront. 
We shall assume that stability w.r.t.~perturbations of the metric is equivalent to stability w.r.t.~perturbations of the wavefront. Siino and Koike \cite{Siino:2004xe} do not explain how their mathematical notion of stability relates to either of these notions of stability.\footnote{
Evidence that their notion of stability does correspond to stability w.r.t.~perturbations of the metric (although without imposing any equations of motion) is provided by results on the cut locus in Riemannian geometry. Recall Lemma \ref{lem:Hendcutlocus} relates $\cH_{\rm end}$ to the cut locus of $H_\star$. Now for a generic compact 3d Riemannian manifold it has been proved \cite{buchner} that the cut locus (of a point, and presumably also a hypersurface) consists of the same type of points as listed in Table \ref{SKtable}. Adding a trivial time direction gives a class of 4d Lorentzian product manifolds for which the Lorentzian cut locus has the structure of Table \ref{SKtable}, and for which the structure is stable w.r.t.~perturbations of the spatial metric.
}
In this section we shall discuss a different approach to this problem based on earlier work in the literature. The reader uninterested in this issue may wish to skip ahead to Section \ref{sec:A3} where we describe the geometry of $\cH$ near an $A_3$ caustic.

Recall that $\cW$ is defined by null geodesics emanating orthogonally to a late time cross-section of the horizon. Although $\cW$ is not smooth, the geodesic flow is smooth in phase space $T^\star M$ (the cotangent bundle of spacetime). These null geodesics generate a $(d-1)$-dimensional smooth submanifold $\cW'\subset T^\star M$ whose projection to spacetime is the non-smooth submanifold $\cW$. We call $\cW'$ the {\it lifted wavefront}. 
At generic points, the projection map restricted to $\cW'$ has maximal rank $(d-1)$ in which case $\cW$ is locally an immersed submanifold (which may exhibit self-intersections, at which it fails to be an embedded submanifold). A caustic is the image of a point at which the rank of this (smooth) map is less than $(d-1)$. Catastrophe theory provides a classification of the possible behaviour near such points, assuming an appropriate notion of stability. 

The classification of wavefront singularities is an application of the classification of stable {\it Legendrian singularities} obtained by Arnol'd and collaborators (reviewed in \cite{arnold,arnold_caustics}). There have been several attempts to apply this classification to  caustics in a general curved spacetime \cite{Friedrich:1983vi,hasse,Low:1998fc}. The work of 
\cite{Friedrich:1983vi,hasse} aims at a classification of stable singularities of a big wavefront. As we shall explain, this work has not yet been fully justified mathematically. The work of \cite{Low:1998fc} provides a classification of stable singularities of a small wavefront. This is on firmer ground mathematically. Therefore we shall discuss this work first.

The approach of \cite{Low:1998fc} uses the space of future-directed null geodesics $\cN$. This is obtained from phase space by taking certain quotients. It can be shown that $\cN$ is a contact manifold of dimension $2d-3$. A lifted wavefront gives a smooth Legendrian submanifold $\cW'\subset \cN$ \cite{Low:1998fc}. Conversely, any such Legendrian submanifold is a lifted wavefront. A small wavefront is the image of a Legendrian map from $\cW'$ to a Cauchy surface $\Sigma$. The Arnol'd classification of stable Legendrian singularities can be used to determine the generic (i.e., stable) behaviour of singularities of a small wavefront. For $d=4$, this implies that generically the singularities of this map can only be those of type $A_2$ (cusp) or $A_3$ (swallowtail) in the Arnol'd classification. A small wavefront with an $A_3$ caustic point and two lines of $A_2$ caustic points is shown on the left of Fig. \ref{fig:A3}.
Here ``generic'' should be understood to include the choice of $\Sigma$: there may be special instants of time at which non-generic singularities occur; these are associated with caustic perestroikas. We shall show below that $A_2$ singularities cannot occur on the part of a small wavefront that corresponds to a horizon cross-section (as on the right of Fig. \ref{fig:A3}), and therefore a stable singularity of a horizon cross-section must be of type $A_3$. 

We now turn to approaches based on the big wavefront \cite{Friedrich:1983vi,hasse}.\footnote{For a review of how the Arnol'd classification relates to big wavefronts see \cite{Ehlers:1999bq}.} In \cite{Friedrich:1983vi}, a contact manifold is obtained by taking a quotient of the fibres of phase space, giving a projectified cotangent bundle $PT^*M$. This is a contact manifold whose base space is the spacetime manifold $M$. The big wavefront corresponds to a Legendrian submanifold of this contact manifold, so one can again apply the classification of stable Legendrian singularities \cite{Friedrich:1983vi}. However, as noted in \cite{hasse}, there is a problem: 
while a wavefront lifts to a Legendrian submanifold of $PT^*M$, a generic Legendrian submanifold of $PT^*M$ does not correspond to a wavefront because a generic point of $PT^*M$ corresponds to a non-null covector. Thus a ``generic'' perturbation of the wavefront, viewed as a Legendrian submanifold, does not give another wavefront. Conversely, perturbations that do remain within the family of wavefronts are non-generic from the Legendrian perspective. In other words, ``stable as a wavefront'' is a weaker condition than ``stable as a Legendrian submanifold of $PT^*M$'', so a stable big wavefront singularity might correspond to an unstable Legendrian singularity and therefore lie outside the Arnol'd classification.\footnote{Ref. \cite{hasse} presented a theorem that was claimed to fix this problem but this claim has been withdrawn \cite{Perlick2000,Perlick:2004tq}.} However, in flat spacetime one can exploit the additional symmetries to obtain a classification of stable big wavefront singularities and this is in agreement with the notion of stability as a Legendrian submanifold \cite{izumiya}. Based on this, and since a curved spacetime is locally flat, it seems reasonable to expect that the behaviour near a generic big wavefront singularity will be qualitatively identical in curved spacetime and in flat spacetime. Therefore we shall proceed on the assumption that stable big wavefront singularities are indeed stable in the Legendrian sense.

For $d=4$, the big wavefront singularities that are stable in the Legendrian sense are those of type $A_2$, $A_3$, $A_4$ and $D_4^\pm$ in Arnol'd's classification. An $A_4$ or $D_4^\pm$ caustic is point-like and so does not intersect a generic Cauchy surface. Therefore caustics on a small wavefront generically will be of type $A_2$ or $A_3$, in agreement with the discussion above. 

In order to relate this discussion to the classification of Siino and Koike \cite{Siino:2004xe}, we first note that $A_3$ points are not isolated but form lines in spacetime (see below). Such a line can intersect another (smooth) section of $\cH$ transversally: this gives a point of type $(A_3,A_1)$ in the classification of \cite{Siino:2004xe}. If we can show that caustics of type $A_2$, $A_4$ and $D_4^\pm$ generically cannot occur on $\cH$ then we recover the results of \cite{Siino:2004xe}.  
Siino and Koike work with a ``Fermat potential'' which is asserted to be minimized on $\cH$. So maybe an $A_2$, $A_4$ or $D_4^\pm$ caustic corresponds to an extremum of this potential that fails to be a  minimum. The Appendix of \cite{Chrusciel:2002mi} gives a rigorous argument for why a horizon satisfies a Fermat principle. This argument shows that the minimization property arises from the achronality of the horizon. Combining these ideas suggests that we should aim to show that $A_2$, $A_4$ and $D_4^\pm$ caustics always violate achronality and are therefore excluded on a future horizon $\cH$. 

In Section \ref{sec:a2} and Appendix \ref{app:a2small} we shall use the canonical form of an $A_2$ caustic to demonstrate that indeed achronality is violated near an $A_2$ point on a big wavefront, thus proving that an $A_2$ caustic cannot occur on $\cH$. In Appendix \ref{app:a2small} we show that an $A_2$ singularity on a horizon cross-section would also violate achronality of $\cH$. Thus achronality excludes $A_2$ caustics on horizons. An $A_4$ or $D_4^\pm$ caustic point on a big wavefront has several 2d sets of $A_2$ points emanating from it. In order for an $A_4$ or $D_4^\pm$ caustic to occur on $\cH$, these $A_2$ points would have to belong to the part of the big wavefront that is not part of $\cH$ (as happens for an $A_3$ caustic: Fig \ref{fig:A3}). In Appendix \ref{app:a4d4} we sketch an argument showing that this is not possible. Hence achronality excludes $A_4$ and $D_4^\pm$ caustics on $\cH$.

In Section \ref{sec:A3} we shall study in detail the horizon geometry near an $A_3$ singularity on a horizon. As mentioned above, $A_3$ points form lines in spacetime. Given a time function $\tau$, generically a Cauchy surface of constant $\tau$ will intersect such a line transversally, which gives a small wavefront with an isolated $A_3$ singularity. However, just as we saw with creases and corners, there may be a special value of $\tau$ for which the Cauchy surface is tangent to the $A_3$ line. This results in a qualitative change in the features of the small wavefront: either a pair of $A_3$ points that merge, or the nucleation of a pair of $A_3$ points. Following Arnol'd, we shall call these processes $A_3$ perestroikas. In Section \ref{sec:elements} we shall describe how a generic black hole merger can be decomposed into a sequence of crease and $A_3$ perestroikas. 

In Section \ref{sec:A3A1} we shall describe the horizon geometry near an $(A_3,A_1)$ caustic and show that there are three possible perestroikas associated with such a caustic. 

We would like to contrast our approach below with that of Ref.~\cite{Friedrich:1983vi}, which presents examples of big wavefronts in Minkowski spacetime exhibiting the various types of caustics. This work makes use of inertial coordinates, i.e., the coordinates are adapted to properties of the metric. In our approach, we consider a general metric and use coordinates for which the caustic takes its canonical form, i.e., coordinates are adapted to the form of the caustic rather than to symmetries of the metric. The fact that the big wavefront is null gives some information about the metric components in these coordinates. This turns out to be enough to establish, for example, the results about achronality mentioned above. 

\subsection{$A_2$ caustics}\label{sec:a2}

Given a big wavefront with a generic singularity, one can apply a diffeomorphism, i.e., choose smooth coordinates, to bring the wavefront to a canonical form in a neighbourhood of the singularity (see e.g. Chapter 21 of \cite{arnold}). In a (finite) neighbourhood of an $A_2$ singularity on a big wavefront we can introduce coordinates $(x,y,z^A)$, $A=2, \ldots, d$ (with $d$ the spacetime dimension), such that the wavefront is the surface given by values $(x,y,z^A)$ for which the cubic polynomial $f(p)=p^3+xp +y$ has degenerate roots. This is the surface defined by the map
$(p,z^A) \mapsto (-3p^2,2p^3,z^A)$ and the $A_2$ point is $(0,0,0)$. Since the dependence on $z^A$ is trivial, an $A_2$ singularity is not isolated, instead 
there is a codimension-$2$ submanifold of $A_2$ points $(0,0,z^A)$. A cross-section of constant $z^A$ is a curve with a cusp at $x=y=0$.

We shall now determine some properties of the metric tensor in these coordinates. We do this by imposing the condition that the wavefront is a null hypersurface. The tangent vectors to our wavefront are $\partial/\partial z^A$ and $\partial/\partial x - p \partial/\partial y$. A covector $n$ is normal to the wavefront if it is orthogonal to these tangent vectors, which implies $n_A=0$ and $n_x = p n_y$. We can set $n_y=1$ so the normal to the wavefront is $n=dy+ p dx$. We now impose the condition that this is null, which is
\begin{equation}
    0=g^{yy} + 2p g^{xy} + p^2 g^{xx}
\end{equation}
where this equation must hold on the wavefront, i.e., at points of the form $(-3p^2,2p^3,z^A)$. The metric and the coordinates are smooth so the dependence of $g^{\mu\nu}$ on $(x,y,z^A)$ must be smooth. We can now expand the above equation in $p$. At order $p^0$ and $p^1$ this gives
\begin{equation}
    0=g^{yy}(0,0,z^A) = g^{xy}(0,0,z^A)
\end{equation}
Thus at the caustic $(0,0,z^A)$ we see that $dy$ is null (it is normal to the wavefront there) and $dx$ is orthogonal to $dy$. The latter implies that $dx$ must be either spacelike, or null and parallel to $dy$. But $dx$ and $dy$ are linearly independent so $dx$ cannot be parallel to $dy$. Hence $dx$ is spacelike at the caustic. It follows that the caustic set $x=y=0$ is a null submanifold. To see this, note that $V^a$ is tangent to this submanifold iff $V \cdot dy=0$ and $V \cdot dx=0$. The former condition implies that $V$ cannot be timelike but $V^a = (dy)^a$ satisfies both conditions hence there is a null tangent vector. We have $n^a(0,0,z^A) = (dy)^a(0,0,z^A) = g^{Ay}(\partial/\partial z_A)^a$. Since $n^a$ is tangent to the null geodesic generators of the big wavefront, it follows that a generator passing through a caustic point is everywhere tangent to the set of caustic points.  

We shall now demonstrate that this big wavefront violates achronality in any neighbourhood of an $A_2$ point. In the tangent space at any $A_2$ point we can consider the plane with normal $dx$. Since $dx$ is spacelike, this plane is timelike. Furthermore, it is spanned by $\partial/\partial y$ and $\partial/\partial z^A$ since these are clearly orthogonal to $dx$. Hence at an $A_2$ point $(0,0,z^A)$ there exists a timelike vector of the form $V=a\partial/\partial y + b^A \partial/\partial z^A$.  We can assume $a \ne 0$ because the set of timelike vectors is open. By continuity this vector is also timelike in a neighbourhood of $(0,0,z^A)$. By rescaling we can set $a=1$. Now starting at the point $(-3p^2,2p^3,z^A)$ for some $p<0$, which lies on the big wavefront, we can follow the integral curve of $V$ a parameter distance $4 |p|^3$ to reach the point $(-3p^2,-2p^3,z^{A'})$ for some $z^{A'}$. This point also lies on the wavefront. Thus our integral curve connects two distinct points of the wavefront lying on opposite sides of the cusp. For small enough $|p|$ this curve is timelike. This violates achronality. Also, for any neighbourhood $O$ of $(0,0,z^A)$, by taking $|p|$ small enough, this timelike curve lies entirely in $O$. It follows that $\mathcal{H}$, viewed as part of a big wavefront, cannot possess an $A_2$ singularity.

This argument was for the big wavefront. Similarly we can show that if there exists an $A_2$ singularity on a small wavefront then the corresponding big wavefront cannot be achronal. This argument is given in Appendix \ref{app:a2small}.
The reason for considering the small wavefront separately is that, as explained above, the classification of singularities of the small wavefront is more rigorously established than the classification for the big wavefront.

We could also consider the possibility of a wavefront that intersects itself, with one sheet of the intersection possessing an $A_2$ singularity. We consider this from the perspective of the small wavefront. Generically such an intersection will be transversal and so a line (on the small wavefront) of $A_2$ caustic points will emerge from the intersection. The above arguments are local so they can applied to a point on this line to show that the resulting big wavefront cannot be achronal. Thus such intersections cannot arise on a cross-section of $\cH$.

\subsection{$A_3$ caustics}

\label{sec:A3}

We start by considering an $A_3$ singularity on a small wavefront in $d=4$ spacetime dimensions. Consider a Cauchy surface $\Sigma$ intersecting the big wavefront $\cW$, so the small wavefront is $W \equiv \Sigma \cap \cW$. If $W$ possesses an $A_3$ caustic then there exist (smooth) coordinates $(x,y,z)$ on $\Sigma$ such that the $A_3$ point is at $(0,0,0)$ and, in a finite neighbourhood of this point, $W$ is the surface where the quartic polynomial $f(p)=p^4-yp^2 - xp +z$ has degenerate roots (Chapter 21 of \cite{arnold}). This can be parameterized by $(p,q)$ (taking values in a neighbourhood of $(0,0)$) as the map 
\begin{equation}
\label{A3map}
    (p,q) \mapsto ( 4 p^3-2qp, q, 3 p^4 -qp^2).
\end{equation}
The $A_3$ point is at the origin $(0,0,0)$, with $p=q=0$. The surface $W$ has the ``swallowtail'' structure shown on the left in Fig. \ref{fig:A3}. The $z$-axis points downwards in this figure. The Jacobian $\partial(x,y,z)/\partial(p,q)$ has rank $2$ except at $q = 6p^2$ where it has rank $1$. For $q=p=0$ this gives the $A_3$ point, for $q>0$, $p \ne 0$ it gives two lines of $A_2$ points with coordinates $(-8p^3,6p^2,-3p^4)$ ($p>0$ or $p<0$). The surface $W$ has a transverse self-intersection at $p^2=q/2$, corresponding to the line $(0,y,y^2/4)$, $y>0$. The $A_2$ singularities lie beyond this intersection line as shown in Fig. \ref{fig:A3}. 

We are interested in a big wavefront defined in terms of a future horizon as explained in Section \ref{sec:smooth_late}. In this case, the horizon cross-section $H = \Sigma \cap \cH$ is a subset of the small wavefront $W$. Horizon generators cannot extend beyond a self-intersection, and we have seen that $H$ cannot contain an $A_2$ singularity. Therefore $H$ is obtained by discarding the part of $W$ that lies beyond the self-intersection, i.e., the part containing the $A_2$ lines. This is the region $q> 2p^2$. Discarding this region gives the surface shown on the right of Fig.~\ref{fig:A3} with a crease (the self-intersection) that ends at the $A_3$ point. The angle between the two planar sections meeting at the crease tends to $\pi$ at the $A_3$ point. 
  
Locally the surface $H$ is a graph over the $(x,y)$ plane, i.e., it is given by an equation of the form $z=Z(x,y)$ where $Z$ is determined implicitly by the above equations. Using $(x,y)$ as coordinates on $H$, the $A_3$ point is at $(0,0)$ and the crease is $(0,y)$ with $y>0$. The function $Z$ is continuous everywhere and smooth except on the line $(0,y)$ with $y \ge 0$. Away from this line we find $dZ = pdx + p^2 dy$ which is continuous at $(0,0)$ because $p \rightarrow 0$ as $(x,y) \rightarrow (0,0)$.
Hence $Z$ is a $C^1$ function except along the crease (where it is not differentiable). A calculation gives
\begin{equation}
    \left( \frac{\partial p}{\partial x}\right)_y = \frac{1}{12p^2 -2y} \qquad \left( \frac{\partial p}{\partial y}\right)_x = \frac{2p}{12p^2 -2y}.
\end{equation}
This shows that $p$, regarded as a function of $(x,y)$, is not differentiable at $(0,0)$. Hence $Z$ is not twice differentiable at the $A_3$ point. In summary, $H$ is not differentiable at the crease (as expected) and $H$ is continuously differentiable, but not twice differentiable at the $A_3$ point.

We shall now discuss the behaviour of the big wavefront near an $A_3$ singularity, and deduce the corresponding behaviour of the horizon $\cH$. In this case the results of Arnol'd {\it et al.}~show that we can introduce smooth coordinates $(w,x,y,z)$ in spacetime such that, in a finite neighbourhood of an $A_3$ singularity, the big wavefront takes the form $(w,p,q) \mapsto (w,x(p,q),y(p,q),z(p,q))$ where $(x(p,q),y(p,q),z(p,q))$ are given by \eqref{A3map}. Clearly this is simply a product of a line with the (small wavefront) $A_3$ surface just discussed. However, this product structure does not extend to the metric tensor. In particular it is not always correct to interpret $w$ as a time coordinate and $(x,y,z)$ as spatial coordinates. The $A_3$ point of interest is at $(0,0,0,0)$ but, since the $w$-dependence is trivial, the big wavefront possesses a line of $A_3$ points, with tangent $\partial/\partial w$. Using our results for the small wavefront $A_3$ surface, we can see that the big wavefront is differentiable on the $A_3$ line. Hence the results of \cite{Beem:1997uv} (see Section \ref{sec:endpoint_general}) imply that exactly one generator enters the horizon at an $A_3$ caustic point.  

We can use the fact that the wavefront is null to constrain the form of the metric in these coordinates, just as we did for an $A_2$ big wavefront singularity. The normal to the wavefront is $dz - pdx -p^2 dy$ (exactly the same calculation as for the small wavefront). Imposing the condition that this is null on the big wavefront gives
\begin{equation}
 g^{zz} - 2p g^{xz} - 2p^2 g^{yz} + p^2 g^{xx} + 2p^3 g^{xy} + p^4 g^{yy} = 0
\end{equation}
where this equation must hold at points with coordinates $(w, 4 p^3-2qp, q, 3 p^4 -qp^2)$. Using the fact that the coordinates and metric are smooth, we can expand the above equation in $p$ and equate powers of $p$. Equating coefficients of $p^0$ and $p^1$ gives (using $y=q$)
\begin{equation}
    g^{zz}(w,0,y,0) = 0 \qquad g^{xz}(w,0,y,0)=-y g^{zz}_{,x}(w,0,y,0).
\end{equation}
Going to order $p^2$ it is easiest to start by setting $q=0$ (i.e. $y=0$), which gives $g^{yz} = g^{xx}/2$ at $(w,0,0,0)$. Combining with the above equations, we learn that on the $A_3$ line $(w,0,0,0)$ we have
\begin{equation}
\label{A3metric}
    g^{zz} = g^{xz}=0 \qquad g^{yz} = \frac{1}{2}g^{xx}.
\end{equation}
At higher order in $p$ one obtains further conditions involving derivatives of the metric components but we shall not need these. These equations imply that $dz$ is null on the $A_3$ line. Indeed $dz$ is null, and normal to the big wavefront, in the entire $(w,y)$-plane. On the $A_3$ line we also have that $dx$ is orthogonal to $dz$ so $dx$ must be either spacelike, or null and parallel to $dz$. But $dx$ and $dz$ are linearly independent so $dx$ cannot be parallel to $dz$. Hence $dx$ is spacelike at the caustic, i.e., $g^{xx}(w,0,0,0)>0$ and hence (by the final equation above) $g^{yz}(w,0,0,0)>0$. On the $A_3$ line we now have
\be
\label{dzA3}
(dz)^a = g^{y z}(\partial/\partial y)^a + g^{wz} (\partial/\partial w)^a \qquad g^{yz}>0.
\ee
This null vector is normal to the big wavefront, and must therefore be tangent to the (unique) generator through $(w,0,0,0)$. Since $g^{yz}>0$, $(dz)^a$ points {\it into} an $A_3$ point from the region $y<0$ where the big wavefront is smooth. Since we know that our $A_3$ point must be a {\it past} endpoint of a generator of $\cH$, it follows that this generator must have future-directed tangent $-(dz)^a$, which points out of the $A_3$ point towards the smooth region of the big wavefront. We also have $0=dz \cdot \partial/\partial w$, hence at an $A_3$ point, $\partial/\partial w$ must be either spacelike or null and tangent to $(dz)^a$ but equation \eqref{dzA3} shows the latter is not true hence $\partial/\partial w$ must be spacelike on the $A_3$ line, i.e., this line is spacelike. (Similarly $\partial/\partial x$ is spacelike on the $A_3$ line.) The tangent plane to the big wavefront (or $\cH$) at an $A_3$ point is normal to $dz$ and hence spanned by $\{\partial/\partial w, \partial/\partial x, \partial/\partial y\}$ or, equivalently, by $\{\partial/\partial w, \partial/\partial x, (dz)^a\}$. 

Emanating from the $A_3$ line is a section of the crease submanifold (so $A_3$ points belong to the closure of the crease submanifold). This is given by points with coordinates $(w,0,y,y^2/4)$ with $y>0$. On the big wavefront we also have two 2d submanifolds of $A_2$ points, with coordinates $(w,-8p^3,6p^2,-3p^4)$ ($p>0$ or $p<0$). As discussed for $H$, these do not belong to the horizon $\cH$, which is constructed by discarding points lying beyond the self-intersection of $\cW$. So $\cH$ corresponds only to the part of $\cW$ with $q \le 2p^2$. The set of tangent vectors to the crease set is spanned by $\partial/\partial w$ and   $\partial/\partial y + (y/2) \partial/\partial z$. In the limit where we approach the $A_3$ line, this tends to the 2-plane spanned by $\partial/\partial w$ and $\partial/\partial y$ or, equivalently, by $\partial/\partial w$ and $(dz)^a$, so this limiting 2-plane is null, and tangent to $\cH$. Locally, the union of the $A_3$ line and the crease submanifold has the structure of a smooth 2d manifold with boundary. At the $A_3$ line, the tangent plane to this manifold with boundary is tangent to $\cH$. (This has been seen previously in examples \cite{Shapiro:1995rr}.) 

We shall now discuss the interpretation of an $A_3$ singularity of $\cH$ w.r.t.~a time foliation. As usual, let $\tau$ be a time function with Cauchy surfaces of constant $\tau$ denoted as $\Sigma_\tau$. For a generic value of $\tau$, such a surface will intersect the $A_3$ line transversally, i.e., $(d\tau)_w \ne 0$. Without loss of generality, assume this intersection occurs at $(0,0,0,0)$ and has $\tau=0$. By the implicit function theorem, we can solve the equation $\tau(w,x,y,z)=w'$ for $w$ for small values of $(w',x,y,z)$. The solution $w(w',x,y,z)$ depends smoothly on $(w',x,y,z)$. We can now change to new coordinates $(w',x,y,z)$. This transformation does not affect the equations determining $\cW$ so all of the above analysis is still valid with $w$ replaced by $w'$. Dropping the prime, we have shown that for this $\tau$ we can perform a change of coordinates that preserves the canonical form of $\cW$ and simplifies the time function to $\tau =w$, recovering a result of \cite{arnold:1976}. It now follows that, for this $\tau$, the small wavefront and the horizon cross-section $H$ have exactly the structure discussed at the start of this section: on $H$ there is an isolated $A_3$ point with a crease emerging from it.  

This analysis was for generic values of $\tau$. However, just as we discussed for corners, there will exist special values of $\tau$ for which $\Sigma_\tau$ is tangent to the $A_3$ line. This corresponds to a qualitative change in the features of the small wavefront. Following the terminology of Arnol'd, we shall refer to this as an {\it $A_3$ perestroika.} We can shift $\tau$ so that the perestroika occurs at $\tau=0$ and we can choose local coordinates as above so that the $A_3$ point on $\Sigma_0$ is at $(0,0,0,0)$. The fact that $\Sigma_0$ is tangent to the $A_3$ line implies that $\partial_w\tau=0$ at $(0,0,0,0)$. Since $-(dz)^a$ is future-directed and null at the $A_3$ point and $-(d\tau)^a$ is future-directed and timelike, we must have $(-dz) \cdot (-d\tau) < 0$ which (using \eqref{dzA3} and $\partial_w\tau=0$) gives $g^{yz} \partial_y \tau<0$ and hence $\partial_y \tau<0$ near the $A_3$ point. Consider a curve extending from this $A_3$ point into the crease submanifold. Such a curve can be written $(w(s),0,y(s),y(s)^2/4)$ where $w(0)=y(0)=0$ and $y(s)>0$ for $s>0$. At $s=0$ we have $d\tau/ds = \partial_y \tau<0$. Hence $\tau<0$ on the crease set in a neighbourhood of this $A_3$ point. In particular, creases are absent near this $A_3$ point for $\tau \ge 0$. Thus an $A_3$ perestroika describes a process in which a (section of) crease {\it disappears}.

\begin{figure}[t]
    \centering
    \includegraphics[width=\textwidth]{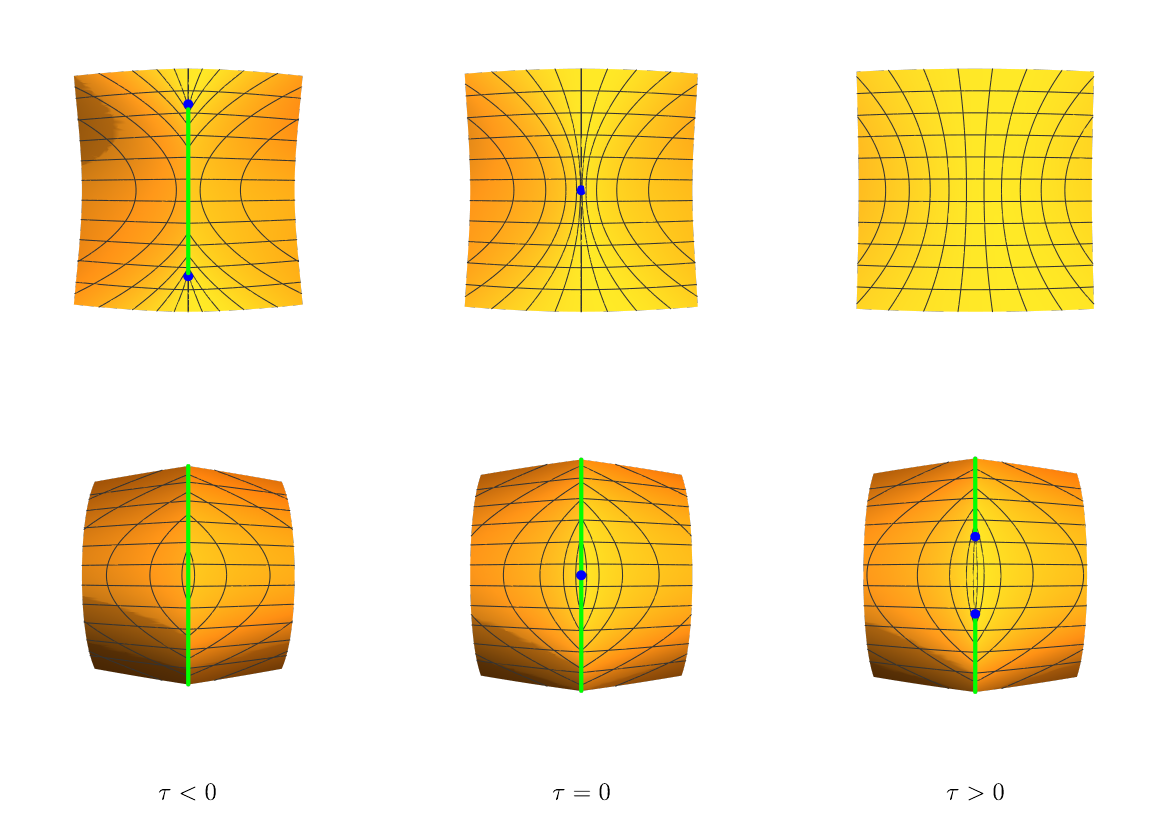}
    \caption{The two types of $A_3$-perestroika. The creases have been marked by green lines and the $A_3$ caustics by blue points. The figures show the horizon cross-section $\Sigma_\tau \cap \cH$ near the perestroika. \textit{Top:} A section of crease bounded by two $A_3$ points shrinks to zero size as $\tau\to 0$. For $\tau>0$ the surface is smooth. \textit{Bottom:} A section of crease nucleates two $A_3$ points at $\tau=0$, which then move away from each other, leaving a smooth surface between them. For $\tau>0$ locally there are now two creases, each bounded at one end by an $A_3$ caustic.}
    \label{fig:a3_perestroika}
\end{figure}

Near the origin we can expand\footnote{
Arnol'd shows that in this case one can change coordinates, preserving the canonical form of $\cW$, to bring the time function to the form $\tau = -y \pm w^2$ \cite{arnold:1976} (if $\partial_y \tau<0$). However we shall not need to do this.}
$\tau=a_i x^i + b_{ij} x^i x^j + 2 b_{wi} w x^i + b_{ww} w^2 + \ldots$ where $x^i=(x,y,z)$ and $a_y < 0$. Since $A_3$ points have $x^i=0$ they have $\tau = b_{ww}w^2 + {\cal O}(w^3)$. Generically $b_{ww} \ne 0$. If $b_{ww}<0$ then no $A_3$ points are present for $\tau>0$ (the surface $\Sigma_0$ ``curves up'' from the $A_3$ line). A single $A_3$ point is present at $\tau=0$. Two $A_3$ points are present for $\tau<0$, with $w \sim \pm \sqrt{-\tau}$ so the distance between them shrinks as $\sqrt{-\tau}$. On the horizon cross-section $H$, emanating from each $A_3$ point is a crease. There are two possibilities: (1) the $A_3$ points are connected locally by a single crease; (2) the $A_3$ points are not connected locally by a single crease. In case (1), the crease perestroika describes a finite section of crease, with $A_3$ endpoints, which shrinks to zero and vanishes at $\tau=0$. This is shown in the top row of Fig. \ref{fig:a3_perestroika}. Case (2) would describe a pair of creases, each with an $A_3$ endpoint, that merge at the origin to form a single section of crease. But this is excluded because we showed above that creases are not present near the origin for $\tau>0$. If $b_{ww}>0$ then one obtains the time reversed versions of (1) and (2). In this case, it is (1) that is excluded and (2) describes a process in which a section of crease nucleates a pair of $A_3$ points which move apart, with separation scaling as $\sqrt{\tau}$ and no crease between them (since no crease is present near the origin for $\tau>0$). In other words it describes the decay of a section of crease via $A_3$ nucleation. This is shown in the bottom row of Fig. \ref{fig:a3_perestroika}. 

In summary, there are two types of crease perestroika: one describes the disappearance of a finite section of crease with $A_3$ endpoints, the other describes the nucleation within a section of crease of a pair of $A_3$ points, which subsequently move apart, smoothing out the crease. Both types of perestroika have a smoothing effect on the horizon. For a wavefront in flat spacetime, these perestroikas are well known in the catastrophe theory literature, see e.g., Fig. 63 of \cite{arnold} (for a horizon cross-section $\Sigma_\tau \cap \cH$ we discard the portions of the figure lying beyond the self-intersection). A difference in our case is that there is a preferred direction of time in these perestroikas. This time asymmetry arises because $\cH$ is a {\it future} horizon.

\subsection{Elements of a black hole merger}

\label{sec:elements}

We shall now discuss how the various perestroikas that we have studied arise during the simplest kind of black hole merger that are generic enough to be described by the perestroikas discussed in this paper.\footnote{This section has significant overlap with Section V of Ref. \cite{Cohen:2011cf}. We have included it in order to highlight the role of perestroikas in a merger. We believe the observation at the end of this section is new.} In simple examples of non-axisymmetric mergers \cite{Husa:1999nm,Cohen:2011cf,Bohn:2016soe,Emparan:2017vyp}, the crease submanifold is an infinite strip. The two asymptotic regions of the strip lie on the two separate black hole horizons long before the merger. The two boundaries of this strip are $A_3$ lines. No corners are present in these simple examples. 

\begin{figure}
    \centering
    \input{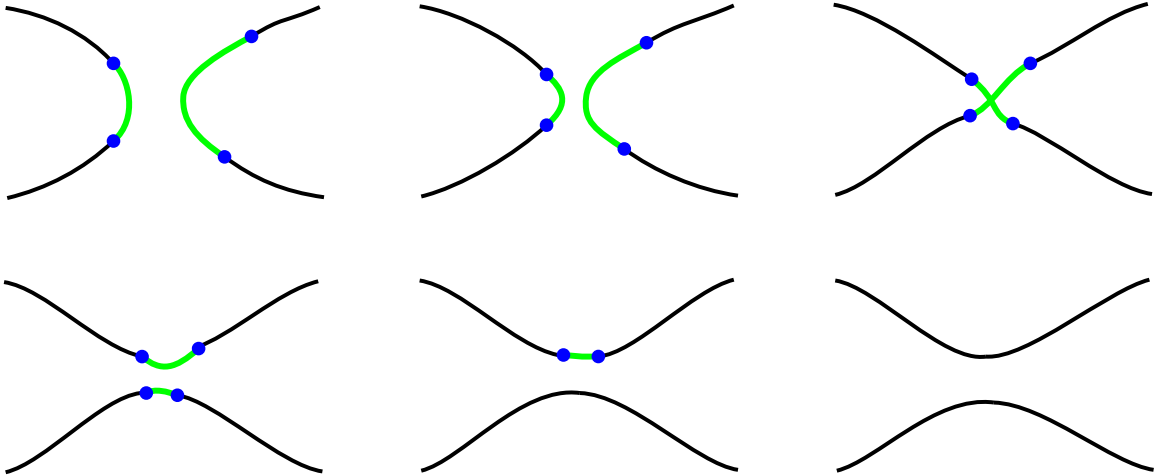}
    \caption{Merger of two black holes through the formation of a ``bridge'', with no holes in it. The horizon cross-section $\Sigma_\tau \cap \cH$ is shown ``from above'', with the black hole regions to the left and right of the curves in the first diagram. The creases are shown in green. The endpoints of the creases are $A_3$ caustics (blue points).}
    \label{fig:merger}
\end{figure}

Consider a time foliation which describes a merger, i.e., $\Sigma_\tau \cap \cH$ is topologically a pair of spheres for large negative $\tau$ and a single sphere for large positive $\tau$. For large negative $\tau$, the intersection of $\Sigma_\tau$ with the crease submanifold is a pair of line segments (creases). The endpoints of these lines are $A_3$ points. So before merger, each black hole horizon exhibits a ``chisel-like'' feature. We'll now describe the simplest possibility for what happens next. See Fig. \ref{fig:merger}, which shows (schematically) the local structure of a horizon cross-section $\Sigma_\tau \cap \cH$. As $\tau$ increases, these intersection lines move towards each other within the crease submanifold and eventually a crease perestroika occurs (top right diagram): the horizon cross-section now becomes connected, so this is the ``instant of merger''. After the merger, the horizon is topologically spherical; there is a thin ``bridge'' connecting the two original black holes, and a finite section of crease runs along each edge of this bridge (bottom left diagram). These finite sections have $A_3$ endpoints. Each of these sections of crease now shrinks. First one vanishes in an $A_3$ perestroika, then the second also vanishes in an $A_3$ perestroika. The horizon is now smooth. The black hole then settles down to equilibrium.

\begin{figure}
    \centering
    \input{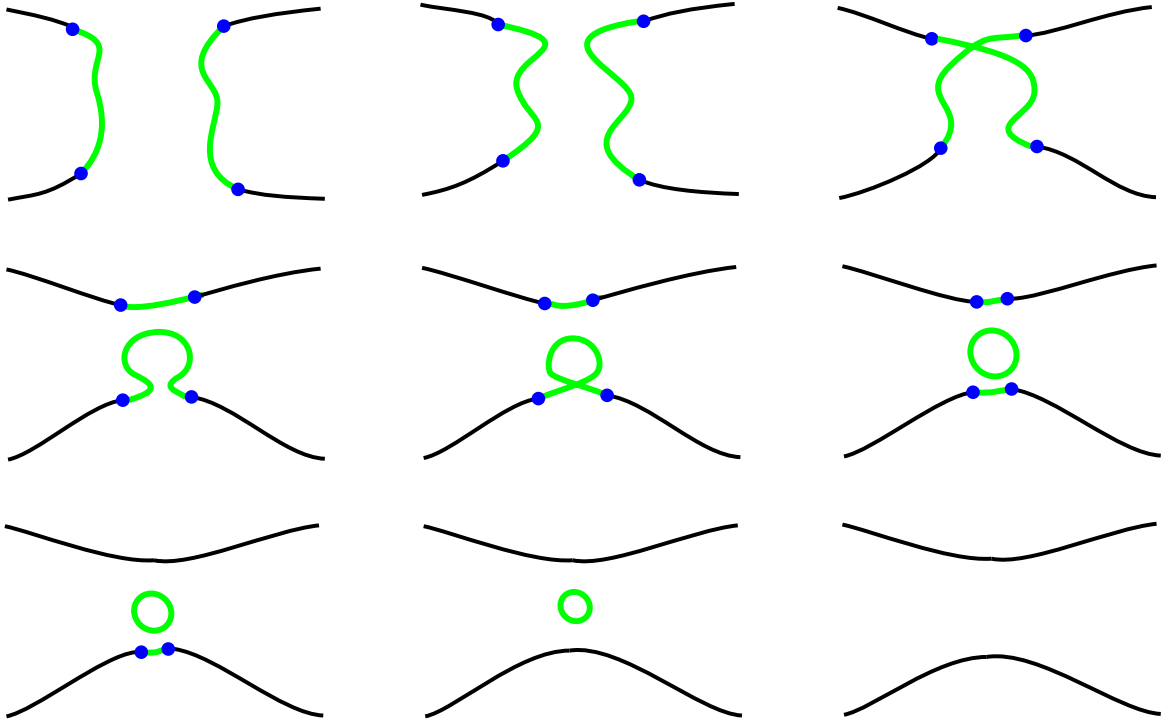}
    \caption{Merger of two black holes through the formation of a ``bridge'', with an intermediate stage of toroidal topology. Same colour scheme as Fig. \ref{fig:merger}. See Fig. 7 of \cite{Cohen:2011cf} for a similar diagram.}
    \label{fig:merger_toroidal}
\end{figure}

A different choice of time function can lead to more complicated behaviour. For example, one can choose a time function so that the crease perestroika occurs close to an $A_3$ line. This implies that the merger point occurs close to an endpoint of the sharp edge of each ``chisel''. A second crease perestroika can subsequently occur close to the other $A_3$ line. The result is the formation of a ``bridge'' with a hole in it. See Fig. \ref{fig:merger_toroidal}. The horizon has toroidal topology, with a crease running around the inner edge of the hole. There is also a pair of (very short) finite creases, with $A_3$ endpoints, running along the two outer edges of the bridge. These creases subsequently shrink and vanish in $A_3$ perestroikas. The hole in the torus shrinks and vanishes in an elliptic crease perestroika, leaving a horizon of spherical topology. This is the behaviour seen in examples of \cite{Bohn:2016soe,Emparan:2017vyp}. 

In these processes, the ``instant of merger'' is always described by a crease perestroika, never by an $A_3$ perestroika. Indeed, neither of the two possible $A_3$ perestroikas (Fig. \ref{fig:a3_perestroika}) describes a merger of two locally disconnected sections of horizon. This appears to contradict statements about some (non-axisymmetric) examples in the literature (e.g. in Ref. \cite{Husa:1999nm}) where it is asserted that the instant of merger is a merger of caustic points. We believe that, in such examples, the crease perestroika occurs very close to an $A_3$ line, leading to this confusion.

\subsection{$(A_3,A_1)$ caustics}

\label{sec:A3A1}

The $(A_3,A_1)$ caustic is an isolated caustic point that arises when a smooth section of a wavefront intersects an $A_3$ line transversally. We can describe the wavefront locally near such a caustic as follows. Introduce coordinates $(w,x,y,z)$ adapted to the $A_3$ caustic as explained in Section \ref{sec:A3}, with the $(A_3,A_1)$ point at the origin. Now consider a smooth null hypersurface $\cN$ passing through this point, with equation $f(w,x,y,z)=0$ where $df$ is null and $f(0,0,0,0)=0$. We choose $f$ so that $(df)^a$, which is tangent to the generators of $\cN$, is future-directed. 

We now adjust our coordinates to simplify $f$. Since $\cN$ intersects the $A_3$ line transversally, we have $(df)_w \ne 0$ at the origin. By the implicit function theorem the equation $f(w,x,y,z)=-w'$ admits a smooth solution $w(w',x,y,z)$ for $(w',x,y,z)$ in a neighbourhood of the origin. We then use $(w',x,y,z)$ as new coordinates. This does not affect the canonical form of the $A_3$ surface. Dropping the prime, we have shown that we can choose coordinates so that $f=-w$.

Locally the big wavefront is the union of $\cN$ and the big wavefront of the $A_3$ caustic described in Section \ref{sec:A3}, with equation $z=Z(x,y)$.  
As usual, we construct $\cH$ from the big wavefront by discarding parts that correspond to extending null geodesics to the past beyond an intersection or caustic. So we start by excluding the part of the $A_3$ big wavefront with $q > 2p^2$, just as in Section \ref{sec:A3}. 

Two generators enter $\cH$ at the $(A_3,A_1)$ point: one is the generator of $\cN$ through the origin, with tangent vector $W^a \equiv -(dw)^a$ and the other is the usual generator that enters at an $A_3$ point with tangent $V^a \equiv -(dz)^a$ there. Since $0 > W \cdot V$ we have $W^z>0$ and $V^w>0$. Hence the ``$A_3$ generator'' that enters at the origin has increasing $w$, so it lies in the region $w>0$. Since $\cN$ is the surface $w=0$ we must discard the region $w<0$ of the $A_3$ big wavefront since it lies beyond the intersection with $\cN$. Similarly, the generator of $\cN$ that enters at the origin has increasing $z$. This implies that it has $z>Z(x,y)$. Therefore we must discard the region $z<Z(x,y)$ of $\cN$ since it lies beyond the intersection with the $A_3$ big wavefront. We've now shown that, in a finite neighbourhood of the $(A_3,A_1)$ point, $\cH$ is the union of $\{w=0,z \ge Z(x,y)\}$ and $\{w>0,z=Z(x,y)\}$. 

\begin{figure}
    \centering
    \includegraphics[trim={0 0 0 2.7cm},clip]{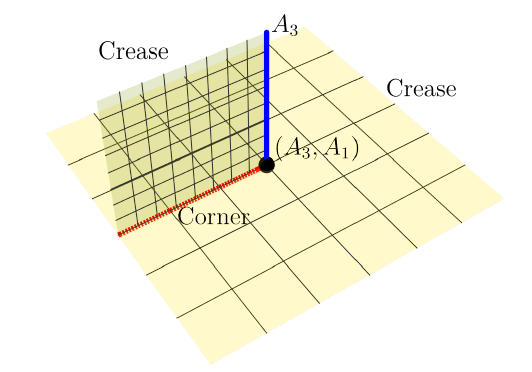}
    \caption{Structure of $\cH_{\rm end}$ set near an $(A_3,A_1)$ point. The $A_3$ line is in blue, the corner submanifold in red and the $(A_3,A_1)$ point in black. The remaining (yellow) surfaces form the crease submanifold.}
    \label{fig:a3a1_crease_set_structure}
\end{figure}

Now we can describe the structure of $\cH_{\rm end}$ near an $(A_3,A_1)$ point (this is also described in \cite{Siino:2004xe}). The $(A_3,A_1)$ point is at the origin and has $N(p)=2$. Emanating from this is a line of $A_3$-points $\{(w,0,0,0):w>0\}$ which have $N(p)=1$. The crease submanifold ($N(p)=2$) is (locally) a disjoint union of two connected components. The first component corresponds to the intersection of $\cN$ with the smooth part of the surface $z=Z(x,y)$. This is the set $\{(0,x,y,Z(x,y))\} \backslash \{(0,0,y,y^2/4),y \ge 0\}$. The second component arises from the crease submanifold associated with the $A_3$ wavefront, away from its intersection with $\cN$. This is the set $\{(w,0,y,y^2/4):w>0,y>0\}$. Finally we have the corner submanifold ($N(p)=3$) which is the intersection of $\cN$ with the $A_3$ crease submanifold, i.e., the line $\{(0,0,y,y^2/4):y>0\}$. 
The structure of $\cH_{\rm end}$ is shown in Fig. \ref{fig:a3a1_crease_set_structure} where the $z$-direction is suppressed and the $w$-direction is vertical (see also Fig.~4 of \cite{Siino:2004xe}). This is, of course, a {\it local} description of $\cH_{\rm end}$ near an $(A_3,A_1)$ point. For an example of how $\cH_{\rm end}$ might behave globally (with a connected crease submanifold) see Fig.~6 of \cite{Siino:2004xe}.\footnote{This figure shows the crease set, not $\cH_{\rm end}$, so it does not include $A_3$ points.}

\begin{figure}[t!]
    \centering
    \includegraphics[width=\textwidth]{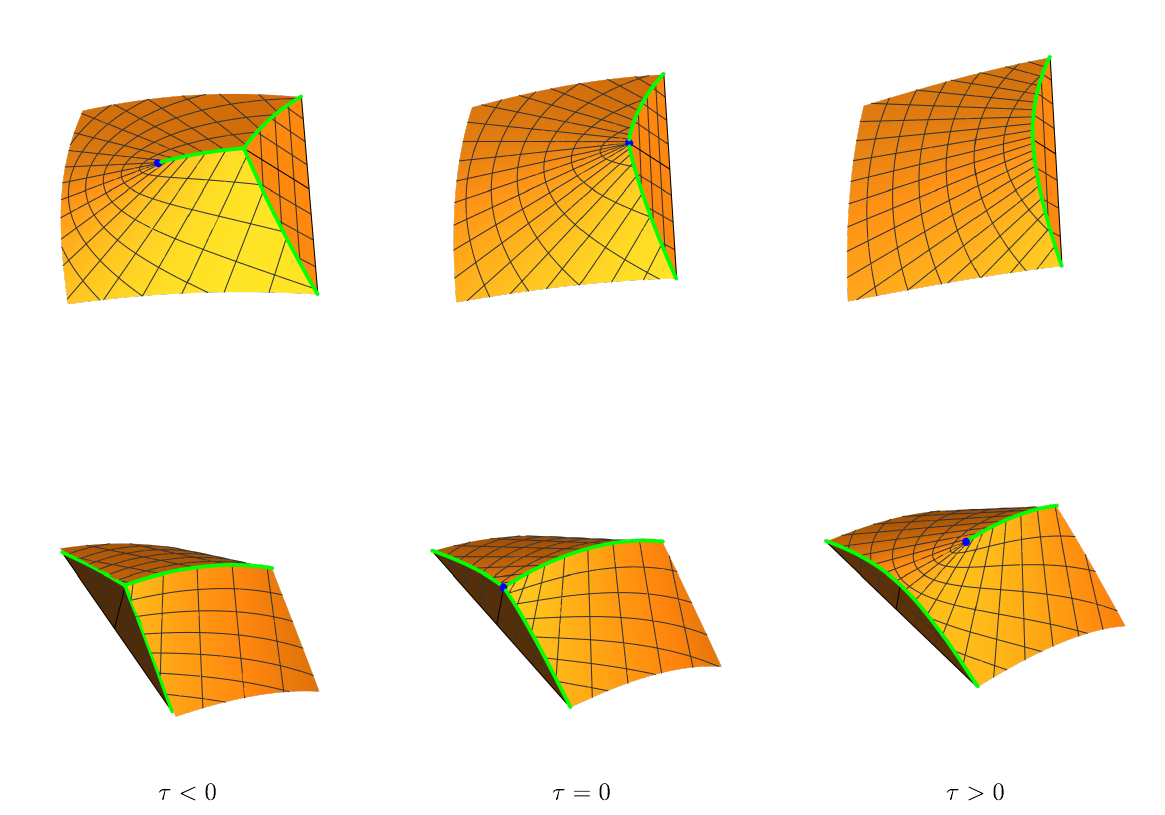}
    \caption{Perestroikas associated with an $(A_3,A_1)$ caustic point. Creases are shown in green and $A_3$ points in blue (these are $(A_3,A_1)$ points at $\tau=0$). \textit{Top:} For $\tau<0$, the horizon cross-section contains a corner with three creases emerging from it, one of which terminates at an $A_3$ point. As $\tau\to0-$, this crease shrinks, forming an $(A_3,A_1)$ point at $\tau=0$. For $\tau>0$, the two remaining creases have merged and smoothed out into a single crease. The time-reverse of this process is also possible. \textit{Bottom:} For $\tau<0$, three creases meet at a corner. They smooth out at the $(A_3,A_1)$ point at $\tau=0$, and at later times split off into a single crease and another crease emanating from an $A_3$ point. The time-reverse of this process cannot occur on a future horizon.}
    \label{fig:A3A1}
\end{figure}

Next we shall describe the different possible perestroikas associated with an $(A_3,A_1)$ caustic. Let $\tau$ be a time function with $\tau=0$ at the origin. First we investigate whether $A_3$ points and corner points occur for positive or negative $\tau$. For small $w$ we have $\tau(w,0,0,0) \approx (d\tau)_w w$. An $A_3$ point has $w>0$ so such a point is present near the origin for $\tau>0$ if $(d\tau)_w>0$ and for $\tau<0$ if $(d\tau)_w<0$. Similarly, for small $y$ we have $\tau(0,0,y,y^2/4) \approx (d\tau)_y y$ and so a corner is present near the origin for $\tau>0$ if $(d\tau)_y>0$ and for $\tau<0$ if $(d\tau)_y<0$. At the origin we have (using \eqref{dzA3})
\be
 0 > V^a (-d\tau)_a = (dz)^a (d\tau)_a = g^{yz}(d\tau)_y + g^{wz} (d\tau)_w
\ee
We know from Section \ref{sec:A3} that $g^{yz}>0$ and also $g^{wz} = (-dw) \cdot (-dz) = W \cdot V <0$. Hence we cannot have both $(d\tau)_y>0$ and $(d\tau)_w<0$ so, for generic $\tau$, it is not possible that a corner but no $A_3$ point is present near the origin for small $\tau>0$. This leaves three possible cases: (1) corner and $A_3$ point present only for $\tau<0$; (2) corner and $A_3$ point present only for $\tau>0$; (3) corner but no $A_3$ for $\tau<0$, $A_3$ but no corner for $\tau>0$.
Drawing in $3$ crease lines emanating from each corner and $1$ crease line emanting from each $A_3$ point we obtain Fig. \ref{fig:A3A1}: the top row shows case (1), taking the time reverse of this gives case (2) and the bottom row shows case (3).

We shall now demonstrate that each of these three cases is possible by exhibiting a time function that realizes each case. First consider $\tau = w+z$. Recall $dw$ is null on $\cN$ so $g^{ww}=0$ at $w=0$. Using $g_{wz}<0$ we see that $-d\tau$ is timelike and future-directed near the origin so $\tau$ is a time function. Now deform this to $\tau = w+z +\epsilon y$. By continuity this is still a time function (locally) for small $\epsilon$. We have $(d\tau)_w>0$ and $(d\tau)_y = \epsilon$ so by choosing the sign of $\epsilon$ we can realize cases (2) and (3). Next consider $\tau = z-\epsilon y - \epsilon^2 w$ with $\epsilon>0$. At the origin this gives $(d\tau)^2 = -2\epsilon g^{zy} + O(\epsilon^2)$ and so $(d\tau)^a$ is timelike for small enough $\epsilon$. $V \cdot (-d\tau) = (-dz) \cdot (-d\tau) = -\epsilon g^{zy}+O(\epsilon^2)$ is negative for small $\epsilon$ so $(-d\tau)^a$ is future-directed. Hence $\tau$ is a time function near the origin. It has $(d\tau)_y<0$ and $(d\tau)_w>0$ so we have realized case (1).

\section{Black hole entropy}

\label{sec:entropy}

\subsection{Creases}

\label{sec:crease_entropy}

The Bekenstein-Hawking formula for the entropy of a horizon cross-section $H$ is
\be
S_{\rm BH} = \frac{A}{4\ell_P^2} 
\ee
where $A$ is the area of $H$ and $\ell_P = \sqrt{G\hbar}$ is the Planck length. In this section we shall discuss the possibility that a crease makes an additional contribution to black hole entropy of the form
\be
\label{Screase}
 S_{\rm crease} = \frac{1}{\ell_P} {\int_{\rm crease} F(\Omega) dl}
 \ee
where $l$ is the proper length along the crease and $F(\Omega)$ is a dimensionless function of the angle $\Omega$ between the two smooth sections of horizon that meet at the crease ($\Omega$ depends on $l$). We shall discuss the form of $F$ below. Note that $S_{\rm crease}$ vanishes for a stationary black hole since the horizon of such a black hole is smooth. Furthermore, creases do not appear in linearized perturbations of a stationary black hole and so $S_{\rm crease}$ does not affect the first law of black hole mechanics.

To motivate this suggestion, we recall the connection between black hole entropy and the entanglement entropy $S_{ee}$ of quantum fields across a black hole horizon \cite{Bombelli:1986rw,Srednicki:1993im}, as explained in \cite{Susskind:1994sm}. For an entangling surface of area $A$, $S_{ee}$ exhibits an area-law divergence $S_{ee} = CA/\epsilon^2+\ldots$ where $\epsilon$ is an ultraviolet cut-off and $C$ is a constant depending on the renormalization scheme. The effective action for quantum fields in curved spacetime also exhibits a divergence: there is a term proportional to $R$ whose coefficient diverges as $\epsilon^{-2}$. When added to the Einstein-Hilbert action, this implies that the effective Newton constant is given by $1/G = 1/G_B(\epsilon)+{4C/\epsilon^2}$ where $G_B(\epsilon)$ is the ``bare'' Newton constant. In a black hole spacetime, it turns out that this is precisely what is needed to render the {\it generalized} entropy $S_{\rm gen} \equiv S_{\rm BH}+S_{\rm ee}$ well-defined: the $1/\epsilon^2$ terms cancel between the two terms and their sum is equal to $A/(4\ell_P^2)$ to leading order. 

For a smooth entangling surface (in four spacetime dimensions), $S_{ee}$ has a subleading divergence proportional to $\log \epsilon$ \cite{Solodukhin:2008dh}. A similar $\log \epsilon$ also appears in the quantum effective action, where it multiplies terms that are quadratic in curvature. This renormalizes the coefficients of terms in the gravitational effective action that are quadratic in curvature. Once again one finds that this implies that the $\log \epsilon$ terms in the generalized entropy cancel out, with the effect that these terms are replaced by the renormalized couplings \cite{Bousso:2015mna}. 

In the presence of a crease, it has been found that $S_{ee}$ exhibits a stronger subleading divergence proportional to $1/\epsilon$ \cite{Klebanov:2012yf,Myers:2012vs} (earlier work established an analogous result in $3$ spacetime dimensions \cite{Casini:2006hu,Hirata:2006jx}). Specifically, for the case of a crease corresponding to the intersection of two planes in flat space, it is found that $S_{ee} = -f(\Omega)l/\epsilon$ where it is assumed that the dimension along the crease has been compactified with length $l$. The function $f(\Omega)$ depends on the quantum field theory in question. In examples it is found that $f$ is always positive, has a simple pole at $\Omega=0$ and then monotonically decreases, vanishing at $\Omega=\pi$. Positivity and monotonicity of $f$ are consequences of the subadditivity property of entanglement entropy \cite{Hirata:2006jx}.

Given that these divergences arise from local short-distance effects, it seems plausible that for an entangling surface with a crease whose opening angle $\Omega$ varies along the crease, the entanglement entropy will diverge as $S_{ee} =  -\epsilon^{-1}I_{\rm crease}$ where $I_{\rm crease} = \int_{\rm crease} f(\Omega) dl$. In the case of a black hole horizon, if the $1/\epsilon$ terms are to cancel out in the generalized entropy then a similar term must be present in the black hole entropy. The simplest way this could happen is if there is a term in the bare black hole entropy proportional to $I_{\rm crease}$ with coefficient depending on $G=G_B(\epsilon)$ in a suitable way. 
In more detail, note that $(-4CG_B(\epsilon))^{-1/2}=\epsilon^{-1}[1-\epsilon^2/(4CG)]^{1/2}$. So, after expanding in $\epsilon$, to cancel the $1/\epsilon$ term in $S_{\rm ee}$ we can include the bare gravitational term $(-4CG_B(\epsilon))^{-1/2} I_{\rm crease}$ with the result that the generalized entropy contains the term $g(\epsilon) I_{\rm crease}$ where $g(\epsilon) = \epsilon^{-1} \left\{ [1-\epsilon^2/(4CG)]^{1/2}-1 \right\} = -\epsilon/(8CG) + \ldots$. This vanishes as $\epsilon \rightarrow 0$ but it is unclear whether taking $\epsilon \rightarrow 0$ is the correct thing to do as it requires a UV complete theory of gravity. If one keeps $\epsilon$ non-zero then we see that the generalized entropy contains the term \eqref{Screase} with $F = \ell_P g(\epsilon) f$. Note that $g(\epsilon)<0$ so $F$ is negative.

We shall now discuss whether this term is consistent with the generalized second law of thermodynamics. We restrict to the regime of small $\ell_P$. Classically, the area spanned by horizon generators cannot decrease. Furthermore, by definition, new generators enter $\cH$ at a crease. Thus one expects that the area of cross-sections of $\cH$ is strictly increasing when a crease is present. Since $\ell_P$ is small, the resulting increase in $S_{\rm BH}$ usually dominates any change in $S_{\rm crease}$ and so the second law is respected. However, the fact that $f(\Omega)$ has a pole at $\Omega=0$ implies that $S_{\rm crease}$ {\it might} become become important in a process where $\Omega \rightarrow 0$. We have seen that this happens at the pinch point of a crease perestroika. Consider the ``collapse of a hole in the horizon'' perestroika. In this case, we saw that $\Omega \sim \sqrt{-\tau}$ and the circumference of the crease also scales as $\sqrt{-\tau}$. Thus $S_{\rm crease}$ remains non-zero as $\tau \rightarrow 0-$ and then jumps discontinuously to zero for $\tau>0$. Since $S_{\rm BH}$ is continuous, the generalized entropy is also discontinuous. In order for the discontinuity to respect the second law, the residue of $F(\Omega)$ at $\Omega=0$ must be non-positive, which is consistent with our argument above that $F$ is negative.

Next consider the ``flying saucer nucleation'' perestroika. In this case, $\Omega \sim \sqrt{\tau}$ and the circumference of the crease also scales as $\sqrt{\tau}$. So again $S_{\rm crease}$ changes discontinuously at $\tau=0$ but with the opposite sign to before. This suggests that the generalized second law requires that the residue of $F(\Omega)$ at $\Omega=0$ should be non-negative. Combining with the result of the previous paragraph, this implies that the residue of this pole must vanish, i.e., there is no pole at $\Omega=0$. Since this pole was one of the few specific predictions made by this idea, it seems that these arguments have ruled out the possibility of a crease term in the generalized entropy. However, this overlooks the fact that $S_{\rm BH}$ and $S_{\rm crease}$ are just the first two terms in an expansion in $\ell_P$ so we should also expect higher order terms to be present. For a Planck-sized horizon, these higher order terms might be important. For example, there might be a term of the form $\ell_P^2/A$ which is subleading for a large black hole but not for a Planckian sized flying saucer. So flying saucer nucleation cannot be used to rule out a term of the form \eqref{Screase}.  

\subsection{Gauss-Bonnet term}

\label{sec:gauss_bonnet}

In an effective field theory (EFT) approach to gravity, one adds higher derivative corrections to the gravitational Lagrangian. The leading corrections are terms quadratic in curvature (here we assume a parity symmetry) so the action is
\begin{equation}
\label{EFT}
    I = \frac{1}{16\pi G} \int d^4 x \sqrt{-g} \left( -2\Lambda + R + \alpha \ell^2 R^2 + \beta \ell^2 R_{ab}R^{ab} + \frac{1}{2} \gamma \ell^2 L_{\rm GB} + \ldots \right)
\end{equation}
where $\ell$ is a length scale associated with UV physics, $\alpha,\beta,\gamma$ are dimensionless constants and $L_{\rm GB}$ is the Euler-density associated with the Gauss-Bonnet invariant:
\begin{equation}
    L_{\rm GB} = \delta^{abcd}_{efgh} R_{ab}{}^{ef}R_{cd}{}^{gh}.
\end{equation}
In vacuum, the $R^2$ and $R_{ab} R^{ab}$ terms can be eliminated via a field redefinition so we focus on the Gauss-Bonnet term. In 4d this term is topological, i.e., it does not affect the equations of motion. Nevertheless, various arguments indicate that this term does make a contribution to black hole entropy \cite{Jacobson:1993xs,Iyer:1994ys}. This contribution is 
\begin{equation}
\label{GBRicci}
 S_{\rm GB} = \gamma \int_H d^2 x \sqrt{\mu} R[\mu]
\end{equation}
where $H$ is a cross-section of $\cH$, with induced metric $\mu_{AB}$ and $R[\mu]$ is the induced Ricci scalar. Here, and henceforth, we have taken the UV scale $\ell$ to be the Planck length $\ell_P = \sqrt{G\hbar}$. The total entropy is then given by adding the Bekenstein-Hawking term:\footnote{
This section is intended to be independent of the suggestion of the previous section so we shall not include the term \eqref{Screase} in the entropy. If we did include this term then its scaling with $\ell_P$ suggests that it would dominate $S_{\rm GB}$ when creases are present, which would only strengthen our arguments below that $S_{\rm GB}$ cannot be excluded using the second law.}
\be
\label{fullentropy}
 S = \frac{A}{4\ell_P^2} + S_{GB}
\ee
where $A$ is the area of $H$. We shall discuss two ways of interpreting this formula. The first is to treat it as an {\it exact} expression, with no other terms present. We shall call this the ``pure GB'' interpretation, a candidate for the entropy of a black hole in Einstein gravity with a Gauss-Bonnet term but no higher order terms, and neglecting any possible additional contributions to black hole entropy from matter fields. The second interpretation, motivated by EFT, is to regard the terms written above as just the first two terms in a series, with the next terms having coefficients proportional to $\ell_P^2$.  (In this section we shall ignore the possibility of a crease term \eqref{Screase} in the entropy.)

For a smooth 2-manifold, the integral in \eqref{GBRicci} evaluates to $4\pi \chi$ where $\chi = 2-g$ is the Euler number with $g$ the genus of $H$. Hence for a smooth horizon cross-section we have
\begin{equation}
\label{SGBchi}
S_{\rm GB} = 4\pi \gamma \chi \qquad \qquad {\rm smooth \,\,\, horizon}.
\end{equation}
Consider a black hole formed in spherically symmetric gravitational collapse. In this case, a {\it smooth} horizon forms immediately and so \eqref{SGBchi} holds. $\chi$ jumps from $0$ to $2$ at the instant the horizon forms. If $\gamma<0$ then \eqref{fullentropy} would exhibit an $O(1)$ discontinuous decrease at the instant the horizon forms. In the ``pure GB'' interpretation, this violates the second law so the second law requires $\gamma \ge 0$ \cite{Sarkar:2010xp}. In the EFT interpretation this argument seems less reliable because it is sensitive to the form of the higher order corrections to \eqref{fullentropy}. If these become $O(1)$ for a Planckian sized black hole then the argument no longer works.\footnote{
See also \cite{Chatterjee:2013daa} which explains why another argument against \eqref{fullentropy} fails in EFT.} 

Now we discuss \eqref{fullentropy} for more general dynamical processes. We have seen that, generically the horizon is {\it not} smooth in a dynamical process (it is not even differentiable at a crease). Therefore it is not obvious how to make sense of the RHS of \eqref{GBRicci}. One approach is to  ``regulate'' $S_{\rm GB}$, defining it by taking a limit of smooth surfaces that converge to $H$ \cite{Sarkar:2010xp}. With this definition, \eqref{SGBchi} holds even for non-smooth horizons. One can then argue as follows that $S_{\rm GB}$ violates the second law of black hole mechanics if $\gamma>0$ \cite{Sarkar:2010xp}.

Consider a merger of two topologically spherical black holes to form another topologically spherical black hole. At the instant of merger, $\chi$ jumps from $4$ to $2$ so to avoid a discontinuous decrease in entropy, $\gamma$ must be non-positive. This argument works for both the ``pure GB'' interpretation and the EFT interpretation. In the latter case the argument assumes that we can neglect higher order corrections to \eqref{fullentropy} if the black holes are large enough. The conclusion is that the second law implies $\gamma \le 0$. In particular, for the ``pure GB'' interpretation, we've already seen that $\gamma \ge 0$ so the only possibility compatible with the second law is $\gamma=0$, i.e., the Gauss-Bonnet terms is apparently excluded by the second law.

This argument relies on assuming that \eqref{SGBchi} is valid for non-smooth horizons, which was justified by regulating $S_{\rm GB}$ by taking a limit of smooth surfaces. However, as briefly noted in \cite{Sarkar:2010xp}, it is possible that non-smooth features of the horizon may play an important role. We shall now argue that this is indeed the case. The new idea is that, by looking at the various types of non-smooth behaviour that the horizon can exhibit, we shall see that $S_{\rm GB}$ {\it does not need regulating}. Without regulating, it is not topological (for non-smooth $H$). This implies that, in a black hole merger, it does not exhibit the discontinuous behaviour just discussed, and so the above argument that it violates the second law when $\gamma>0$ no longer applies. 

When we say that $S_{\rm GB}$ does not need regulating, what we mean is that the integral on the RHS of \eqref{GBRicci} exists as an improper Riemann integral. To justify this claim, we shall discuss each of the different types of generic non-smooth behaviour that $H$ can exhibit. Here we assume that the black hole belongs to the class defined in Section \ref{sec:smooth_late}, in particular that the horizon is smooth at late time. 

First consider a crease or corner. Here the horizon cross section is locally piecewise smooth, so there is no difficulty defining the integral in \eqref{GBRicci}: $R[\mu]$ is discontinuous but the discontinuity is bounded, so the integral converges as a Riemann integral. Second consider a caustic point on $H$. We know that a generic caustic point is of type $A_3$ (or the closely related $(A_3,A_1)$). We calculate the intrinsic and extrinsic curvature of $H$ near such a point in Appendix \ref{app:A3curvature}. We find that $R[\mu]$ diverges at an $A_3$ point on $H$. However, we show that this divergence is integrable: if we excise a small region around the $A_3$ point and the crease emanating from it then the integral \eqref{GBRicci} converges as the size of this region is shrunk to zero. In other words, this integral exists as an improper Riemann integral. This holds both for $A_3$ points on a generic horizon cross-section, and for the pinch point associated with an $A_3$ perestroika (as studied in Section \ref{sec:A3}). Therefore, generically, the integral \eqref{GBRicci} exists without any need to regulate it. 

As an example, consider the nucleation of a ``flying saucer'' horizon, as described in Section \ref{sec:crease_perestroikas}, see Fig. \ref{fig:pinch_diagrams}. $H$ is topologically spherical and looks like the intersection of two smooth surfaces. $R[\mu]$ is smooth on each section and remains bounded as $\tau \rightarrow 0+$ (where $\tau$ is a time function with the nucleation occuring at $\tau=0$). Thus $S_{\rm GB}$ scales in the same way as the area of the surface, i.e., it is $O(\tau)$. In particular it is continuous at $\tau=0$, unlike the `regulated'' version of $S_{\rm GB}$. The Bekenstein-Hawking entropy is also proportional to $\tau$ but accompanied by the very large factor $\ell_P^{-2}$. Hence, in the EFT interpretation of \eqref{fullentropy}, the first term dominates $S_{\rm GB}$ and one cannot deduce anything about the sign of $\gamma$ from flying saucer nucleation. The same applies to the crease perestroika describing the closing up of a hole in the horizon (note that both of these processes increase $\chi$).

Another interesting (but non-generic) case to consider is an axisymmetric merger of two non-spinning black holes. We described the behaviour around the instant of merger in Section \ref{sec:crease_perestroikas}, see the lower row of Fig. \ref{fig:axisymmetric_pinch_diagrams}. The black hole horizons before the merger exhibit conical singularities. (Recall that these are caustics, but of a non-generic type.) A compact $2$-manifold that is smooth except at conical singularities satisfies
\cite{troyanov}
\begin{equation}
 \int d^2 x \sqrt{\mu}R[\mu] = 4\pi \chi + 2 \sum_i (\theta_i-2\pi) 
\end{equation}
where $\theta_i$ is the angle at the $i$th conical singularity (i.e., the ratio of circumference to radius for a small circle around the singularity). We can now substitute the above result in \eqref{GBRicci}. Before the merger, each black hole has a single conical singularity and from Section \ref{sec:crease_perestroikas} we know that $\theta \sim \sqrt{-\tau}$ (where $\tau$ is a time function, and the merger occurs at $\tau=0$). Hence, for each black hole, just before the merger, the above expression evaluates to $4\pi +O(\sqrt{-\tau})$ and so the sum of the contributions from each black hole approaches $8\pi$ as $\tau \rightarrow 0-$. This matches precisely with the contribution $8\pi$ of the smooth black hole that exists just after the merger. Hence $S_{\rm GB}$ is continuous at the merger, unlike what happens for the regulated version of $S_{\rm GB}$. (A similar argument applies to the nucleation of ``spindle'' sections of the horizon as discussed in \ref{sec:crease_perestroikas}.
In this case each section of spindle has two conical singularities so the above formula evaluates to $O(\sqrt{\tau})$ for small positive $\tau$. So again we have continuity at $\tau=0$.)

We have shown that $S_{\rm GB}$ is continuous in an axisymmetric merger. However, for $\gamma>0$ it is rapidly decreasing, as $\sqrt{-\tau}$ as $\tau \rightarrow 0-$. One might worry that, for a very short time, this rapid decrease might dominate over the slower increase in entropy coming from the Bekenstein-Hawking term. If so then one would have a violation of the second law for $\gamma>0$. Balancing $A/\ell_P^2$ against $\sqrt{-\tau}$ and assuming $\dot{A} = O(1)$, one sees that \eqref{fullentropy} decreases for $|\tau| \sim \ell_P^4$, and the size of this decrease is of order $\ell_P^2$. Thus, for the ``pure GB'' interpretation this argument implies that, even without regulation, $S_{\rm GB}$ violates the second law if $\gamma>0$. However, since the decrease in the entropy is comparable to the size of the higher order $O(\ell_P^2)$ terms in \ref{fullentropy} this argument is inconclusive if we adopt the EFT interpretation of \eqref{fullentropy}.

In summary, previous arguments that including $S_{\rm GB}$ leads to a violation of the second law are based on the ``regulated'' version of equation \eqref{GBRicci}, i.e., equation \eqref{SGBchi}. We have argued that equation \eqref{GBRicci} does not actually require regulating. If one does not regulate then, in the ``pure GB'' interpretation, a more refined argument still leads to the conclusion that \eqref{fullentropy} violates the second law unless $\gamma=0$. However, in the (more physical) EFT interpretation of \eqref{fullentropy}, the arguments that $S_{\rm GB}$ leads to a violation of the second law are inconclusive (for either sign of $\gamma$).

\subsection{Extrinsic curvature terms in entropy}

If one chooses not to eliminate the $R^2$ and $R_{ab} R^{ab}$ terms in \eqref{EFT}, or one considers properties of entanglement entropy, then various arguments \cite{Jacobson:1995uq,Solodukhin:2008dh,Dong:2013qoa,Wall:2015raa} indicate that the black hole entropy should contain terms quadratic in the extrinsic curvature $k_{ij}$ of the horizon cross-section $H$, viewed as a submanifold of the Cauchy surface $\Sigma$. There are two independent terms:
\be
 S_1 = \int_H d^2 x \sqrt{\mu} k^i_i k^j_j \qquad \qquad S_2 = \int_H d^2 x \sqrt{\mu} k^{ij} k_{ij}
\ee
where, as in the previous section, $\mu_{AB}$ is the induced metric on $H$ and indices $i,j$ are raised with $h^{ij}$, the inverse of the metric $h_{ij}$ on $\Sigma$. 

Are these terms well-defined on a non-smooth horizon? As for the Gauss-Bonnet term, there is no problem in defining the above integrals in the presence of a crease or corner: the horizon is locally piecewise smooth near such structures and $k_{ij}$ is smooth on each smooth piece. In Appendix \ref{app:A3curvature} we calculate $k_{ij}$ near an $A_3$ point on $H$, for a generic Cauchy surface $\Sigma$. We show that, although $k_{ij}$ diverges, the above integrals still exist as improper integrals. However, in the case where $\Sigma$ is a special Cauchy surface associated with the $A_3$ perestroika describing the disappearance of a section of crease with $A_3$ endpoints (top row of Fig. \ref{fig:a3_perestroika}), we find that the divergence is non-integrable and the above integrals are both proportional to $\log \tau$ as $\tau \rightarrow 0+$ (with the perestroika at $\tau=0$). The combination $S_1-S_2$ is finite; by the Gauss-Codacci equation \eqref{gausscodacci} this combination can be written in terms of $S_{\rm GB}$ and an integral involving curvature components of the smooth metric $h_{ij}$. 

In summary, on a generic non-smooth horizon the quantities $S_1$ and $S_2$ will diverge at an $A_3$ perestroika. Only the combination $S_1-S_2$ remains finite. For the theory \eqref{EFT}, the $R^2$ term makes a contribution to the entropy proportional to the integral of $R$ (the spacetime Ricci scalar) over $H$ \cite{Jacobson:1995uq}. Since $R$ is smooth, this contribution is finite. However, the $R_{ab} R^{ab}$ term gives a contribution involving $S_1$ and $S_2$ \cite{Dong:2013qoa,Wall:2015raa} and this is not in the combination $S_1-S_2$ so it diverges at the $A_3$ perestroika. Hence the formulae of \cite{Dong:2013qoa,Wall:2015raa} do not work for a generic non-smooth horizon. This is not necessarily a problem since, e.g., the analysis of \cite{Wall:2015raa} applies only to linear perturbations of stationary black holes. For the case of entanglement entropy \cite{Solodukhin:2008dh}, the coefficients of the terms $S_1$ and $S_2$ are proportional to $\log \epsilon$ where $\epsilon$ is a UV cut-off. The divergence at the $A_3$ perestroika may indicate that for such $H$ there is a new term in the entanglement entropy, intermediate between $\log \epsilon$ and the  $1/\epsilon$ behaviour associated with a crease.

\section{Discussion}

Given that the crease submanifold is (generically) the ``most important'' part of $\cH_{\rm end}$, it would be interesting to study its properties in greater detail. For example: are there any constraints on its topology? Does it have finite area? The latter question can be easily answered in a situation where the black hole area theorem holds: given a horizon cross-section $H$ lying to the future of the crease submanifold, consider the map from $H$ to the crease submanifold obtained by following the generators of $\cH$. By definition, this map is two-to-one so its inverse image is a pair of disjoint sets $H_1,H_2 \subset H$. Following the generators through $H_1$, the area theorem gives $A_{\rm crease} \le A_1$ where $A_{\rm crease}$ is the area of the crease submanifold and $A_1$ the area of $H_1$. Similarly $A_{\rm crease} \le A_2$. Hence $2A_{\rm crease} \le A_1+A_2 \le A_H$ where $A_H$ is the area of $H$. (This is a special case of the ``weighted'' area theorem of \cite{Chrusciel:2000cu}.) So we see that indeed the crease submanifold has finite area. It would be interesting to know what physical significance can be attached to this area. This result is perhaps related to an observation about axisymmetric black hole mergers, where $\cH_{\rm end}$ is a line of caustic points. In examples, this line has been found to have finite length \cite{Hamerly:2010cr,Emparan:2016ylg}.

A possible role for the crease submanifold is in the Bousso entropy conjecture \cite{Bousso:1999xy}. This is an upper bound on the entropy crossing a {\it lightsheet}: a non-expanding null hypersurface generated by a family of geodesics emanating orthogonally from a 2d spacelike surface $\Sigma$. A version of this conjecture was proved by Flanagan, Marolf and Wald \cite{Flanagan:1999jp}: assuming that the entropy of matter is described by an entropy current obeying certain bounds in terms of the energy-momentum tensor, they showed that
\be
 S \le \frac{A-A'}{4G\hbar}
\ee
where $S$ is the entropy of matter crossing a lightsheet extending from $\Sigma$ to another 2d spacelike surface $\Sigma'$ and $A$, $A'$ are the areas of $\Sigma$ and $\Sigma'$. If one does not introduce a second surface $\Sigma'$ then it is natural to terminate the lightsheet emanating from $\Sigma$ where it intersects the null cut locus of $\Sigma$, which is essentially the proposal of \cite{Tavakol:1999as}. At the end of Section \ref{sec:smooth_late} we explained how to define a crease submanifold for a general null cut locus. A simple modification of the arguments of \cite{Flanagan:1999jp} now gives
\be
 S \le \frac{A-2A_{\rm crease}}{4G\hbar}
\ee
where $A_{\rm crease}$ is the area of the intersection of the lightsheet with the crease submanifold of $\Sigma$.\footnote{It is necessary to take this intersection because, as discussed at the end of sec \ref{sec:smooth_late} (and in Lemma \ref{lem:Hendcutlocus}), there are two past/future directed families of null geodesics emanating orthogonally from $\Sigma$ and both play a role in defining the null cut locus; since the lightsheet is defined by only one of these families, the crease submanifold may have a component that does not intersect this lightsheet.} So the crease submanifold plays a role in bounding the amount of entropy that can cross the lightsheet.  

We have introduced the notion of a normal corner point and shown that such points form a submanifold. However, we are unaware of any physically relevant examples of black hole solutions (numerical or otherwise) of the Einstein equation that exhibit horizons with corners. It would be interesting to construct such examples.

We have reviewed the classification of Siino and Koike of endpoints of the horizon generators of a {\it generic} black hole (Table \ref{SKtable}). In Section \ref{sec:caustics} we explained why it is unclear whether or not the notion of genericity used in this classification is the same as genericity w.r.t.~perturbations of the metric. We described an alternative approach towards such a classification. This places the classification on a firmer footing if one restricts to a horizon cross-section but for the full horizon the genericity issue remains an open problem. A Lorentzian analogue of the Riemannian results of \cite{buchner} would go some way towards addressing this problem. This might be possible for a generic globally hyperbolic spacetime. However ideally one would like results for a generic {\it solution} of suitable equations of motion which looks more challenging. 

We used properties of entanglement entropy to motivate the possibility of a crease contribution to black entropy. One could similarly use properties of entanglement entropy (see e.g., \cite{Sierens:2017eun}) to motivate the 
possible existence of a corner contribution to black hole entropy. It might be interesting to study this possibility further. 

Higher derivative theories of gravity typically lead to higher-derivative terms in black hole entropy. We have considered the possible $2$-derivative terms in black hole entropy (in 4d), namely the ``Gauss-Bonnet'' term and terms quadratic in extrinsic curvature. We showed that the former is well-defined on a generic horizon but the latter diverge at an $A_3$ perestroika. This raises the question of what kinds of higher-derivative contributions to black hole entropy can ``make sense'' (i.e., remain finite) on a generic dynamical black hole horizon. A large class of possible terms are those that can be written in terms of components of the (smooth) curvature of spacetime. For example, in $f(R)$ theories the entropy density depends only on the spacetime Ricci scalar \cite{Jacobson:1995uq} and so the entropy is well-defined on a non-smooth horizon. However, in more typical higher derivative theories, extrinsic curvature terms are required if the second law is to be respected by linear \cite{Wall:2015raa} or quadratic \cite{Hollands:2022fkn} perturbations of a stationary black hole. So there is a tension between what is required perturbatively and what makes sense in a fully nonlinear situation.

\subsection*{Acknowledgments}

We are very grateful to Roberto Emparan for comments on a draft. We are also grateful to the following people for helpful discussions and suggestions: Raphael Bousso, Piotr Chru\'sciel, Mihalis Dafermos, Greg Galloway, Sean Hartnoll, Stefan Hollands, Ted Jacobson, Aron Wall. MG is supported by an STFC studentship and a Cambridge Trust Vice-Chancellor's Award. HSR is supported by STFC grant no. ST/T000694/1. 

\appendix

\section{$A_2$ caustic on a small wavefront}

\label{app:a2small}

For the small wavefront, we interpret the canonical coordinates $x^i \equiv (x,y,z^A)$, $A=1,\ldots, d-3$, of Section \ref{sec:a2} as coordinates on a Cauchy surface $\Sigma$. The small wavefront is the surface $(-3p^2,2p^3,z^A)$ lying within $\Sigma$. The $A_2$ points are at $p=0$. Now the generators of the corresponding big wavefront must depend smoothly on the wavefront parameters $(p,z^A)$. In particular $p_\mu = g_{\mu\nu} dx^\mu/d\lambda$ depends smoothly on $p$ where $\lambda$ is an affine parameter along the generators. Projecting to $\Sigma$ we see that $p_i$ must depend continuously on $p$. But $p_i$ is normal to the small wavefront, which is proportional to $\pm (dy + p dx)$. Continuous dependence on $p$ implies that the $\pm$ cannot change sign at the cusp $p=0$. Thus $p_i$ is a non-zero multiple of $n_i$ where $n \equiv dy +p dx$. Without loss of generality we assume it is a positive multiple. By rescaling the affine parameter we can set $p_i = n_i$ on $\Sigma$. Now we introduce Gaussian normal coordinates $(t,x^i)$ such that the metric near $\Sigma$ is
\begin{equation}
    g = -dt^2 + g_{ij}(t,x^k) dx^i dx^j
\end{equation}
with $\Sigma$ the surface $t=0$. The metric depends smoothly on these coordinates. 

Fix a point $q$ on the small wavefront with parameters $(p,z^A)$ where $p>0$. We shall construct a timelike curve from $q$ to the generator with parameters $(-p,z^A)$, so the big wavefront is not achronal. To do this, consider following this generator affine parameter distance $\lambda>0$ to reach a point $r$. Let $P^\mu$ be the future-directed tangent to this generator at $\Sigma$. This is
\begin{equation}
    P^i = g^{ij}(0,-3p^2,-2p^3,z^A) n_j(-p,z^A) \qquad P^t = \sqrt{g_{ij}(0,-3p^2,-2p^3,z^A) P^i P^j }.
\end{equation}
The point $r$ has coordinates
\begin{equation}
    x^\mu_r = (0,-3p^2,-2p^3,z^A)+ \lambda P^\mu  + O(\lambda^2).
\end{equation}
Consider the straight line (in these coordinates) from $q$ to $r$. This has tangent 
\begin{equation}
    V^\mu \equiv x^\mu_r - x^\mu_q = (0,0,-4p^3,0)+\lambda P^\mu + O(\lambda^2).
\end{equation}
We set $\lambda = Cp^3$ where $C>0$. For small $p$ we have
\begin{equation}
    g_{\mu\nu}V^\mu V^\nu = 16 p^6 g_{yy}-8Cp^6 P_y+o(p^6)=8 p^6 (2g_{yy}-C) + o(p^6) 
\end{equation}
where, to leading order, it does not matter at which point along the line $g_{yy}$ is evaluated. By taking $C$ large enough we ensure that $V^\mu$ is timelike. Hence this line is timelike so the big wavefront is not achronal. To exclude an $A_2$ singularity on a cross-section of $\cH$, when we apply the above argument note that $r$ is obtained by following a generator of the big wavefront to the future of the caustic, and hence coincides with a generator of $\cH$, so $r$ must belong to $\cH$, in violation of achronality of $\cH$.

\section{$A_4$ and $D_4^\pm$ caustics}

\label{app:a4d4}

In this Appendix we shall sketch an argument that the presence of an $A_4$ or $D_4^\pm$ caustic on $\cH$ would violate achronality and so such caustics cannot arise as endpoints of horizon generators. 
The basis of the argument is the presence of $A_2$ caustics arbitrarily close to the $A_4$ and $D_4^\pm$ points, in such a way that they cannot be removed by discarding the parts of the big wavefront lying behind creases (as is done for an $A_3$ caustic). Since achronality is violated arbitrarily near to an $A_2$ caustic, it must also be violated by $A_4$ and $D_4^\pm$ caustics. Note that $A_4$ and $D_4^\pm$ caustics are isolated points in spacetime. 

It is easiest to see the presence of these $A_2$ singularities through diagrams of small wavefronts (i.e. cross-sections of the big wavefront). A diffeomorphism can be used to bring a time function $\tau$ to a canonical form \cite{arnold:1976}. Sketches of the (constant $\tau$) small wavefront near caustic points are depicted in Fig.~63 of Arnol'd {\it et al.}~\cite{arnold}. These cross-sections exhibit crease lines and lines of $A_2$ points. It is clear that there exist $A_2$ singularities in any neighbourhood of an $A_4$ or $D_4^+$ point, and that it is not possible to choose a section (bounded by creases) of each small wavefront that does not contain $A_2$ caustics. Thus, unlike the $A_3$ case, we cannot eliminate the $A_2$ caustics by discarding part of the big wavefront lying beyond a crease. 

\begin{figure}
    \centering
    \includegraphics{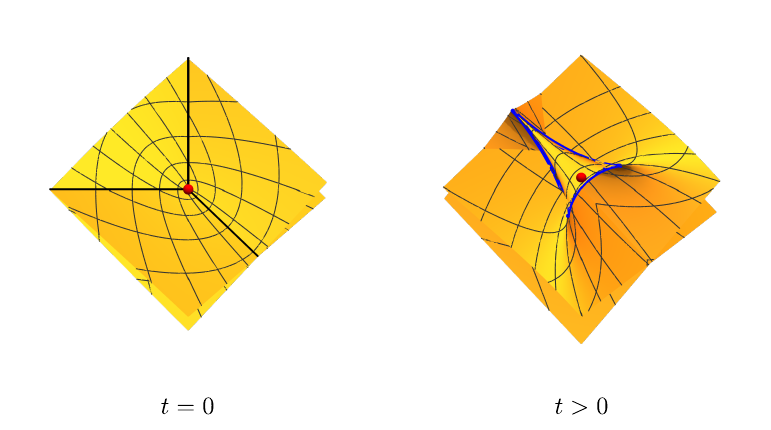}
    \caption{Small wavefronts near a $D_4^-$ caustic in Minkowski spacetime. The time function is the standard Minkowski time $t$. The red dot represents the generator through the $D_4^-$ point. At $t=0$, there are no $A_2$ singularities. For $t>0$, a triangular configuration of $A_2$ lines develops (blue), within which lies the generator through $D_4^-$.}
    \label{fig:d4minus}
\end{figure}

For the $D_4^-$ caustic, Fig.~63 of \cite{arnold} shows that, for $\tau>0$ or $\tau<0$, the small wavefront exhibits a section that is bounded by three $A_2$ lines in a triangular configuration. The triangle shrinks to zero size as $\tau \rightarrow 0$. We aim to show that there is a unique horizon generator entering at the $D_4^-$ caustic point, and this generator belongs to this triangular section of wavefront. Hence if there is a $D_4^-$ caustic on $\cH$ then this section of wavefront also belongs to $\cH$ and so there are $A_2$ singularities on $\cH$, a contradiction. Rather than attempting to prove this in full generality we shall demonstrate this for a $D_4^-$ singularity in Minkowski spacetime. (Since in four dimensions this caustic is a point, we expect that the behaviour of the wavefront in curved spacetime should be locally similar to that in flat spacetime.) Take the following big wavefront discussed in Section 4.7 of~\cite{stewart},
\be
\label{eq:D4-}
   (t,x,y,z)= \left(2p^3-2pq^2+r,-3p^2+q^2+p r, 2p q+q r,r\sqrt{1-p^2-q^2}\right),
\ee
which has a $D_4^-$ point at $(0,0,0,0)$. The parameter $r$ is an affine parameter along the generators of the wavefront. There is a unique generator through the $D_4^-$ point, which has $p=q=0$. Hence, if the $D_4^-$ point occurs on $\cH$, then this must be the horizon generator entering at the $D_4^-$ point. One can also solve for the $A_2$ caustics by finding the subspace of the wavefront where the Jacobian of the map $(p,q,r)\mapsto(t,x,y,z)$ given by~\eqref{eq:D4-} drops by one. For simplicity, we may take the time function $\tau=t$. We find that for small $t>0$, the generator lies inside the triangular region of $A_2$ lines, as illustrated in Fig.~\ref{fig:d4minus}, so a $D_4^-$ singularity cannot occur on an event horizon.

\section{Curvature near $A_3$ caustic}

\label{app:A3curvature}

Let $H=\Sigma \cap \cH$ be a generic cross-section of the horizon with an $A_3$ caustic point. In this section we shall determine the behaviour of the extrinsic and intrinsic curvature near this point. 

As explained in Section \ref{sec:A3} we can introduce coordinates $x^i \equiv (x,y,z)$ on $\Sigma$ so that the $A_3$ point is at $(0,0,0)$ and $H$ is given by equations \ref{A3map}. $(x,y)$ can be used as coordinates on $H$. In these coordinates, the $A_3$ point is at $(0,0)$ and the crease is $(0,y)$ with $y>0$. If we remove the subset $(0,y)$ with $y \ge 0$ then we obtain a smooth manifold on which $(p,q)$ can be used as coordinates, with $q<2p^2$. In terms of $(p,q)$, the crease corresponds to $p\ne 0,q \rightarrow 2p^2$ and the $A_3$ point is $(p,q) \rightarrow (0,0)$. If we write $(p,q)$ in terms of $(x,y)$ then we have $q=y$ and $p(x,y)$ is continuous at $(0,0)$ but discontinuous (changing sign) across the crease. 

The tangent vectors to the smooth part of $H$ are 
\be
 \frac{\partial}{\partial p} = \Delta \left( \frac{\partial}{\partial x}+ p \frac{\partial}{\partial z} \right) \qquad \frac{\partial}{\partial q} = -2p \frac{\partial}{\partial x}+\frac{\partial}{\partial y}-p^2 \frac{\partial}{\partial z}
\ee
where we define
\be
 \Delta = 12p^2 -2q. 
\ee
This quantity is positive everywhere on $H$ (including the crease) except at the $A_3$ point, where it vanishes. Using the above expressions we can determine the unit normal to $H$:
\be
 n = \alpha \tilde{n} \qquad \tilde{n} = dz - pdx -p^2 dy
\ee
where $\alpha>0$ is chosen to make $n$ a unit vector w.r.t.~the induced metric $h_{ij}$ on $\Sigma$. Note that $\tilde{n}$ and $\alpha$ are continuous, but not differentiable, at the $A_3$ point. 

Let $X$ be tangent to the smooth part of $H$. From $X \cdot n=0$ we have $X^z = pX^x + p^2 X^y$. We can also write $X = X^p \partial_p + X^q \partial_q$ so plugging in the above expressions for $\partial_p$ and $\partial_q$ gives
\be
\label{XiXp}
 X^x = \Delta X^p -2 p X^q \qquad X^y = X^q \qquad X^z = p \Delta X^p - p^2 X^q.
\ee
Now let $k_{ij}$ be the extrinsic curvature of $H$ viewed as a surface in $\Sigma$ and let $X,Y$ both be tangent to $H$. We have
\be
 X^i Y^j k_{ij} = X^i Y^j D_i n_j=\alpha X^i Y^j D_i \tilde{n}_j= \alpha X^i Y^j \partial_i \tilde{n}_j -\Gamma^k_{ij}X^i Y^j n_k
\ee
where $D_i$ is the covariant derivative defined by $h_{ij}$ on $\Sigma$. The Christoffel symbols are smooth and so the final term is continuous at the $A_3$ point. Substituting our expression for $\tilde{n}$ gives
\be
 X^i Y^j k_{ij} = -\alpha Y^x X^i \partial_i p - \alpha Y^y X^i \partial_i p^2 + \ldots = -\alpha X^p (Y^x + 2p Y^y) + \ldots
\ee
where the ellipses indicate terms depending smoothly on $(p,q)$. Such terms are continuous at the $A_3$ point and bounded at the crease. Finally, using \eqref{XiXp} to write $X^p$ in terms of $X^x,X^y$ gives
\be
 X^i Y^j k_{ij}=-\frac{\alpha}{\Delta} (X^x + 2p X^y)(Y^x+2pY^y) + \ldots
\ee
and so we have isolated the part of $k_{ij}$ that diverges at the $A_3$ point:
\be
 k_{ij} = -\frac{\alpha}{\Delta} m_i m_j + \ldots \qquad \qquad m = dx+2pdy.
\ee
Note that $m$ is continuous at the $A_3$ point. We now have (raising indices with $h^{ij}$)
\be
 k^{ij}k_{ij} = \frac{\alpha^2}{\Delta^2} (m^i m_i)^2 + O(1/\Delta) \qquad k^i_i k^j_j = \frac{\alpha^2}{\Delta^2} (m^i m_i)^2 + O(1/\Delta).
\ee
The Ricci scalar of the induced metric $\mu_{AB}$ on $H$ is determined by the Gauss-Codacci equation:
\be
\label{gausscodacci}
 R[\mu] = R-2R_{ij} n^i n^j + k^i_i k^j_j - k^{ij}k_{ij} 
\ee
where $R_{ij}$ and $R$ are the Ricci tensor and Ricci scalar of $h_{ij}$. Since these are smooth and $n^i$ is continuous we obtain
\be
 R[\mu] = O(1/\Delta).
\ee
Thus the divergence in $R[\mu]$ at the $A_3$ point is milder than that in $k^{ij}k_{ij}$ and $k^i_i k^j_j$. Now let's examine the volume element using $(p,q)$ as coordinates on the smooth part of $H$. Since $\partial_p = O(\Delta)$, the induced metric on $H$ is
\be
 \mu_{pp} = h(\partial_p,\partial_p) = O(\Delta^2) \qquad \mu_{pq} = O(\Delta) \qquad \mu_{qq} = O(1)
\ee
and hence $\mu \equiv \det \mu_{AB} = O(\Delta^2)$. Combining these results we see that $\sqrt{\mu} R[\mu]$ extends continuously to the $A_3$ point, and has a finite discontinuity at the crease. Thus we can define the integral \eqref{GBRicci} by removing from $H$ a small region surrounding the $A_3$ point and crease, and then taking the limit as the size of this region is shrunk to zero, i.e., the integral exists as an improper Riemann integral. 
 
The terms $\sqrt{\mu} k^{ij}k_{ij}$ and $\sqrt{\mu} k^i_i k^j_j$ diverge as $1/\Delta$ at the $A_3$ point. However, this divergence is integrable:
\be
 \int dp \int^{2p^2}_{q_{\rm min}} dq \frac{1}{12p^2 - 2q} = \int dp [-\frac{1}{2} \log(12p^2-2q)]^{2p^2}_{q_{\rm min}} \sim \int dp \log |p|={\rm finite}
\ee
(recall we are only interested in integrability near the $A_3$ point $p=q=0$). Hence $\int_H d^2 x \sqrt{\mu} k^{ij}k_{ij}$ and $\int_H d^2x \sqrt{\mu} k^i_i k^j_j$ are also well-defined. 

These results hold for a horizon cross-section $\Sigma_\tau \cap \cH$ for a {\it generic} value of $\tau$. However, we saw in Section \ref{sec:A3} that for special values of $\tau$ an $A_3$ perestroika will occur. Assume this happens at $\tau=0$. Arnol'd shows that one can use a diffeomorphism that preserves \eqref{A3map} to bring the time function to the form $\tau = -y \pm w^2$ \cite{arnold:1976} (here we used the result from Section \ref{sec:A3} that $\partial_y \tau<0$ at the $A_3$ point to fix the sign of the $y$ term). As explained in Section \ref{sec:A3} we can use $(w,x,z)$ as coordinates on $\Sigma_\tau$. We can use $(p,r)$ as parameters on $H$ where $w=r$ and eliminating $y=q$ gives $q =q(\tau,r) = -\tau \pm r^2$ (still with $q \le 2p^2$). $x(p,r)$ and $z(p,r)$ are given by substituting $q=q(\tau,r)$ in \eqref{A3map}. We now have tangent vectors
\be
  \frac{\partial}{\partial p} = \Delta \left( \frac{\partial}{\partial x}+ p \frac{\partial}{\partial z} \right) \qquad \frac{\partial}{\partial r} = \frac{\partial}{\partial w}\mp 4 p r \frac{\partial}{\partial x} \mp 2p^2 r \frac{\partial}{\partial z}
\ee
with
\be
\Delta = 12p^2 -2q(\tau,r).
\ee
Repeating the calculations above now leads to
\be
 k_{ij} = -\frac{\alpha}{\Delta} m_i m_j + \ldots \qquad \qquad m = dx \pm 4pr dw.
\ee
We now have $R[\mu] = O(1/\Delta)$ and $\mu = O(\Delta^2)$ exactly as above, so $\sqrt{\mu} R[\mu]$ is continuous at the $A_3$ point and bounded at the crease and so its integral is well-defined. $\sqrt{\mu} k^{ij}k_{ij}$ and $\sqrt{\mu} k^i_i k^j_j$ still diverge as $1/\Delta$ but now this divergence is {\it not} integrable at $\tau=0$. For example, choose the lower sign and take $\tau>0$. The range of $r$ is unrestricted (as creases are absent for $\tau>0$ in this case) and we have
\be
\int dp dr \Delta^{-1} =  \int\frac{dp dr}{12p^2 +2 r^2 +2\tau} \sim \int_{R\ge 0} \frac{RdR}{R^2 + \tau} \sim \log \tau
\ee
which diverges as $\tau \rightarrow 0+$.

\end{document}